\documentclass[12pt]{article}

\usepackage{color,graphicx}
\usepackage{amsmath,amssymb}

  \renewenvironment{thebibliography}[1]{%
    \begin{oldthebibliography}{#1}%
      \setlength{\parskip}{0ex}%
      \setlength{\itemsep}{0ex}%
      \small
  }%
  {%
    \end{oldthebibliography}%
  }

\newcommand{\Q}{{\cal Q}}
\newcommand{\xB}{x_{\rm B}}
\newcommand{\GeV}{{\rm GeV}}
\newcommand{\cffF}{{\cal F}}
\newcommand{\cffH}{{\cal H}}
\newcommand{\cffE}{{\cal E}}
\newcommand{\cfftH}{\widetilde{\cal H}}
\newcommand{\cfftE}{\widetilde{\cal E}}
\newcommand{\cffbE}{\overline{\cal E}}
\newcommand{\cffFeff}{{\cal F}_{\rm eff}}
\newcommand{\cffHeff}{{\cal H}_{\rm eff}}
\newcommand{\cffEeff}{{\cal E}_{\rm eff}}
\newcommand{\cfftHeff}{\widetilde{\cal H}_{\rm eff}}
\newcommand{\cfftEeff}{\widetilde{\cal E}_{\rm eff}}
\newcommand{\cffFT}{{\cal F}_{\rm T}}
\newcommand{\cffHT}{{\cal H}_{\rm T}}
\newcommand{\cffET}{{\cal E}_{\rm T}}
\newcommand{\cfftHT}{\widetilde{\cal H}_{\rm T}}
\newcommand{\cfftET}{\widetilde{\cal E}_{\rm T}}
\newcommand{\cffreim}{sub-CFF}
\newcommand{\cffsreim}{sub-CFFs}

\font\cmss=cmss12 
\def\1{\hbox{{1}\kern-.25em\hbox{l}}}
\def\bfZ{\relax{\hbox{\cmss Z\kern-.4em Z}}}

\renewcommand\Re{\operatorname{\mathfrak{Re}}}
\renewcommand\Im{\operatorname{\mathfrak{Im}}}

\newcommand{\Asin}[2]{A^{\sin(#1\phi)}_{\rm #2}}
\newcommand{\Acos}[2]{A^{\cos(#1\phi)}_{\rm #2}}
\newcommand{\Asinsin}[2]{A^{\sin(\varphi)\sin(#1\phi)}_{\rm #2}}
\newcommand{\Asincos}[2]{A^{\sin(\varphi)\cos(#1\phi)}_{\rm #2}}
\newcommand{\Acossin}[2]{A^{\cos(\varphi)\sin(#1\phi)}_{\rm #2}}
\newcommand{\Acoscos}[2]{A^{\cos(\varphi)\cos(#1\phi)}_{\rm #2}}

\newcommand{\blue}[1]{{\color{blue}{#1}\color{black}}}

\newcounter{comment}

{\refstepcounter{comment}%
\begin{quote}
\ttfamily\small$\blacksquare$ \textbf{\underline{Comment} $\sharp$\thecomment:}}%
{\end{quote}}

\begin{document}
\begin{titlepage}

\centerline{\large \bf HERMES impact for the access of}
\centerline{\large \bf Compton form factors}

\vspace{10mm}

\centerline{K.~Kumeri\v{c}ki$^a$,
            D.~M\"{u}ller$^{b}$, and
            M.~Murray$^c$}

\vspace{8mm}
\centerline{\it $^a$  Department of Physics, University of Zagreb}
\centerline{\it HR-10002 Zagreb, Croatia}

\vspace{4mm}
\centerline{\it $^b$Institut f\"ur Theoretische
Physik II, Ruhr-Universit\"at Bochum}
\centerline{\it  D-44780 Bochum, Germany}

\vspace{4mm}
\centerline{\it $^c$ School of Physics and Astronomy, University of Glasgow}
\centerline{\it G128QQ Glasgow, Scotland, UK}

\vspace{25mm}

\centerline{\bf Abstract}
\vspace{4mm}

\noindent
We utilize the DVCS asymmetry measurements of the H{\sc ermes} collaboration for
access to Compton form factors in the deeply virtual regime and to generalized parton distributions. In particular, the (almost) complete measurement of DVCS observables allows us to map various asymmetries into the space of Compton form factors, where we still rely in this analysis on dominance of twist-two associated Compton form factors. We compare this one-to-one map with local Compton form factor fits and a model dependent global fit.

\vspace{3cm}

\noindent

\vspace*{12mm}


\end{titlepage}

\newpage
\tableofcontents
\newpage

\section{Introduction}

As is well known, at the beginning of the last century Compton
scattering played a crucial role in the debate on the dual nature of
light. Namely,  if light scatters off an electron, energy is
transferred to the electron and so the wavelength of the scattered
light increases. In addition to this Compton effect
revealing the particle aspect of light \cite{Compton:1923zz},
the Compton scattering process has now many applications in material
science, medicine, astro- and particle physics, where in particular,
it is utilized to probe the electron wave function in matter. In
hadronic physics the Compton scattering process is used to reveal
static properties of the nucleon in terms of electric and magnetic
polarizabilities. Its generalization into virtual Compton
scattering \cite{Guichon:1995pu} offers a supplementing window
to electromagnetic form factor studies, which have been intensively performed in the last six decades.  At larger photon virtualities $\Q^2 \gtrsim 1 \GeV^2$, the Compton scattering process probes the partonic content of the nucleon and gives access to the so-called generalized parton distributions (GPDs) \cite{Mueller:1998fv,Radyushkin:1996nd,Ji:1996nm}.  In particular, it was suggested
to measure by means of deeply virtual Compton scattering (DVCS) the quark
orbital angular momentum \cite{Ji:1996ek} and to access the transverse
distributions of partons \cite{Burkardt:2000za,Ralston:2001xs}. With these goals in mind, many
experimental and theoretical activities were started in the field of hard exclusive processes, where we consider the measurements of (DV)CS observables as important as the measurements of electromagnetic form factors.

Measurements of DVCS have taken place at H{\sc era} and Jefferson Lab
since the turn of the millennium, with the aim of understanding the
decomposition of the nucleon spin. The first observed DVCS candidate
events at high energy were announced by the Z{\sc eus} collaboration
using a collider experiment at H{\sc era} in 2000 \cite{Saull:1999kt} and few months later for fixed target kinematics \cite{Amarian_2000vx}.  The
first DVCS measurements in fixed target experiments were
published simultaneously by the H{\sc ermes} and C{\sc las}
collaborations a year later \cite{Airapetian:2001yk,Stepanyan:2001sm}.
Both experiments measured the beam spin asymmetry arising from the
interference term between DVCS and elastic scattering with bremsstrahlung in the scattering amplitude. The latter process, called the
Bethe-Heitler (BH) process, has the same initial and final states as DVCS
($ep\rightarrow ep\gamma$) and at fixed target kinematics typically the interference term in the scattering amplitude is more experimentally accessible than the pure DVCS term \cite{Airapetian:2008aa}. Both H{\sc ermes} and C{\sc las} utilized the fact that the
large BH contribution in the interference term at $\Q^2\sim 2\, \GeV^2$ and $-t\sim 0.1\, \GeV^2$ amplifies the contribution from the more interesting DVCS process.

Since then, large experimental effort has been expended to
measure various observables in the electroproduction of
photons and mesons. The phenomenological challenge is now to
describe these data in terms of GPDs, which requires a flexible model.  Thereby,
we meet a more elementary problem, in particular for DVCS: the
number of Compton form factors (or helicity amplitudes) is usually larger than
the number of observables at a given kinematical point. One
must therefore rely on model assumptions or hypotheses
which means that, independently of the applied method or
framework, a theoretical bias cannot be avoided in analyzing
the present available world data set. This may even lead to a qualitative
misinterpretation of the data, where the apparent influence of any
particular GPD model set may be determined more by the initial
theoretical approach to the problem than by the observed data.

In this paper we will concentrate on the results from the H{\sc ermes} collaboration,
which had both electron and positron beams available and is currently the
experiment that has delivered the most complete set of DVCS observables.
In Sect.~\ref{sec:HERMES-obs} we introduce the DVCS observables and
we give details on the H{\sc ermes} experiment. In
Sect.~\ref{sec:analyze} we consider  the extraction of  CFFs at
given kinematical points from the H{\sc ermes} measurements as a map of random variables. Additionally, we utilise the regression approach
and use the H{\sc ermes} data to access CFFs by least squares fitting.  We also present a global GPD model fit
that additionally includes H{\sc era} collider and
Jefferson Lab measurements and conclude on what we have learned. Finally, we summarize and provide an outlook.

\section{HERMES measurements of DVCS observables}
\label{sec:HERMES-obs}
\subsection{Definitions of photon leptoproduction observables}
\label{sec:observables}

\begin{figure}[t!]
\begin{center}
\includegraphics[width=14cm]{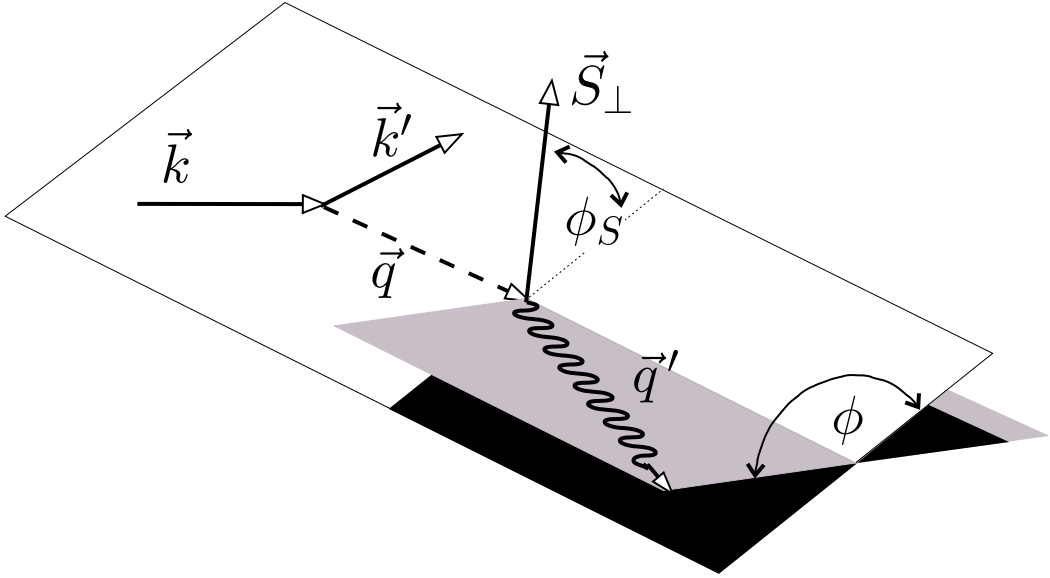}
\end{center}
\caption{\small Reference frame adopted by the H{\sc ermes} collaboration \cite{Airapetian:2011uq}.}
\label{fig:frame-Trento}
\end{figure}

The DVCS process enters as a subprocess in deeply virtual photon leptoproduction and its amplitude can be decomposed in terms of twelve independent helicity amplitudes given in some reference frame. Alternatively, we might parameterize the DVCS amplitude in Lorentz-invariant Compton form factors (CFFs), which are defined in analogy to GPD definitions and are called \cite{Belitsky:2001ns}:
\begin{eqnarray}
\label{cffFs}
\cffF \in \left\{\cffH,\cffE,\cfftH,\cfftE\right\}\,,
\quad
\cffFeff \in \left\{\cffHeff,\cffEeff,\cfftHeff,\cfftEeff\right\}\,,
\quad
\cffFT \in \left\{\cffHT,\cffET,\cfftHT,\cfftET\right\}\,,
\end{eqnarray}
where we separated them into twist-two related CFFs $\cffF$,
twist-three related CFFs $\cffFeff$, and transversity CFFs $\cffFT$.
Unfortunately, different conventions, which differ by power suppressed
contributions, are used in the literature. As noted previously,
the DVCS subprocess is accompanied by the Bethe-Heitler bremsstrahlung process, the amplitude
of which is, at leading order in the QED fine structure constant,
real-valued. Moreover, it is entirely parameterized in terms of the
electromagnetic form factors, which for the nucleon are well-known for
the kinematics of interest. The interference term of both subprocesses
is charge-odd and depends linearly on the CFFs, while, the BH-squared,
depending only on electromagnetic form factors, and DVCS-squared,
given as bilinear form of CFFs, amplitudes are charge-even.  The
differential cross section, for the most general setup, is five-fold
and we write it in terms of the Bjorken scaling variable $\xB$, the
negative virtuality squared of the intermediate photon $\Q^2=-q^2$, the momentum transfer square in the $t$-channel, and two relative azimuthal angles $\phi$, between  lepton scattering plane and reaction plane, and  $\varphi$, between  lepton scattering plane and transverse spin polarization vector, as in \cite{Belitsky:2001ns}:
\begin{eqnarray}
\label{dX^eN2eNgamma}
\frac{d^5\sigma}{d\xB d\Q^2 dt d\phi d\varphi } = \frac{\alpha^3 \xB y^2}{16\pi^2 \Q^4 \sqrt{1+ \frac{4\xB^2 M_p^2}{\Q^2}}} \left[
\frac{|{\cal T}_{\rm BH}|^2}{e^6} + \frac{{\cal I}(\cffF)}{e^6} + \frac{|{\cal T}_{\rm DVCS}|^2(\cffF^\ast,\cffF)}{e^6}
\right]\!\!(\xB,\Q^2,t, \phi, \varphi )\,,
\nonumber\\
\end{eqnarray}
where $\alpha=e^2/4\pi$ is the electromagnetic fine structure constant and $M_p$ is the proton mass.
Here, and in the following, we neglect the electron mass and define the azimuthal angle in an reference frame where the photon travels along the $z$-axis and the
$x$-component of the incoming electron is positive, see Fig.~\ref{fig:frame-Trento}.
This frame has been adopted by the H{\sc ermes} collaboration.
Compared to the BMK conventions in \cite{Belitsky:2001ns},
in which the photon travels in opposite direction of the $z$-axis, for
the frame used by H{\sc ermes} we have the relations
\begin{eqnarray}
\phi^{\mbox{\tiny BMK}} \equiv \phi^{\mbox{\tiny BMK}}_N-\phi^{\mbox{\tiny BMK}}_e  = \pi-\phi
\,,
&&
\phi
\equiv \phi_\gamma
-\phi_e
\\
\varphi^{\mbox{\tiny BMK}} \equiv \Phi^{\mbox{\tiny BMK}}-\phi^{\mbox{\tiny BMK}}_N  =  \varphi
-\pi\,,
&&
\varphi
= \phi_\gamma
-\phi_{\rm S}
\,,
\end{eqnarray}
where $\phi_N$, $\phi_\gamma = \phi_N + \pi$, $\phi_e =0$, and $\Phi \equiv \phi_{\rm S}$  are the azimuthal angles of the nucleon, photon, electron, and the transverse polarization vector, respectively.

The Fourier expansion of the various amplitude squares may be written  in analogy to the notation of \cite{Belitsky:2001ns} as
\begin{eqnarray}
\label{|T_{BH}|^2}
|{\cal T}_{\rm BH}|^2 &\!\!\!=\!\!\!& \frac{e^6}{-t\, \xB^2\, w(\phi)}
\left\{\hat{c}_0^{\rm BH} +\sum_{n=1}^2 \hat{c}_n^{\rm BH} \cos(n \phi) + \hat{s}_1^{\rm BH} \sin\phi\right\}\,,
\\
\label{T_{I}}
{\cal I}&\!\!\!=\!\!\!& \frac{\pm e^6}{-t\,\xB^2\, w(\phi)}
\left\{ \hat{c}_0^{\cal I} + \sum_{n=1}^3 \left[ \hat{c}_n^{\cal I} \cos(n \phi)+\hat{s}_n^{\cal I} \sin(n \phi) \right]\right\}\,,
\\
\label{|T_{DVCS}|^2}
|{\cal T}_{\rm DVCS}|^2 &\!\!\!=\!\!\!& \frac{e^6}{y^2 \Q^2\, \xB^2} \left\{ \hat{c}_0^{\rm DVCS} + \sum_{n=1}^2\left[
\hat{c}_n^{\rm DVCS} \cos(n \phi) + \hat{s}_n^{\rm DVCS}\right]\right\}\,.
\end{eqnarray}
Here, the $+$ $(-)$ overall sign of the interference term (\ref{T_{I}}) refers to an electron (positron) beam,
$1/w(\phi) \propto 1/{\cal P}_1(\phi){\cal P}_2(\phi)$ is the product of  scaled BH
propagators given in \cite{Belitsky:2001ns},  which is expanded in the even harmonics of $\phi$ up the second order,
\begin{eqnarray}
w(\phi) = 1 + w_1 \cos(\phi) + w_2 \cos(2\phi)\,.
\end{eqnarray}
Note that we included a generic kinematical overall factor $\sqrt{-t/(y^2 \Q^2)}$ from the interference term (\ref{T_{I}}) in the definition of the Fourier coefficients. Moreover, if the differential cross section (\ref{dX^eN2eNgamma}) is
weighted with $w(\phi)$, its Fourier
expansion w.r.t.~the azimuthal angles becomes finite.

Having both electrons and positrons with which to make measurements, one can access the interference and DVCS-squared terms by means of cross section differences and sums
\begin{eqnarray}
d\sigma_I(\phi,\varphi) \stackrel{{\rm LO}_{\rm QED}}{\equiv} d\sigma_{\rm odd}(\phi,\varphi)= \frac{1}{2}\left[ d\sigma_+ - d\sigma_-\right](\phi,\varphi)\,,
\quad
d\sigma_{\rm even}(\phi,\varphi) = \frac{1}{2}\left[ d\sigma_+ + d\sigma_-\right](\phi,\varphi)\,.
\end{eqnarray}
Measuring such charge-odd cross sections with different lepton
and nucleon polarizations, one can, in principle, reach an almost
complete decomposition in terms of the twelve CFFs
(\ref{cffFs}). Namely, in the deeply virtual regime, the first and
second order harmonics of the weighted differential cross section
differences are dominated by linear combinations of twist-two and
twist-three related CFFs $\cffF$ and $\cffFeff$ respectively, while
the third harmonic is related to the transversity CFFs $\cffFT$. In the
first two cases the number of possible observables allows access to both the imaginary and real parts of the corresponding form factors, while in the latter case
only one combination of $\Re \cffFT$ and three combinations of $\Im \cffFT$ can be accessed. Moreover, in this approximation the constant terms are governed only by twist-two associated CFFs $\cffF$ and, hence, such measurements allow for an experimental consistency check of the underlying formalism, given in some approximation.

The charge-even combination may serve also for an experimental consistency check or may be used to access the desired CFFs in a different manner. Most importantly, in single spin-flip experiments, the BH term drops out in the considered order of QED and so such measurements give direct access to bilinear CFF combinations. In double spin-flip experiments one needs to subtract the BH-squared term, which may be feasible at small $\xB$ where an effective `pomeron' behavior
may overwhelm the contributions of the BH subprocess.

Thus far, only H{\sc era} experiments at D{\sc esy} had both charges
of lepton beams at hand. The H{\sc ermes} experiment achieved
measurements with polarized electron and positron beams for
longitudinal and transversal target spin polarizations and thus the following asymmetries where extracted.
\begin{itemize}
\item Single beam-spin asymmetries in the charge-odd and charge-even sectors:
\begin{eqnarray}
\label{A_LUI}
A_{\rm LU,I}(\phi) &\!\!\!=\!\!\!&
\frac{
\left[d\sigma^{\rightarrow}_+-d\sigma^{\leftarrow}_+\right]- \left[d\sigma^{\rightarrow}_- -d\sigma^{\leftarrow}_-\right]
}{
d\sigma^{\rightarrow}_+ +d\sigma^{\leftarrow}_+ +d\sigma^{\rightarrow}_- +d\sigma^{\leftarrow}_-
}\,,
\\
\label{A_LUDVCS}
A_{\rm LU,DVCS}(\phi) &\!\!\!=\!\!\!&
\frac{
\left[d\sigma^{\rightarrow}_+-d\sigma^{\leftarrow}_+\right] +\left[ d\sigma^{\rightarrow}_- -d\sigma^{\leftarrow}_-\right]
}{
d\sigma^{\rightarrow}_+ +d\sigma^{\leftarrow}_+ +d\sigma^{\rightarrow}_- +d\sigma^{\leftarrow}_-
}\,,
\end{eqnarray}
where $\rightarrow$ ($\leftarrow$) denotes electron ($-$) or positron ($+$) polarization along (opposite to) the beam direction.
\item Beam-charge asymmetry:
\begin{eqnarray}
\label{A_C}
A_{\rm C}(\phi) = \frac{d\sigma_+-d\sigma_-}{d\sigma_+ + d\sigma_-}\,.
\end{eqnarray}
\item Single and double longitudinal target-spin asymmetries with a positron beam:
\begin{eqnarray}
\label{A_UL+}
A_{{\rm UL},+}(\phi)  &\!\!\!=\!\!\!&
\frac{
d\sigma^{\Leftarrow}_+-d\sigma^{\Rightarrow}_+
}{
d\sigma^{\Leftarrow}_+ + d\sigma^{\Rightarrow}_+
}\,,
\\
\label{A_LL+}
A_{{\rm LL},+}(\phi) &\!\!\!=\!\!\!& \frac{
\left[d\sigma^{\leftarrow\Rightarrow}_+ + d\sigma^{\rightarrow\Leftarrow}_+\right]
-
\left[d\sigma^{\rightarrow\Rightarrow}_+ + d\sigma^{\leftarrow\Leftarrow}_+\right]
}{
d\sigma^{\leftarrow\Rightarrow}_+ + d\sigma^{\rightarrow\Leftarrow}_+
+
d\sigma^{\rightarrow\Rightarrow}_+ + d\sigma^{\leftarrow\Leftarrow}_+
}\,.
\end{eqnarray}
Note that, contrary to H{\sc ermes} notation, $\Leftarrow$
($\Rightarrow$) denotes proton polarization opposite to (along) the positron beam direction.
\item Single transverse target-spin asymmetries in the charge-odd and charge-even sectors:
\begin{eqnarray}
\label{A_UTI}
A_{\rm UT, I}(\phi,\varphi) &\!\!\!=\!\!\!&
\frac{
\left[d\sigma^{\Uparrow}_+-d\sigma^{\Downarrow}_+\right]-\left[d\sigma^{\Uparrow}_- -d\sigma^{\Downarrow}_-\right]
}{
d\sigma^{\Uparrow}_+ +d\sigma^{\Downarrow}_+ +d\sigma^{\Uparrow}_- +d\sigma^{\Downarrow}_-
}\,,
\\
\label{AUT_DVCS}
A_{\rm UT, DVCS}(\phi,\varphi) &\!\!\!=\!\!\!&
\frac{
\left[d\sigma^{\Uparrow}_+-d\sigma^{\Downarrow}_+\right] + \left[ d\sigma^{\Uparrow}_- -d\sigma^{\Downarrow}_- \right]
}{
d\sigma^{\Uparrow}_+ +d\sigma^{\Downarrow}_+ +d\sigma^{\Uparrow}_- +d\sigma^{\Downarrow}_-
}\,.
\end{eqnarray}
\item Double longitudinal beam and transverse target-spin asymmetries in the charge-odd and charge-even sectors:
\begin{eqnarray}
\label{A_LTI}
A_{\rm LT, I}(\phi,\varphi) &\!\!\!=\!\!\!&
\frac{
\left[d\sigma^{\rightarrow\Uparrow}_+ +d\sigma^{\leftarrow\Downarrow}_+ \right]
-\left[d\sigma^{\rightarrow\Downarrow}_++d\sigma^{\leftarrow\Uparrow}_+ \right]
\!-\!
\left[d\sigma^{\rightarrow\Uparrow}_- +d\sigma^{\leftarrow\Downarrow}_- \right]
+\left[d\sigma^{\rightarrow\Downarrow}_- +d\sigma^{\leftarrow\Uparrow}_- \right]
}{
d\sigma^{\rightarrow\Uparrow}_+ + d\sigma^{\leftarrow\Downarrow}_+ + d\sigma^{\rightarrow\Downarrow}_+
+
d\sigma^{\leftarrow\Uparrow}_+  + d\sigma^{\leftarrow\Downarrow}_-
+
d\sigma^{\rightarrow\Uparrow}_- + d\sigma^{\rightarrow\Downarrow}_-
+
d\sigma^{\leftarrow\Uparrow}_-
},
\nonumber\\
\\
\label{ALTBHDVCS}
A_{\rm LT,even}(\phi,\varphi) &\!\!\!=\!\!\!&
\frac{
\left[d\sigma^{\rightarrow\Uparrow}_+ +d\sigma^{\leftarrow\Downarrow}_+ \right]
-\left[d\sigma^{\rightarrow\Downarrow}_++d\sigma^{\leftarrow\Uparrow}_+ \right]
+
\left[d\sigma^{\rightarrow\Uparrow}_- +d\sigma^{\leftarrow\Downarrow}_- \right]
-\left[d\sigma^{\rightarrow\Downarrow}_- +d\sigma^{\leftarrow\Uparrow}_- \right]
}{
d\sigma^{\rightarrow\Uparrow}_+ + d\sigma^{\leftarrow\Downarrow}_+ + d\sigma^{\rightarrow\Downarrow}_+
+
d\sigma^{\leftarrow\Uparrow}_+  + d\sigma^{\leftarrow\Downarrow}_-
+
d\sigma^{\rightarrow\Uparrow}_- + d\sigma^{\rightarrow\Downarrow}_-
+
d\sigma^{\leftarrow\Uparrow}_-
}.
\nonumber\\
\end{eqnarray}
\end{itemize}
All these observables can be theoretically expressed in terms of
weighted cross sections. However, since the denominator given by the
unpolarized cross section depends on the azimuthal angle $\phi$, they
are given by a series of $\phi$ harmonics rather than finite sums.  Nevertheless, in the kinematics where the
BH process dominates, the zeroth and first harmonics of these
asymmetries are roughly determined by the linear combinations of
twist-two associated CFFs $\cffF$ that enter the interference
term. Note also that, except for the longitudinal target spin
asymmetries (\ref{A_UL+}) and (\ref{A_LL+}), the denominators are
expressed by the charge-even cross section. Utilizing the charge
asymmetry (\ref{A_C}), we find in general that an asymmetry measured
with a positron (or electron) beam can be expressed by charge-odd (interference
term) and charge-even (a possible squared BH term plus a DVCS-squared
term) expression:
\begin{eqnarray}
\label{A_UL-relations}
A_{{\cdots,+}}(\phi) = \frac{A_{\rm \cdots,I}(\phi)}{1+ A_{\rm C}(\phi)} + \frac{A_{{\rm \cdots,BH}}(\phi)+ A_{{\rm \cdots,DVCS}}(\phi)}{1+ A_{\rm C}(\phi)}\,.
\end{eqnarray}
Assuming that the BH amplitude overwhelms the DVCS one, we may drop the DVCS induced part and we  approximately have
\begin{eqnarray}
\label{A_UL+-2-A_ULI}
A_{\rm \cdots,I}(\phi) \approx A_{{\cdots,+}}(\phi)\left[1+ A_{\rm C}(\phi)\right] - A_{{\rm \cdots,BH}}(\phi)\,,
\end{eqnarray}
where the BH-associated asymmetry $A_{\rm \cdots,BH}(\phi)$ drops out in single-spin asymmetries.
We add that the double-spin asymmetry in the charge-odd sector is defined in terms of cross sections as
\begin{eqnarray}
\label{A_LLI}
A_{\rm LL,I}(\phi) &\!\!\!=\!\!\!&\frac{
\left[d\sigma^{\rightarrow\Rightarrow}_+ + d\sigma^{\leftarrow\Leftarrow}_+\right] -
\left[d\sigma^{\leftarrow\Rightarrow}_+ + d\sigma^{\rightarrow\Leftarrow}_+\right]-
\left[d\sigma^{\rightarrow\Rightarrow}_- + d\sigma^{\leftarrow\Leftarrow}_-\right] +
\left[d\sigma^{\leftarrow\Rightarrow}_- + d\sigma^{\rightarrow\Leftarrow}_-\right]
}{
d\sigma^{\rightarrow\Rightarrow}_+ + d\sigma^{\leftarrow\Leftarrow}_+ + d\sigma^{\leftarrow\Rightarrow}_+ + d\sigma^{\rightarrow\Leftarrow}_+
+
d\sigma^{\rightarrow\Rightarrow}_- + d\sigma^{\leftarrow\Leftarrow}_- + d\sigma^{\leftarrow\Rightarrow}_- + d\sigma^{\rightarrow\Leftarrow}_-
}\,.
\nonumber\\
\end{eqnarray}

Let us finally remind the reader here that
asymmetries are expanded in an infinite Fourier series. Hence, the
method for the extraction of CFFs, outlined above for cross section
differences, has to be modified.  For illustration let us consider
here the lepton beam-spin (\ref{A_LUI}) and beam-charge  (\ref{A_C})
asymmetries that offer access to the imaginary and real parts of
CFFs respectively. Substituting the cross section
(\ref{dX^eN2eNgamma}) and the harmonic expansions
(\ref{|T_{BH}|^2},\ref{T_{I}},\ref{|T_{DVCS}|^2}) into (\ref{A_LUI})
and (\ref{A_C}), we find the azimuthal angle dependencies of the
charge-odd electron beam spin and charge asymmetry respectively:
\begin{eqnarray}
\label{A_{LU,I}(phi)-spin0}
A_{\rm LU,I}(\phi) =\frac{\hat{s}^{\cal I}_1 \sin(\phi) + \hat{s}^{\cal I}_2 \sin(2\phi)}{
\sum_{n=0}^2 \hat{c}^{\rm BH}_n \cos(n \phi) + \frac{-t}{y^2\Q^2} w(\phi) \sum_{n=0}^2 \hat{c}^{\rm DVCS}_n  \cos(n \phi) },
\phantom{\Bigg|}\\
\label{A_{C}(phi)-spin0}
A_{\rm C}(\phi) =\frac{\hat{c}^{\cal I}_0+\hat{c}^{\cal I}_1 \cos(\phi) + \hat{c}^{\cal I}_2 \cos(2\phi)+ \hat{c}^{\cal I}_3 \cos(3\phi)}{
\sum_{n=0}^2 \hat{c}^{\rm BH}_n \cos(n \phi) + \frac{-t}{y^2\Q^2} w(\phi) \sum_{n=0}^2 \hat{c}^{\rm DVCS}_n  \cos(n \phi) },
\phantom{\Bigg|}
\end{eqnarray}
where the odd and even Fourier coefficients $\hat{s}_n^{\cal I}$ and $\hat{c}_n^{\cal I}$  of the interference term are linear functions of the imaginary and real parts of the CFFs and the even DVCS Fourier coefficients $\hat{c}^{\rm DVCS}_n $ are bilinear in all CFFs.  In the $1/\Q$ expansion, the first harmonics of the interference term
dominate and are governed by the CFFs $\cal F$, the second harmonics are relatively suppressed by $\sqrt{(t_{\rm min}-t)/\Q^2}$ and determined by the CFF ${\cal F}_{\rm eff}$
while the third harmonic of the charge asymmetry (\ref{A_{C}(phi)-spin0}) is given by the CFFs ${\cal F}_{\rm T}$ and, to leading-twist accuracy, is predicted by the gluon transversity GPD. In reality, however,
all harmonics of the interference term are functions of all twelve CFFs
\cite{Belitsky:2010jw,Belitsky:2012ch}.  Note that the third odd harmonic is absent and hence
the imaginary part of the transversity CFF combination cannot be accessed from
the interference term. In principle, this missing information is contained in
the DVCS-squared term.
Besides the mixing of the various CFFs for a given harmonic, the projection
\begin{eqnarray}
\label{Asin{n}{LU,I}}
\Asin{n}{LU,I} &\!\!\! = \!\!\!& \frac{1}{\pi}\int_{-\pi}^\pi\! d\phi\, \sin(n \phi) A_{\rm LU,I}(\phi)\,,
\quad
\\
\label{Acos{n}{C}}
\Acos{n}{C}  &\!\!\! = \!\!\!& \frac{1}{\pi}\int_{-\pi}^\pi\! d\phi\,  \cos(n \phi) A_{\rm C}(\phi)\;\; \mbox{for}\;\;  n > 0\,,  \qquad
\Acos{0}{C} = \frac{1}{2\pi}\int_{-\pi}^\pi\! d\phi\,   A_{\rm C}(\phi)\,,
\end{eqnarray}
yields an additional contamination due to the $\phi$ dependence in the denominator.

\subsection{Experimental details}
\label{sec:HERMES}

Exclusive photon events  at H{\sc ermes} were selected if having
exactly one photon and one lepton track detected within the acceptance of the spectrometer.
The event selection was subject to the kinematic constraints
$$1\, \GeV^{2} < \Q^2 < 10\, \GeV^{2}\,, \; 0.03 < \xB < 0.35\,,\;
-t < 0.7\, \GeV^2\,, \;  W^2 > 9 \, \GeV^2\,, \; \mbox{and} \;  \nu < 22\, \GeV\,,$$
where $W$ is the invariant mass of the
$\gamma^{*}p$ system and $\nu$ is the energy of the virtual photon in the target
rest frame. The polar angle between the directions of the virtual and real photons was required to be within the limits 5~mrad $<$ $\theta_{\gamma^{*}\gamma}$ $<$ 45~mrad.

An event sample was selected requiring
that the squared missing-mass $M_{\textrm{X}}^{2}=(q+M_{p}-q')^{2}$
of the $e\,p \rightarrow\, e'\,\gamma\, \textrm{X}$ measurement
corresponded to the square of the proton mass, $M_{p}$, within the limits of the
energy resolution of the H{\sc ermes} spectrometer (mainly the calorimeter). Recall that $q$ is the
four-momentum of the virtual photon, $p$ is the initial four-momentum
of the target proton and $q'$ is the four-momentum of the produced
photon. The ``exclusive region'' was defined as $-$($1.5$\,GeV)$^{2} <
M_{\textrm{X}}^{2}$ $<$ (1.7\,GeV)$^{2}$. This exclusive region was shifted by up to
0.17\,GeV$^{2}$ for certain subsets of the data in order to reflect
observed differences in the distributions of the electron and positron
data samples. A systematic uncertainty contribution is assigned for
this effect. This event sample selection technique was used as it
allows for the most complete set of DVCS observable measurements to be
considered; although results using the measurement of the recoiling
proton have been released by H{\sc ermes} \cite{Airapetian:2012pg},
this technique is only currently available for beam-spin DVCS
measurements and was never utilised with a polarised target. Other systematic uncertainty contributions arise from potential misalignment of the spectrometer, acceptance and smearing effects, and the inclusion in the data set from misidentified semi-inclusive deep inelastic scattering events. The latter is mitigated somewhat using a correction procedure. The sizes of the systematic uncertainties are estimated using Monte Carlo techniques.

The data sample in the exclusive region contains events not only involving the production of real photons in which the proton remains intact, but also
events involving the excitation of the target proton to a $\Delta^+$
resonant state (``associated production''). This is a consequence of
using a missing-mass event sample selection technique as noted
above; the calorimeter resolution for measuring the produced photon does not allow separation of the resonant events from the rest of the data sample.
No systematic uncertainty is assigned for the contribution from these
events; they are treated as part of the signal. A Monte Carlo
calculation based on the parameterisation from ref.~\cite{Brasse:1976bf} is
used to estimate the fractional contribution to the event sample from resonant
production in each kinematic bin; the uncertainty on this estimate
cannot be adequately quantified because no sufficiently precise measurements have been made in the H{\sc ermes} kinematic region. The results of the estimate, called the associated fractions and labelled ``Assoc. fraction'', are shown in refs.~\cite{Airapetian:2008aa,Airapetian:2011uq,Airapetian:2009aa,Airapetian:2010ab,Airapetian:2012mq}. The method used to perform this estimation is described in detail in ref.~\cite{Airapetian:2008aa}.

\begin{figure}[t]
\begin{center}
\includegraphics[width=17cm]{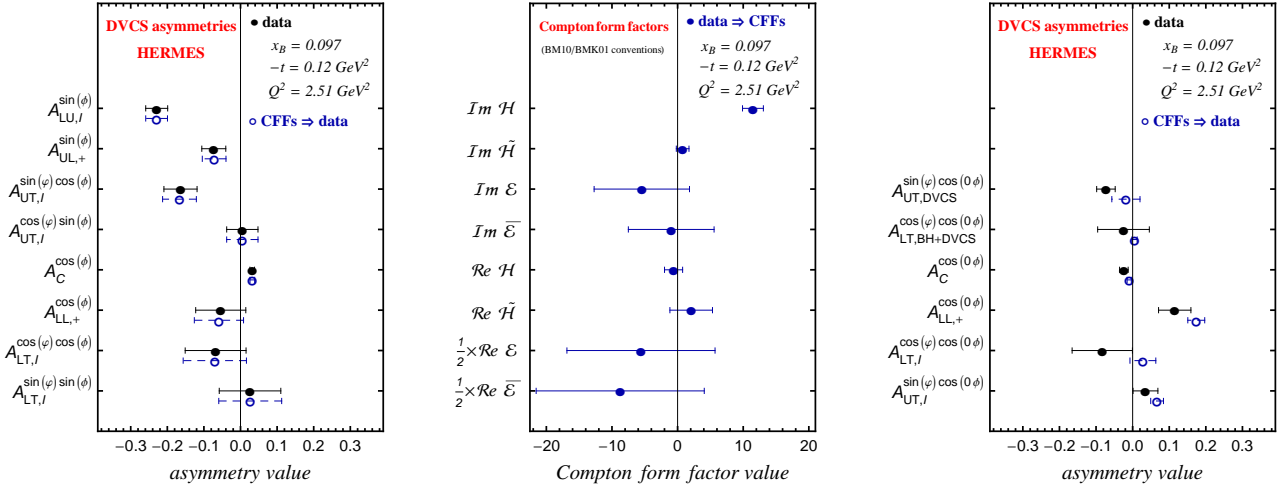}
\end{center}
\vspace{-4mm}
\caption{\small H{\sc ermes} measurements (solid circles) of interference dominated asymmetries in the twist-two sector (left) and BH/DVCS-squared dominated  as well as charge odd twist-tree  asymmetries (right) at the overall mean values $\xB=0.097$, $t=-0.119\,\GeV^2$, and $\Q^2=2.51\,\GeV^2$. In the middle panel  the resulting CFFs from a linearized  map of the interference dominated asymmetries are displayed. The empty circles, shown in the left and right panels, arise from the inverse linearized map of extracted CFFs back to asymmetries.}
\label{fig:A_...-HERMES}
\end{figure}

\begin{table}[t]
\begin{center}
\begin{tabular}{|c||c|c|c|c||c|c|c|c||c|c|c|c|}
  \hline
   bin no. & 1 & 2 & 3 & 4 & 5 & 6 & 7 & 8 & 9 & 10 & 11 & 12 \\
   \hline
  $-t\; [\GeV^2]$ &0.03& 0.1\phantom{0}& 0.2\phantom{0}& 0.42& 0.1\phantom{0}& 0.1\phantom{0}& 0.13& 0.2\phantom{0}& 0.08& 0.1\phantom{0}& 0.13& 0.19 \\
  $\phantom{-}\xB\phantom{[\GeV^2]}$  & 0.08& 0.1\phantom{0}& 0.11& 0.12& 0.05& 0.08& 0.12& 0.2\phantom{0}& 0.06& 0.08& 0.11& 0.17 \\
  $\phantom{-}\Q^2\;  [\GeV^2]$  & 1.9\phantom{0}& 2.5\phantom{0}& 2.9\phantom{0}& 3.5\phantom{0}& 1.5\phantom{0}& 2.2\phantom{0}& 3.1\phantom{0}& 5.0\phantom{0}& 1.2\phantom{0}& 1.9\phantom{0}& 2.8\phantom{0}& 4.9\phantom{0} \\
  \hline
\end{tabular}
\end{center}
\vspace{-2mm}
\caption{\label{tab:means}\small
Kinematical values of three times four H{\sc ermes} bins from ref.~\cite{Airapetian:2008aa}, labeled as $\#1,\cdots,\#12$.
}
\end{table}
The H{\sc ermes} measurements were presented in terms of four or six
bins times three one dimensional projection in $-t$, $\xB$, and
$\Q^2$ (see Tab.~\ref{tab:means}). The projections in
the kinematic variables are each correlated; the $\Q^2$ and
$x_\mathrm{B}$ projections are very highly correlated as a consequence
of the experimental design. For our analyses we employ the four bin
data, where the four-binned version of the six-binned data for the
beam spin and charge asymmetries from the H{\sc ermes} measurements,
published in ref.~\cite{Airapetian:2012mq}, were provided to us by the
H{\sc ermes} collaboration.
We neglect the small differences in the kinematical
means of different asymmetry measurements, which are very much within
the experimental uncertainty and take the values that are
listed in Tab.~\ref{tab:means}, labeled as bin $\#1,\cdots,
\#12$. Alternatively,  we take the kinematical values that are given
in the publications, assuming that the asymmetry values do not
vary within the small changes in the mean kinematic values.
The selected set of {\em fourteen} observables out of the {\em thirty-four} asymmetry measurements,  which we will use, are shown
in Fig.~\ref{fig:A_...-HERMES} as solid circles at the overall mean values $\xB=0.097$, $t=-0.119\,\GeV^2$, and $\Q^2=2.51\,\GeV^2$.

\section{Analysis of HERMES data}
\label{sec:analyze}

In Sect.~\ref{sec:analyze-cffs2obs} we first illustrate which information one may extract from the H{\sc ermes} measurements, cf.~Fig.~\ref{fig:A_...-HERMES}, if one assumes that effective twist-three and transverse photon helicity flip effects are absent. Employing the BMK formulae set \cite{Belitsky:2001ns}, we will also show that data are not in contradiction with this hypothesis.  The outline of the remainder is as following: in Sect.~\ref{sec:regression-method} we give a review of the methods used to analyze DVCS data and introduce the (non-)linear mapping method. In Sect.~\ref{sec:regression} this method
is then applied to the H{\sc ermes} data and  in
Sect.~\ref{sec:CFFfit} we analyze the same data with local CFF
fits. Finally, in Sect.~\ref{sec:global} we present a global CFF fit
with a simple GPD model used previously in global fits to DVCS data
off unpolarized protons. Finally, we discuss the implications of our analyses for GPD model building.

\subsection{Asymmetries in terms of CFFs}
\label{sec:analyze-cffs2obs}

Relying on the dominance of the BH-squared term over the DVCS-squared term and approximating the BH-squared term by its dominant constant harmonic, the following linear combinations of CFFs enter in the various asymmetries (here we set the minimal value of $-t$ to zero):
\begin{itemize}
\item The first harmonics of single beam spin  (\ref{A_LUI}) and beam charge (\ref{A_C}) asymmetries,
\begin{eqnarray}
\label{A_LUI^sinphi-BMK}
\Asin{1}{LU,I} \propto -\frac{y\sqrt{1-y}}{2-2y+y^2}\sqrt{\frac{-t}{y^2 \Q^2}}\times \xB\Im \left[
{\cal C}^{\rm I}_{\rm unp}(\cffF)
 +\cdots
\right],
\\
\label{A_C^cosphi-BMK}
\Acos{1}{C} \propto -\frac{\sqrt{1-y}}{2-y} \sqrt{\frac{-t}{y^2 \Q^2}} \times \xB\Re \left[{\cal C}^{\rm I}_{\rm unp}(\cffF) +\cdots
\right],
\end{eqnarray}
are {\it approximately} given by the imaginary and real part of the linear combination
\begin{eqnarray}
\label{C_unp^I}
{\cal C}^{\rm I}_{\rm unp}(\cffF) \approx F_1 {\cal H} -\frac{t}{4M_p^2} F_2 {\cal E} + \frac{\xB}{2} (F_1 + F_2) \widetilde {\cal H}\,.
\end{eqnarray}
\item The first harmonics of single longitudinal target spin  asymmetry (\ref{A_UL+}) and double  longitudinal target spin  asymmetry (\ref{A_LL+})
\begin{eqnarray}
\label{A_UL+^sinphi-BMK}
\Asin{1}{UL,+} &\!\!\! \propto\!\!\! & - \frac{\sqrt{1-y}}{2-y} \sqrt{\frac{-t}{y^2 \Q^2}} \times \xB\Im \left[
{\cal C}^{\rm I}_{\rm LP}(\cffF) +\cdots
\right] +  \mbox{${\rm DVCS}^2$-term}\,,
\\
\label{A_LL+^cosphi-BMK}
\Acos{1}{LL,+}  &\!\!\! \propto\!\!\! & -
\frac{y\sqrt{1-y}}{2-2y+y^2} \sqrt{\frac{-t}{y^2 \Q^2}}\times \xB\Re\left[ {\cal C}^{\rm I}_{\rm LP}(\cffF)+\cdots
\right] + \mbox{${\rm BH}^2$-term} +  \mbox{${\rm DVCS}^2$-term}\,,
\nonumber\\
\end{eqnarray}
are {\it approximately} governed by the imaginary and real part of the linear combination
\begin{eqnarray}
\label{C^I_LP}
{\cal C}^{\rm I}_{\rm LP}(\cffF)  \approx F_1 \widetilde{\cal H}- \left(\!\frac{\xB}{2} F_1  +\frac{t}{4M_p^2} F_2\!\right) \frac{\xB}{2}\cfftE  +\frac{\xB }{2}(F_1 + F_2) {\cal H}\,.
\end{eqnarray}
\item The first even harmonic of the single transverse-target spin asymmetry (\ref{A_UTI})
and first odd harmonic of the double transverse-target spin asymmetry (\ref{A_LTI}) in the charge-odd sector,
\begin{eqnarray}
\label{A_UTI^sinvarphicosphi-BMK}
\Asincos{1}{UT,I} &\!\!\!\propto\!\!\!&
-\frac{\sqrt{1-y}}{2-y}\sqrt{\frac{M^2_p}{y^2 \Q^2}} \times \xB \Im \left[{\cal C}^{\rm I}_{{\rm TP}-}(\cffF) + \cdots \right],
\\
\label{A_LTI^sinvarphisinphi-BMK}
\Asinsin{1}{LT,I} &\!\!\!\propto\!\!\!&
+\frac{y\sqrt{1-y}}{2-2y+y^2}\sqrt{\frac{M^2_p}{y^2 \Q^2}} \times \xB\Re \left[{\cal C}^{\rm I}_{{\rm TP}-}(\cffF) + \cdots \right],
\end{eqnarray}
are {\it approximately} dominated by the imaginary and real part of the linear combination
\begin{eqnarray}
\label{C^I_TP-}
{\cal C}^{\rm I}_{{\rm TP}-}(\cffF)  \approx \frac{-t}{4M_p^2} \left[2F_2 \cffH-  2F_1 \cffE + \xB (F_1+F_2) \frac{\xB}{2}  \cfftE\right]  +\cdots\,.
\end{eqnarray}
\item  The first odd harmonic of the single transverse-target spin asymmetry (\ref{A_UTI}) and first even harmonic of the double transverse-target spin
asymmetry (\ref{A_LTI}) in the charge-odd sector,
\begin{eqnarray}
\label{A_UTI^cosvarphisinphi-BMK}
\Acossin{1}{UT,I} &\!\!\!\propto\!\!\!& + \frac{\sqrt{1-y}}{2-y} \sqrt{\frac{M^2_p}{y^2 \Q^2}} \times \xB\Im
\left[
{\cal C}^{\rm I}_{{\rm TP}+}(\cffF) +\cdots
\right],
\\
\label{A_LTI^cosvarphicosphi-BMK}
\Acoscos{1}{LT,I} &\!\!\!\propto\!\!\!& +  \frac{y \sqrt{1-y}}{2-2 y+y^2}\sqrt{\frac{M^2_p}{y^2 \Q^2}}\times \xB \Re
\left[
{\cal C}^{\rm I}_{{\rm TP}+}(\cffF) +\cdots
\right],
\end{eqnarray}
are {\it approximately} dominated by the imaginary and real part of the linear combination
\begin{eqnarray}
\label{C^I_TP+}
{\cal C}^{\rm I}_{{\rm TP}+}(\cffF)  \approx \frac{-t}{4M_p^2} \left[
2 F_1 \frac{\xB}{2}\widetilde{\cal E}-2F_2 \widetilde{\cal H} \right] +\cdots \,.
\end{eqnarray}
\end{itemize}

As one realizes from the relations among observables and CFFs listed
above, there are six linear combinations of CFFs: three for the
imaginary parts
(\ref{A_LUI^sinphi-BMK},\ref{A_UTI^sinvarphicosphi-BMK},\ref{A_UTI^cosvarphisinphi-BMK})
and three for the real parts
(\ref{A_C^cosphi-BMK},\ref{A_LTI^sinvarphisinphi-BMK},\ref{A_LTI^cosvarphicosphi-BMK}). Unfortunately,
the longitudinally polarized target double spin flip experiment
has been performed only with positron beam and, hence,  the
asymmetries (\ref{A_UL+^sinphi-BMK}) and (\ref{A_LL+^cosphi-BMK}) are
contaminated by twist-three contributions from the
DVCS-squared term. Moreover, the latter term also depends  on the first harmonic of the BH-squared term. Since the DVCS-squared contributions are expected to be relatively small and the first BH harmonic is suppressed for H{\sc ermes} kinematics,
the single spin and double spin flip asymmetries may be used to access the imaginary and real parts of the CFF combination (\ref{C^I_LP}).  To get rid of these contaminations,
we would need longitudinal single and double spin flip measurements in the charge odd sector which were unfortunately not performed%
\footnote{For the history of the H{\sc ermes} experiment see \blue{http://www-hermes.desy.de/hedt/seminar.html}.}.

\begin{figure}[t]
\begin{center}
\includegraphics[width=17cm]{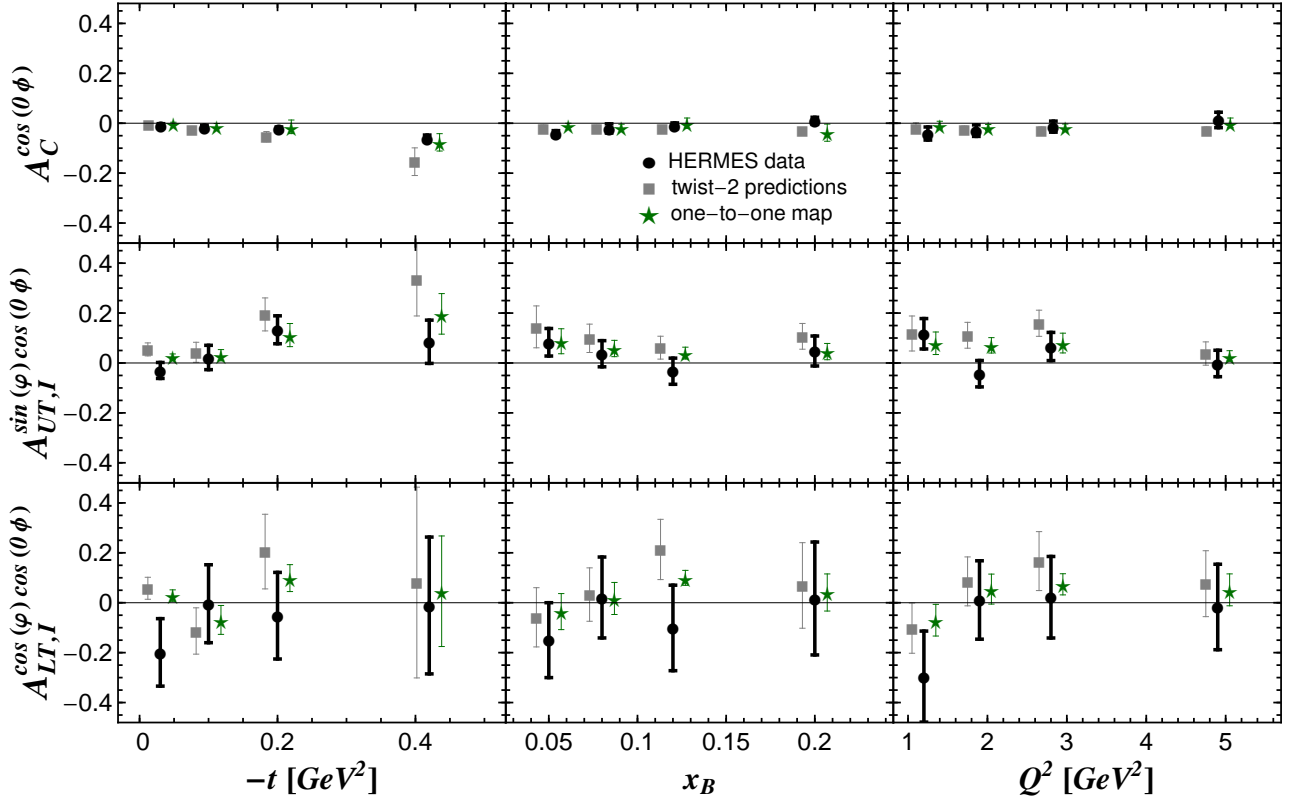}
\end{center}
\vspace{-4mm}
\caption{\small Expectations (squares, slightly shifted to the left) from the approximations (\ref{eqs:pred_cos0phi}--\ref{eqs:pred_coscos0phi}) and
values extracted from the one-to-one map (stars, slightly shifted to the right) are compared to the H{\sc ermes}  measurements  (circles)  of $\Acos{0}{C}$ (top), $\Asincos{0}{UT,I}$ (middle), and $\Acoscos{0}{LT,I}$ (down) asymmetries  in 12 bins by means.}
\label{fig:pred_cos0phi}
\end{figure}

\begin{figure}[t]
\begin{center}
\includegraphics[width=17cm]{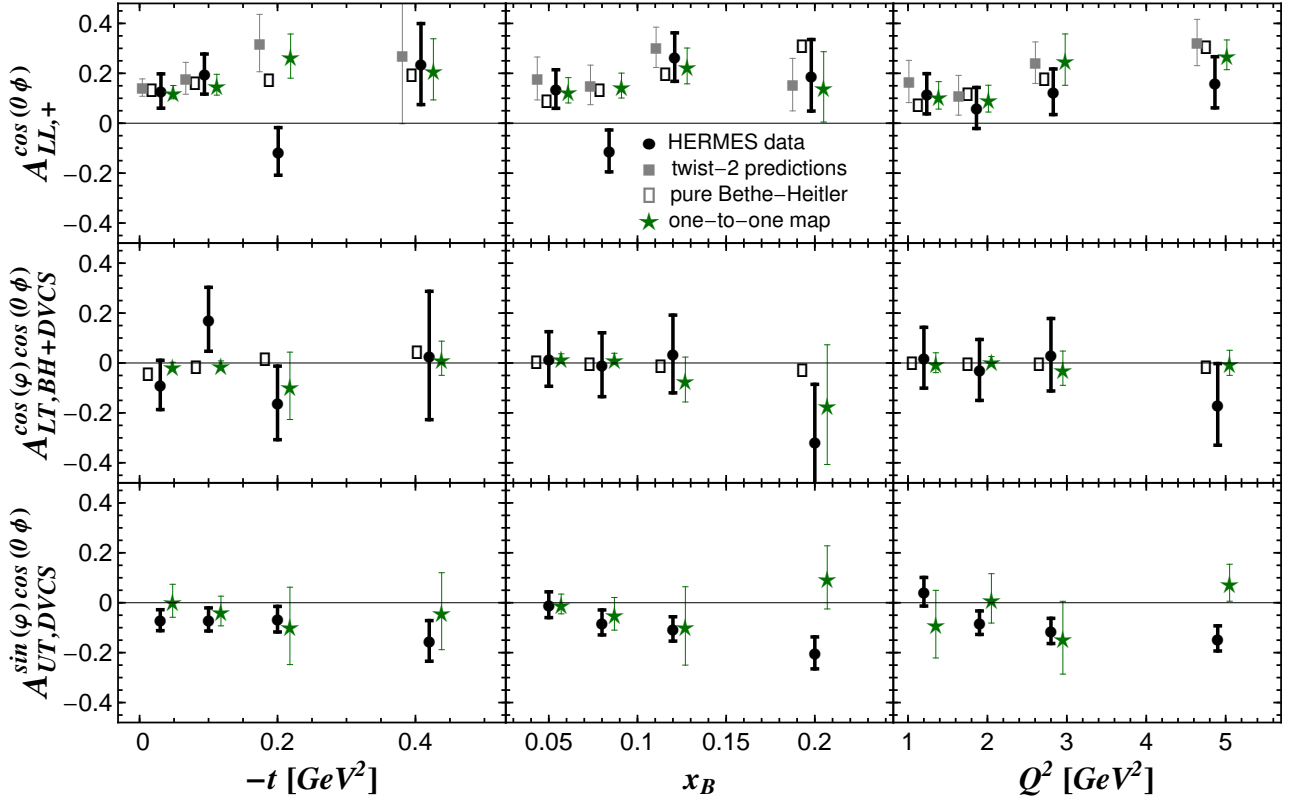}
\end{center}
\vspace{-4mm}
\caption{\small
H{\sc ermes} measurements of the double spin flip asymmetries
$\Acos{0}{LL,+}$ (top row) and $\Acoscos{0}{LT,BH+DVCS}$ (middle row),
as well as of the transverse single proton spin asymmetry
$\Asincos{0}{UT,DVCS}$ (bottom row) in the charge even sector are displayed as circles. The measurements are
compared with pure BH contributions (empty rectangles, slightly shifted to the left), expectations from the approximation (\ref{eqs:pred_LL+cos0phi})
(squares, shifted to the left) and values extracted from the one-to-one map (stars, slightly shifted to the right).}
\label{fig:pred_BH}
\end{figure}

Further twist-two dominated observables are the single transverse proton spin asymmetry  and the double longitudinal-transverse spin asymmetry in the charge even sector:
\begin{itemize}
\item The zeroth harmonic of the single transverse target spin asymmetry, which is governed by the bilinear combination
\begin{eqnarray}
\label{A_UTDVCS^sinvarphicos0phi-BMK}
\Asincos{0}{UT,DVCS} &\!\!\!\propto\!\!\!& +\frac{1-y}{4(2-y)}\, \frac{(-t)^{3/2}}{M_p\,y^2\Q^2}\times \xB^2 \Im \left[\cffH \cffE^\ast -\frac{\xB}{2} \cfftH \cfftE^\ast\right],
\end{eqnarray}
\item The zeroth harmonic of the double longitudinal-beam and transverse target spin asymmetry, which is dominated by the bilinear combination
\begin{eqnarray}
\label{A_LTDVCS^cosvarphicos0phi-BMK}
\Acoscos{0}{LT,BH+DVCS} &\!\!\!\propto\!\!\!& + \frac{y(1-y)}{4(2-2y+y^2)} \frac{(-t)^{3/2}}{M_p\,y^2\Q^2} \times \xB^2 \Re \left[\cfftH \cffE^\ast -\frac{\xB}{2} \cffH \cfftE^\ast\right] +{\rm BH}^2\,.
\end{eqnarray}
\end{itemize}
The asymmetry in (\ref{A_UTDVCS^sinvarphicos0phi-BMK}) is a small
quantity if the CFFs $\cffH$ and $\cfftH$ are nearly in phase with
$\cffE$ and  $\cfftE$, respectively. Interestingly, in asymmetry
(\ref{A_LTDVCS^cosvarphicos0phi-BMK}), CFFs with different parity are combined.
We emphasize that for this longitudinal-transverse double spin asymmetry (\ref{A_LTDVCS^cosvarphicos0phi-BMK}) the BH-squared term is, at H{\sc ermes}, kinematically suppressed:
\begin{eqnarray}
\Acoscos{0}{LT,BH+DVCS} &\!\!\! \approx \!\!\!&  \Acoscos{0}{LT,DVCS} \,.
\end{eqnarray}
Similarly to the single spin asymmetry  $\Asincos{0}{UT,DVCS}$, this
double spin asymmetry is therefore sensitive to the
DVCS-squared term; this is demonstrated by the empty rectangles
  in the middle row of Fig.~\ref{fig:pred_BH} that show the pure BH contribution.
As mentioned above, the constant terms for charge odd asymmetries,
which are relatively suppressed by $1/\Q$ w.r.t.~the first harmonic, may
also be dominated by twist-two associated CFFs; however,
they may suffer from a larger contamination of effective twist-three
CFFs, which are neglected here completely.  Here we have available the real part of two linear
combinations and the imaginary part of one:
\begin{itemize}
\item
The zeroth harmonic of the beam charge asymmetry (\ref{A_C})
\begin{eqnarray}
\label{A_C^cos0phi-BMK}
\Acos{0}{C} \propto \frac{-t}{y\Q^2} \times \xB\Re \left[
{\cal C}^{\rm I}_{\rm unp}(\cffF) - \frac{(1-y)\xB}{2-2 y+y^2}  (F_1+F_2)\cfftH +\cdots
\right].
\end{eqnarray}
\item
The zeroth harmonic  of the single and double transverse target flip asymmetries (\ref{A_UTI}),
and (\ref{A_LTI})
\begin{eqnarray}
\label{A_UTI^sinvarphicos0phi-BMK}
\Asincos{0}{UT,I}\propto + \frac{\sqrt{-t}M_p}{y \Q^2}\times \xB \Im \left[
{\cal C}^{\rm I}_{{\rm TP}-}(\cffF) + \frac{(1-y)\xB}{2-2 y+y^2}\,  \frac{-t}{2M_p^2}
(F_1+F_2)\frac{\xB}{2}\cfftE  \!+\! \cdots \right],
\\
\label{A_UTI^cosvarphicos0phi-BMK}
\Acoscos{0}{LT,I}\propto - \frac{\sqrt{-t}M_p }{y \Q^2} \times \xB\Re \left[
{\cal C}^{\rm I}_{{\rm TP}+}(\cffF) + \frac{(1-y)\xB}{2-2 y+y^2}\,  \frac{-t}{2M_p^2}
(F_1+F_2)\frac{\xB}{2}\cfftE  \!+\! \cdots
\right].
\end{eqnarray}
\end{itemize}
Apart from  terms that are kinematically suppressed at small
$\xB$ and large $y$, we realize that these asymmetries
are already expressed in terms of CFF combinations (\ref{C_unp^I},
\ref{C^I_TP-}, \ref{C^I_TP+}) that determine the
first harmonics. Hence these three asymmetries, measured by {\sc
  hermes}, may roughly expressed in terms of the first harmonics:
\begin{eqnarray}
\label{eqs:pred_cos0phi}
\Acos{0}{C} &\!\!\! \sim \!\!\!& -\frac{\sqrt{-t}}{\Q}\, \frac{2-y}{\sqrt{1-y}}\,  \Acos{1}{C}\,,
\label{eqs:pred_sincos0phi}\\
\Asincos{0}{UT,I} &\!\!\! \sim \!\!\!& -\frac{\sqrt{-t}}{\Q}\, \frac{2-y}{\sqrt{1-y}}\,  \Asincos{1}{UT,I} \,,
\\
\label{eqs:pred_coscos0phi}
\Acoscos{0}{LT,I} &\!\!\! \sim \!\!\!& -\frac{\sqrt{-t}}{\Q}\, \frac{2-y}{\sqrt{1-y}}\,  \Acoscos{1}{UT,I} \,,
\end{eqnarray}
where a fourth relation of the same form exists for the double longitudinal spin asymmetry, see (\ref{A_LL+^cosphi-BMK})
and (\ref{A_LLI^cos0phi-BMK}) below.
Note that the accuracy of these crude relations can be
drastically improved by taking the higher BH harmonics into consideration.
The relations among the charge odd asymmetries
(\ref{eqs:pred_cos0phi}--\ref{eqs:pred_coscos0phi}) offer an
experimental consistency check of the underlying formalism, which is shown in Fig.~\ref{fig:pred_cos0phi}. It can be seen that
the expectations (squares) of our rough approximations
(\ref{eqs:pred_cos0phi}--\ref{eqs:pred_coscos0phi}) are mostly
satisfied, where the largest deviation of $\sim 2\sigma$ appears in
the $10^{\textrm{th}}$ bin of $\Asincos{0}{UT,I}$. Unfortunately, the double spin asymmetries $\Acoscos{0}{LT,I}$ and $\Acoscos{1}{LT,I}$ suffer from limited statistics.

Another experimental test of the BH dominance is provided by:
\begin{itemize}
\item
The zeroth harmonic of the longitudinally double spin flip asymmetry (\ref{A_LL+})
\begin{eqnarray}
\label{A_LL+^cos0phi-BMK}
\Acos{0}{LL,+} &\!\!\! = \!\!\!&  \Acos{0}{LL,BH} + \Acos{0}{LL,I} + \Acos{0}{LL,DVCS}\,,
\end{eqnarray}
which is decomposed in its charge odd part,
\begin{eqnarray}
\label{A_LLI^cos0phi-BMK}
\Acos{0}{LL,I} &\!\!\! \propto \!\!\!&
+ \frac{2-y}{2-2y+y^2}\, \frac{-t}{y\Q^2} \times \xB\Re\left[
{\cal C}^{\rm I}_{\rm LP}( \cffF )  +\frac{(1-y)\xB}{(2-y)^2}\, (F_1+F_2)\cffH+\cdots
\right],
\end{eqnarray}
and the charge even part, where the DVCS induced asymmetry part can be safely dropped,
\begin{eqnarray}
\label{A_LLBHDVCS^cos0phi-BMK}
\Acos{0}{LL,BH+DVCS} &\!\!\! \approx \!\!\!&  \Acos{0}{LL,BH} \,.
\end{eqnarray}
\end{itemize}
For H{\sc ermes} measurements, the relations (\ref{A_UL+^sinphi-BMK}, \ref{A_LL+^cos0phi-BMK}--\ref{A_LLBHDVCS^cos0phi-BMK}) yield the expectation
\begin{eqnarray}
\label{eqs:pred_LL+cos0phi}
\Acos{0}{LL,+} &\!\!\! \approx \!\!\!&   \Acos{0}{LL,BH} -
\sqrt{\frac{-t}{y^2\Q^2}}\, \frac{2-y}{\sqrt{1-y}} \left[\Acos{1}{LL,+}-\Acos{1}{LL,BH}\right],
\end{eqnarray}
which is visualized in the upper panels of
Fig.~\ref{fig:pred_BH}.
It can be seen that the expectations
(squares) of our rough approximation (\ref{eqs:pred_LL+cos0phi})
are satisfied, except in the third and sixth bins. Note that if
we take these bins literally, we have to conclude that in these two cases the DVCS
amplitude overwhelms the BH amplitude and so the sign expected from
the BH term changes. Perhaps more realistically, we may view these two bins simply as a statistical fluctuation on the  $\sim 2 \sigma$ level.

\subsection{Methods for analyzing DVCS data}
\label{sec:regression-method}

The amplitudes of the harmonics for the various asymmetries, extracted by the H{\sc ermes} collaboration,
can be analyzed in various manners, e.g., one may consider  any of the following strategies.
\begin{enumerate}
\item[\phantom{ii}i.]  CFFs are locally extracted from a complete set of observables  by a map.
\item[\phantom{i}ii.]  CFFs are locally extracted by means of least squares,  likelihood, or neural network fits.
\item[iii.] Comparing measurements with model predictions.
\item[\phantom{i}iv.] Global model fits based on the least squares method, maximum-likelihood estimation, or neural networks.
\end{enumerate}
The term {\em local} refers here to a three dimensional
kinematical data point (bin) given by a  $\xB$, $t$, and $\Q^2$ value.
The first method has been proposed and discussed in
\cite{Belitsky:2001ns}, however, it has not been applied so
far. Local extraction of the imaginary and real parts of CFFs (which, for brevity, we will call \cffsreim) by means of
least squares fits were
applied to H{\sc ermes} and JLAB data
\cite{Guidal:2008ie,Guidal:2010ig,Guidal:2010de}. However, although
called model-independent, these analysis are biased, too, by utilizing
the hypothesis that seven \cffsreim\ determine all harmonics and satisfy
some model constraints. Both methods require that the observables
have been measured in a common set of kinematic bins, where it is
assumed that mean values of the kinematical variables are
identical for all of the measurements. This, however, can be only
approximately true. The kinematical condition can be relaxed if one
employs neural networks. Presently, they have been employed to access the imaginary and real part of CFF $\cffH$ from H{\sc ermes} data on unpolarized protons \cite{Kumericki:2011rz}.

Model predictions for DVCS have existed for over a decade. In the beginning, it was rather popular to use
Radyuskin's double distribution ansatz (RDDA) for GPDs, given at $t=0$ \cite{Rad97}.
In this method, one uses information on parton distribution functions (PDFs) and form factors to
build the model, while the GPD arises from a skewing operation
that is governed by a profile function and the width of this
concave function is controlled by a ``$b$''
parameter. Such models, e.g. given in \cite{GoePolVan01},  have been privately distributed as
numerical code, referred according to the author names of the paper \cite{VanGuiGui98} as VGG, and it is often used to compare data
with model predictions. However, when this is done, authors often
 neglect to give full details of the code version and
precise details of the underlying model, e.g. the set
of PDFs that are used to constrain the GPDs. The same
underlying double-distribution ansatz was also used in the BMK model \cite{Belitsky:2001ns},
in the numerical code of Freund-McDermott \cite{Freund:2001rk},
and the Goloskokov-Kroll model \cite{Goloskokov:2005sd,Goloskokov:2007nt}, used in an adjusted
hand-bag model framework for the description of deeply virtual meson production.  Certainly there are differences in all these models, however,
since these models imply $H$ dominance and $H$ is inherited from a
unpolarized PDF parametrization and sometimes from the $t$-dependence
of electromagnetic nucleon form factors, it is not very  surprising
that all these models applied to DVCS off unpolarized protons
show similar features in a LO analysis. When describing to some extent present DVCS measurements with a
RDDA based GPD model, one may use a small skewness effect, use in a LO
analysis NLO PDF parameterizations and neglect GPD evolution (important in H1/ZEUS Collider kinematics).
Aiming to describe DVCS data, it has been also suggested to build ``flexible'' GPD parameterization
by violating their spectral (or polynomiality) property \cite{Freund:2002qf} (see also \cite{Goldstein:2010gu}, where the claim of satisfying polynomiality
is simply not true).
This is an entirely unacceptable recipe (see comments in \cite{Diehl:2007jb} and \cite{Kumericki:2008di}) and so
the phenomenological aspects of such models cannot be discussed.


Since it was widely realized in the past that most variations of
the RDDA usually overestimate the size of beam spin asymmetries
and do not describe some other observables, it was necessary to build flexible GPD models which can be used in numerical fast fitting routines \cite{Kumericki:2007sa,Kumericki:2009uq}, where GPDs are modelled  in terms of their conformal moments \cite{Mueller:2005ed} (related representations were suggested in \cite{Shuvaev:1999fm,Polyakov:2002wz,Kirch:2005tt}). Thereby, one tries initially to parameterise the partonic degrees of freedom in such a manner that they are accessible from the experimental measurements.  We will provide more details in Sect.~\ref{sec:global}.

We emphasize that one major impediment to the understanding of DVCS
data in terms of one  specific double-distribution ansatz is the lack
of exclusivity in the data. As noted in section
\ref{sec:HERMES}, there is an estimated average of $12\%$
contamination of the H{\sc ermes} data by processes involving an
intermediate $\Delta$-resonance. While H{\sc ermes} has published a
beam-helicity measurement from a very pure
sample~\cite{Airapetian:2012pg} that indicates that this
contamination only acts as a small dilution of the asymmetry
magnitude at higher values of $-t$ (e.g.~$-t >0.25\,\GeV^2$), there is sufficient uncertainty
that one cannot make any definitive statements on the specific form of a double-distribution ansatz.

In the next section we outline the method of mapping the asymmetries to CFFs. Thereby we consider the asymmetries as normally distributed random variables and consider the extraction problem of CFFs simply as a map of random variables, where the functional dependence of asymmetries on CFFs is provided by the theory. In fact, for
normally distributed random variables the formalism we employ is very well known from (non-)linear regression and we need only to discuss the maps of means and variances rather than probability distributions.

\subsubsection{Maps of normally distributed random variables}
\label{sec:maps}

Let us first clarify what we call the twist-two dominance hypothesis and the additional approximation that arise from a linearization procedure.
For instance, the first harmonic of the beam spin asymmetry (\ref{A_{LU,I}(phi)-spin0}) in the charge odd sector reads
\begin{eqnarray}
\label{A_{LU,I}(phi)-spin0-approx}
\Asin{1}{LU,I} &\!\!\!\backsimeq\!\!\!& \frac{1}{\pi}\int_{-\pi}^\pi\!d\phi\,\sin(\phi)
\\
&&\!\!\!\!\!\!\times  \frac{\hat{s}^{\rm I}_1 \sin(\phi) + 0 \times \hat{s}^{\rm I}_2 \sin(2\phi)}{
\sum_{n=0}^2 \hat{c}^{\rm BH}_n \cos(n \phi) + \frac{-t}{y^2\Q^2} w(\phi) \hat{c}^{\rm DVCS}_0  +0 \times \frac{-t}{y^2\Q^2} w(\phi) \sum_{n=1}^2 \hat{c}^{\rm DVCS}_n  \cos(n \phi) }
\Bigg|_{{\cal H}_{\rm eff} = {\cal H}_{\rm T} =0}\,,
\nonumber
\end{eqnarray}
where $0\times \cdots$ indicate the expressions that are induced by
twist-three and transversity contributions and are neglected. Note
that the power suppressed admixture of $\cffH$ in the second odd
harmonic is consequently neglected, too, and that we also alter the
definition of the $\hat{c}^{\rm DVCS}_0$ coefficient, since power
suppressed twist-three square terms and presumably-small transversity square terms are set to zero, cf.~formula (2.18) in \cite{Belitsky:2010jw}.  In this approximation the Fourier transform (\ref{A_{LU,I}(phi)-spin0-approx}) can be evaluated in terms of elementary functions,
\begin{eqnarray}
\label{A_{LU,I}(phi)-spin0-approx1}
\Asin{1}{LU,I} &\!\!\! \backsimeq\!\!\!  &  \frac{N \hat{s}^{\rm I}_1}{\hat{c}^{\rm BH}_0}\times \frac{1}{b}\left(
\sqrt{\frac{a^2-4 b \left(1+b+\sqrt{(1+b)^2-a^2}\right)}{a^2-8 (1-b) b}}-1\right),
\end{eqnarray}
where
\begin{eqnarray}
\label{coef-N}
N= \frac{\hat{c}^{\rm BH}_0}{\hat{c}^{\rm BH}_0  +\frac{-t}{y^2\Q^2}  \hat{c}^{\rm DVCS}_0}
\end{eqnarray}
is considered as an overall normalization factor and the coefficients
\begin{eqnarray}
\label{coef-ab}
a = \frac{\hat{c}^{\rm BH}_1 +\frac{-t}{ y^2\Q^2}  w_1 \hat{c}^{\rm DVCS}_0}{\hat{c}^{\rm BH}_0  +\frac{-t}{y^2\Q^2}  \hat{c}^{\rm DVCS}_0}\; \stackrel{{\rm BH} > {\rm DVCS}}{\approx} \;
\frac{\hat{c}^{\rm BH}_1}{\hat{c}^{\rm BH}_0} \quad\mbox{and}\quad
b = \frac{\hat{c}^{\rm BH}_2  +\frac{-t}{y^2\Q^2}  w_2 \hat{c}^{\rm DVCS}_0}{\hat{c}^{\rm BH}_0  +\frac{-t}{y^2\Q^2}   \hat{c}^{\rm DVCS}_0} \;
\stackrel{{\rm BH} > {\rm DVCS}}{\approx} \;  \frac{\hat{c}^{\rm BH}_2}{\hat{c}^{\rm BH}_0}
\end{eqnarray}
arise from the higher harmonics of the denominator in (\ref{A_{LU,I}(phi)-spin0-approx}). In our case of interest
they satisfy the condition $|a|>|b|$, they are small quantities, and the indicated approximation (\ref{coef-ab}) can be considered as justified in
a BH dominated regime.  Consequently, in such an approximation the DVCS-squared term enters only in the overall normalization factor $N$.
Analogous formulae can be also obtained for even harmonics, where we restrict ourselves here to the two lowest harmonics.

Before we discuss the proton case, let us provide a pedagogical example with a spin-zero target. In this case we are dealing
with three CFFs ${\cal H}$, ${\cal H}_{\rm eff}$, and ${\cal H}_{\rm T}$.
Moreover, we assume that the second and third harmonics are compatible
with zero, which suggest that the CFFs ${\cal H}_{\rm eff}$ and ${\cal H}_{\rm T}$, associated with partonic twist-three
and transversity processes,  can be neglected. Note that this approximation may induce an
ambiguity in the phenomenological treatment; instead of neglecting
$\cffH_{\rm eff}$ it may be alternatively justified to neglect the twist-three CFF
$\cffH_3$, i.e., one sets $\cffH_{\rm eff}\approx -\xB \cffH$.
We relate the dominant asymmetries to the twist-two associated CFF ${\cal H}$ by two linearized equations
\begin{eqnarray}
\label{A2CFF}
\Asin{1}{LU,I} \approx N \mbox{$c$}^{-1}_{\Im} {\cal H}^{\Im}
\quad\mbox{and}\quad
\Acos{1}{C} \approx N \mbox{$c$}^{-1}_{\Re}  {\cal H}^{\Re} \,,
\end{eqnarray}
where we use for later convenience  the shorthands ${\cal H}^{\Im}= \Im {\cal H}$ and ${\cal H}^{\Re}= \Re {\cal H}$. The coefficients
\begin{eqnarray}
\label{c_^{-1}}
c_{\Im}^{-1} = \frac{\partial\Asin{1}{LU,I}}{\partial {\cal H}^{\Im} }\Bigg|_{{\cal F}=0}
\quad\mbox{and}\quad
c_{\Re}^{-1} = \frac{\partial \Acos{1}{C}}{\partial {\cal H}^{\Re} }\Bigg|_{{\cal F}=0}
\end{eqnarray}
are calculated from (\ref{A_{LU,I}(phi)-spin0},\ref{Asin{n}{LU,I}})  and  (\ref{A_{C}(phi)-spin0},\ref{Acos{n}{C}}) and are given as elementary functions of the Fourier coefficients $s^{\rm I}_1$, $c^{\rm I}_1$, and $c^{\rm BH}_n$. In this procedure, we set the DVCS-squared term in the denominator to zero which, however, appears in the normalization
factor $N$.
This overall factor can be considered as a bilinear function of the
twist-two associated CFF $\cffH $ or equivalently as a function of
the asymmetries. To a good approximation, it can be also
expressed by the ratio of the BH and DVCS cross sections
\begin{eqnarray}
\label{eq:N}
0 \lesssim  N(\mbox{\boldmath $A$}) \approx \frac{\int_{-\pi}^\pi\! d\phi\,w(\phi)d\sigma_{\rm BH}(\phi)}{\int_{-\pi}^\pi\! d\phi\, w(\phi)\left[d\sigma_{\rm BH}(\phi)+ d\sigma_{\rm DVCS}(\phi)\right]} \lesssim 1.
\end{eqnarray}
The solution of the linearized equations (\ref{A2CFF}) is immediately obtained and, with our twist-two dominance assumption, the imaginary and real part of the leading twist-two associated CFF reads
\begin{eqnarray}
\label{As2CFFs-spin0-tw2}
\Im \cffH = \frac{\mbox{$c$}_{\Im} }{N(\mbox{\boldmath $A$})} \Asin{1}{LU,I}
\quad\mbox{and}\quad
\Re \cffH = \frac{\mbox{$c$}_{\Re}}{N(\mbox{\boldmath $A$})}\Acos{1}{C}\,,
\end{eqnarray}
where $c_{\Im}$ and $c_{\Re}$ represent now two numbers for a given
kinematical point. The normalization factor is approximately given as
\begin{eqnarray}
\label{N-spin0-tw2}
N\approx \frac{1}{1+ \frac{k}{4} |\cffH|^2 } \quad\mbox{with}\quad
 \frac{k}{4} |\cffH|^2 =
 \frac{\int_{-\pi}^\pi\! d\phi\,w(\phi)d\sigma_{\rm DVCS}(\phi)}{\int_{-\pi}^\pi\! d\phi\, w(\phi) d\sigma_{\rm BH}(\phi)}\Bigg|_{\cffH_{\rm eff}=\cffH_{\rm T}=0}
\,,
\end{eqnarray}
where $k$ is a known kinematical factor. Plugging the solution (\ref{As2CFFs-spin0-tw2}) into the normalization (\ref{N-spin0-tw2}) yields a cubic equation
in $N$ that has two non-trivial solutions:
\begin{eqnarray}
\label{N-spin0-tw2-sol}
N(\mbox{\boldmath $A$})\approx \frac{1}{2} \left(1 \pm \sqrt{1-k\,c_{\Im}^2 \left(\Asin{1}{LU,I}\right)^2 -k\, c_{\Re}^2\left(\Acos{1}{C}\right)^2 }\right).
\end{eqnarray}
The solution with the positive root satisfies the boundary
condition $N(\mbox{\boldmath $A$}=\mbox{\boldmath $0$})=1$ and is
the one to take if the BH amplitude overwhelms the DVCS
one. Obviously, the solution with the negative root satisfies the boundary
condition $N(\mbox{\boldmath $A$}=\mbox{\boldmath $0$})=0$ and it is the one to take if the unpolarized DVCS cross section is larger than the BH one.
Finally, error propagation is done in the common manner. If the experimental errors are uncorrelated, we may write the relative standard error in the following form
\begin{eqnarray}
\frac{(\delta \Im\cffH)^2}{|\Im\cffH|^2} &\!\!\! = \!\!\!&\frac{\left(\delta\Asin{1}{LU,I}\right)^2}{\left(\Asin{1}{LU,I}\right)^2} + \frac{(\delta N)^2}{N^2}
- 2 \frac{\partial N}{N \partial\Asin{1}{LU,I} } \frac{1}{\Asin{1}{LU,I}} \left(\delta\Asin{1}{LU,I}\right)^2,
\\
\frac{(\delta \Re\cffH)^2}{|\Re\cffH|^2} &\!\!\! = \!\!\!&\frac{\left(\delta\Acos{1}{C}\right)^2}{\left(\Acos{1}{C}\right)^2} + \frac{(\delta N)^2}{N^2}
-2 \frac{\partial N}{N \partial\Acos{1}{C} } \frac{1}{\Acos{1}{C}} \left(\delta\Acos{1}{C}\right)^2\,,
\end{eqnarray}
where the normalization uncertainty,
\begin{eqnarray}
\frac{(\delta N)^2}{N^2} =  \left|\frac{\partial N}{N \partial\Asin{1}{LU,I} } \right|^2 \left(\delta\Asin{1}{LU,I}\right)^2 +
\left|\frac{\partial N}{N \partial\Acos{1}{C} } \right|^2 \left(\delta\Acos{1}{C}\right)^2 + \cdots\,,
\end{eqnarray}
can be easily evaluated by means of (\ref{N-spin0-tw2-sol}) and one may add a model estimate due to the neglected twist-three and other transversity CFFs,
which is indicated by the ellipsis.

The method, i.e., the map of random variables, can be refined by
the inclusion of higher harmonics and it can be extended  to a nucleon target.
According to the experimental observables and the assumptions,
we have generically a set of charge-odd asymmetries, arising from the interference of BH and DVCS amplitudes, that we arrange into an $m$--dimensional vector:
\begin{equation}
\label{def-A}
\left(\mbox{\boldmath $A$}^{\cal I}\right)^\intercal = \left( A_1,\cdots, A_{m} \right)
\,.
\end{equation}
For a complete twist-two DVCS off-the-nucleon analysis, we need $4$
even and $4$ odd harmonics. Inclusion of the twist-three sector
increases the number of harmonics to $16$, while for an
hypothesis-free treatment we need $24$ asymmetries. Surely hypotheses
can be refined, e.g., in \cite{Guidal:2008ie} the assumption $\Im
\widetilde{\cal E}=0$  is made in a twist-two analysis.
Up to an overall normalization, we can formulate a linear mapping problem as follows:
\begin{eqnarray}
\label{F2A-gen}
\mbox{\boldmath $A$}^{\cal I} &\!\!\!=\!\!\!& N\!\left(\mbox{\boldmath $A$}^{\cal I}| \mbox{\boldmath ${\cal G}$}\right)\,  \mbox{\boldmath $c$}^{-1} \cdot \mbox{\boldmath $\cffF$} +  N\!\left(\mbox{\boldmath $A$}^{\cal I}| \mbox{\boldmath ${\cal G}$}\right)\,  \mbox{\boldmath $b$}\cdot \mbox{\boldmath ${\cal G}$}\,,
\end{eqnarray}
where $\mbox{\boldmath $\cffF$}$ is the set of \cffsreim\ (written as an $m$--dimensional vector) that one wants to extract from the measurements
and $\mbox{\boldmath ${\cal G}$}$ is the set of \cffsreim\ that is considered as inaccessible within a given hypothesis.
The inverse $m\times m$ coefficient matrix $ \mbox{\boldmath $c$}^{-1}$ and $m\times(n-m)$ matrix $\mbox{\boldmath $b$}$ of the inhomogeneous term are calculated from the theoretical formulae, e.g., for asymmetries proportional to the interference term, we may use
\begin{eqnarray}
\label{c^{-1}-matrices}
\mbox{\boldmath $c$}^{-1} = \frac{\partial\mbox{\boldmath $A$}^{\cal I}}{\partial \mbox{\boldmath ${\cal F}$}}
\Bigg|_{\mbox{\boldmath ${\cal F}$}=\mbox{\boldmath ${\cal G}$}=0}
\quad\mbox{and}\quad
\mbox{\boldmath $b$} = \frac{\partial\mbox{\boldmath $A$}^{\cal I}}{\partial \mbox{\boldmath ${\cal G}$}}
\Bigg|_{\mbox{\boldmath ${\cal F}$}=\mbox{\boldmath ${\cal G}$}=0}
\,.
\end{eqnarray}

Obviously, the solution of the inhomogeneous problem (\ref{F2A-gen}) provides us the \cffsreim\ $\mbox{\boldmath $\cffF$}$ in dependence on
the $m$ observables $\mbox{\boldmath $A$}^{\cal I}$ and the $(n-m)$ dimensional set of unknown \cffsreim\ $\mbox{\boldmath $\cal G$}$:
\begin{eqnarray}
\label{eq:A2F-generic}
\mbox{\boldmath $\cffF$}=  \frac{1}{N\!\left(\mbox{\boldmath $A$}^{\cal I}| \mbox{\boldmath ${\cal G}$}\right)} \mbox{\boldmath $  c$}\cdot
\mbox{\boldmath $A$}^{\cal I}-  \mbox{\boldmath $  c$ }\cdot \mbox{\boldmath $ b $}\cdot \mbox{\boldmath ${\cal G}$}\,.
\end{eqnarray}
The variation of the solution w.r.t.~both the measurement and the unknown \cffsreim\ can be written in terms of an $m \times m$ matrix
and an $m \times (n-m)$ matrix
\begin{eqnarray}
\label{partial_A F}
\frac{\partial \mbox{\boldmath $\cffF$}}{\partial \mbox{\boldmath $A$}^{\cal I}} &\!\!\! =\!\!\!&
\frac{1}{N\!\left(\mbox{\boldmath $A$}^{\cal I}| \mbox{\boldmath ${\cal G}$}\right)}\left[
\mbox{\boldmath $ c$} - \frac{\mbox{\boldmath $  c$}\cdot \mbox{\boldmath $A$}^{\cal I} }{N\!\left(\mbox{\boldmath $A$}^{\cal I}| \mbox{\boldmath ${\cal G}$}\right)} \otimes \frac{\partial N\!\left(\mbox{\boldmath $A$}^{\cal I}| \mbox{\boldmath ${\cal G}$}\right)}{\partial\mbox{\boldmath $A$}^{\cal I}}\right],
\\
\label{partial_A G}
\frac{\partial \mbox{\boldmath $\cffF$}}{\partial  \mbox{\boldmath $\cal G$}} &\!\!\! =\!\!\!&
-\frac{\mbox{\boldmath $  c$}\cdot \mbox{\boldmath $A$}^{\cal I} }{N^2\!\left(\mbox{\boldmath $A$}^{\cal I}| \mbox{\boldmath ${\cal G}$}\right)}
\otimes \frac{\partial N\!\left(\mbox{\boldmath $A$}^{\cal I}| \mbox{\boldmath ${\cal G}$}\right)}{\partial\mbox{\boldmath $\cal G$}}
 - \mbox{\boldmath $  c$} \cdot \mbox{\boldmath $  b$},
\end{eqnarray}
where $\otimes$ symbolizes the direct product of two vectors, e.g.,
$$
\left\{  \mbox{\boldmath $\cal A $}  \otimes \frac{\partial N}{N \partial \mbox{\boldmath $\cal G$} }\right\}_{ab} =
 \mbox{\boldmath $\cal A $}_a\, \frac{1}{N}\, \frac{\partial N}{\partial \mbox{\boldmath $\cal G$}_b}\,,
 \quad  a,b\in\{1,\cdots,m\}
\,,\;\;  b\in\{1,\cdots,n-m\}
 \,.
$$

In the case that (with a given hypothesis) the number of
\cffsreim\ matches the numbers of observables, our equation
(\ref{F2A-gen}) reduces to a homogenous one, i.e., we can set the
inhomogeneous term in (\ref{eq:A2F-generic}), containing the
matrix $\mbox{\boldmath $b$}$, to zero and the variation of the
solution is given by the Jacobian (\ref{partial_A F}), which can be now written in the form
\begin{eqnarray}
\label{partial_A F-hom}
\frac{\partial \mbox{\boldmath $\cffF$}}{\partial \mbox{\boldmath $A$}^{\cal I}} &\!\!\! =\!\!\!&
\frac{1}{N\!\left(\mbox{\boldmath $A$}^{\cal I}\right)}\left[
\mbox{\boldmath $ c$} - \mbox{\boldmath $\cffF$} \otimes \frac{\partial N\!\left(\mbox{\boldmath $A$}^{\cal I}\right)}{\partial\mbox{\boldmath $A$}^{\cal I}}\right].
\end{eqnarray}
{}From this we can easily transform the (variance-)covariance matrix of the experimental measurements to that of  \cffsreim:
\begin{eqnarray}
\label{eq:cov}
{\rm cov}\!\left(\mbox{\boldmath $\cffF$}\right) =
 \left[\frac{\partial \mbox{\boldmath $\cffF$}}{\partial \mbox{\boldmath $A$}^{\cal I}} \right]
\cdot
 {\rm cov}\left(\mbox{\boldmath $A$}^{\cal I}\right)
 \cdot
 \left[\frac{\partial \mbox{\boldmath $\cffF$}}{\partial \mbox{\boldmath $A$}^{\cal I}} \right]^\intercal \,,
\end{eqnarray}
where the result reads more explicitly as
\begin{eqnarray}
\label{eq:cov-hom}
 {\rm cov}\!\left(\mbox{\boldmath $\cffF$}\right)&\!\!\!=\!\!\!&
 \frac{1}{N^2} \mbox{\boldmath $c$} \cdot{\rm cov}\!\left(\mbox{\boldmath $A$}^{\cal I}\right)\cdot \mbox{\boldmath  $c$}^\intercal
 \\
 && -\frac{1}{N}\!\!\left[\mbox{\boldmath  $c$} \cdot  {\rm cov}\!\left(\mbox{\boldmath $A$}^{\cal I}\right)\cdot
\left(\!\mbox{\boldmath $\cffF$} \otimes \frac{\partial N(\mbox{\boldmath $A$}^{\cal I})}{N \partial\mbox{\boldmath $A$}^{\cal I} }\!\right)^\intercal
 + \left(\!\mbox{\boldmath $\cffF$} \otimes \frac{\partial N(\mbox{\boldmath $A$}^{\cal I})}{N \partial\mbox{\boldmath $A$}^{\cal I} }\!\right) \cdot {\rm cov}\!\left(\mbox{\boldmath $A$}^{\cal I}\right)\cdot  \mbox{\boldmath  $c$}^\intercal \right]
 \nonumber \\
 &&+ \left(\!\mbox{\boldmath $\cffF$} \otimes \frac{\partial N(\mbox{\boldmath $A$}^{\cal I})}{N \partial\mbox{\boldmath $A$}^{\cal I} }\!\right)
\cdot {\rm cov}\!\left(\mbox{\boldmath $A$}^{\cal I}\right)\cdot
\left(\!\mbox{\boldmath $\cffF$}  \otimes
\frac{\partial N(\mbox{\boldmath $A$}^{\cal I})}{N \partial\mbox{\boldmath $A$}^{\cal I} }\!\right)^\intercal\,.
\nonumber
\end{eqnarray}
This representation allows us to discuss separately the overall normalization error given by the last term on the r.h.s.~of this equation.
The normalization factor $N$ is determined as in our toy example from substituting the resultant \cffsreim\ into (\ref{eq:N}), which provides again a
cubic equation for $N$. The correct solution is picked up by the requirement that $N$ is real valued, lies in the interval $0< N < 1$,  and from
the knowledge if the observables arise from a BH $(N>1/2)$ or DVCS $(N<1/2)$ dominated scenario.
If such a  solution does not exist, e.g.~due to some large statistical fluctuation of a mean value, the mapping method is strictly-speaking
not applicable; however, as we will see below in Sect.~\ref{sec:regression} in some of such cases it can be still considered as a useful tool.

If no further experimental information in a given kinematical bin
is available, one may use a model estimate for the remaining
unknown \cffsreim\ $\mbox{\boldmath $\cal G$}$ and propagate the
estimated uncertainties by means of a covariance matrix,
providing us with the sum
\begin{eqnarray}
{\rm cov}\!\left(\mbox{\boldmath $\cffF$}\right) =
 \left[\frac{\partial \mbox{\boldmath $\cffF$}}{\partial \mbox{\boldmath $A$}} \right]
\cdot
 {\rm cov}\left(\mbox{\boldmath $A$}\right)
 \cdot
 \left[\frac{\partial \mbox{\boldmath $\cffF$}}{\partial \mbox{\boldmath $A$}} \right]^\intercal
 +
 \left[\frac{\partial \mbox{\boldmath $\cffF$}}{\partial \mbox{\boldmath $\cal G$}} \right]
\cdot
 {\rm cov}\left(\mbox{\boldmath $\cal G$}\right)
 \cdot
 \left[\frac{\partial \mbox{\boldmath $\cffF$}}{\partial \mbox{\boldmath $\cal G$}} \right]^\intercal\,.
\end{eqnarray}
A more appropriate method would be to constrain the uniformly randomly distributed values of  unknown \cffsreim\ $\mbox{\boldmath $\cal G$}$ and propagate the uncertainties numerically to the errors of the extracted \cffsreim\ $\mbox{\boldmath $\cal F$}$, formally written as
\begin{eqnarray}
\label{eq:A2F-Gweighed}
\mbox{\boldmath $\cffF$} &\!\!\! = \!\!\!&
\Bigg\langle \frac{1}{N\!\left(\mbox{\boldmath $A$}| \mbox{\boldmath ${\cal G}$}\right)} \mbox{\boldmath $c$}\cdot
\mbox{\boldmath $A$}- \mbox{\boldmath $c$}\cdot \mbox{\boldmath $b$}\cdot \mbox{\boldmath ${\cal G}$}
\Bigg\rangle ,
\\
\frac{\partial \mbox{\boldmath $\cffF$}}{\partial \mbox{\boldmath $A$}} &\!\!\! =\!\!\!&
\Bigg\langle
\frac{1}{N\!\left(\mbox{\boldmath $A$}| \mbox{\boldmath ${\cal G}$}\right)}\left[ \mbox{\boldmath $c$} -
\frac{\mbox{\boldmath $  c$}\cdot \mbox{\boldmath $A$}^{\cal I} }{N\!\left(\mbox{\boldmath $A$}^{\cal I}| \mbox{\boldmath ${\cal G}$}\right)} \otimes \frac{\partial N\!\left(\mbox{\boldmath $A$}| \mbox{\boldmath ${\cal G}$}\right)}{\partial\mbox{\boldmath $A$}}\right]
\Bigg\rangle .
\end{eqnarray}

A third possibility is that the remaining unknown \cffsreim\
$\mbox{\boldmath ${\cal G}$}$ could be extracted from asymmetries in the charge odd sector, i.e.,
related to the DVCS-squared term. Let us arrange these asymmetries as an  $n-m$ dimensional vector
\begin{equation}
\left(\mbox{\boldmath $A$}^{\mbox{\tiny DVCS}}\right)^\intercal= \left(A_{m+1},\cdots,A_{n}\right) \;,
\end{equation}
 and let us suppose that we can complete the number of observables, written now in terms of a $n$ dimensional vector
\begin{equation}
\label{def-A-full}
\mbox{\boldmath $A$}^\intercal = \left( A_1,\cdots, A_{m}, A_{m+1},\cdots, A_n \right) \;.
\end{equation}
The charge-even observables are linear or quadratic in the remaining
\cffsreim\ $\mbox{\boldmath ${\cal G}$}$, i.e., they are constrained by
a system of linear constraints,
\begin{eqnarray}
\label{eq:AUTDVCS_linear}
A_i^{\rm DVCS} = N\!\left(\mbox{\boldmath $A$}\right)\, \mbox{\boldmath $b$}^\intercal_{i}(\mbox{\boldmath ${\cal F}$})\cdot \mbox{\boldmath ${\cal G}$}
\,, \quad i\in \{m+1,\cdots, n\}\,,
\end{eqnarray}
such as those accessed by the transverse target spin asymmetry
(\ref{A_UTDVCS^sinvarphicos0phi-BMK}), or quadratic equations,
\begin{eqnarray}
\label{eq:AUTDVCS_quadratic}
 A_i^{\rm DVCS}= N\!\left(\mbox{\boldmath $\cffF$}\right) \left[ {a}_i(\mbox{\boldmath  $\cffF$})+
 \mbox{\boldmath ${b}$}_i^\intercal(\mbox{\boldmath $\cffF$})  \cdot  \mbox{\boldmath ${\cal G}$}+
\mbox{\boldmath ${\cal G}$}^{^\intercal} \cdot  \mbox{\boldmath ${c}$}_i\cdot \mbox{\boldmath ${\cal G}$}
\right], \quad i\in \{m+1,\cdots, n\}\,,
\end{eqnarray}
where the overall normalization can now be considered as a function of the complete set of asymmetries (\ref{def-A-full}).
Substituting the solutions of the linear constraints (\ref{eq:A2F-generic}) into (\ref{eq:AUTDVCS_linear}) and/or (\ref{eq:AUTDVCS_quadratic}),
one will generally end up with $n-m$ quadratic equations
\begin{eqnarray}
\label{eq:ALL2ReEtE}
 A_i^{\rm DVCS}= N\!\left(\mbox{\boldmath $A$}\right) \left[ {\rm a}_i(\mbox{\boldmath $A$}^{\cal I}|N\!\left(\mbox{\boldmath $A$}\right))+
 \mbox{\boldmath $\rm b$}_i^\intercal(\mbox{\boldmath $A$}^{\cal I}|N\!\left(\mbox{\boldmath $A$}\right))  \cdot  \mbox{\boldmath ${\cal G}$}+
\mbox{\boldmath ${\cal G}$}^{^\intercal} \cdot  \mbox{\boldmath $\rm c$}_i(\mbox{\boldmath $A$}^{\cal I}|N\!\left(\mbox{\boldmath $A$}\right))\cdot \mbox{\boldmath ${\cal G}$}
\right],
\end{eqnarray}
where all coefficients explicitly depend only on the charge-odd asymmetries $\mbox{\boldmath $A$}^{\cal I}$ and the normalization factor $N$, which appears
on the r.h.s.~as $N^p$ with $p\in \{-1,0,1\}$. The quadratic constraints (\ref{eq:ALL2ReEtE}) may be analytically solved, too, and the solution may be written for convenience in terms of a matrix equation
\begin{eqnarray}
\label{eq:A2reE}
 \mbox{\boldmath ${\cal G}$} = \frac{1}{N\!\left(\mbox{\boldmath $A$}\right)}\,
\mbox{\boldmath$ c$}_{\mbox{\tiny\rm DVCS}}(\mbox{\boldmath $A$}|N\!\left(\mbox{\boldmath $A$}\right))
 \cdot \mbox{\boldmath $A$}^{\mbox{\tiny DVCS}}\,
\,,
\end{eqnarray}
where $ \mbox{\boldmath$ c$}_{\mbox{\tiny\rm DVCS}}$ is an $(n-m)\times (n-m)$ matrix that explicitly depends on the asymmetries and on the normalization, in general in
a non-linear manner (containing roots).
Of course, the solution (\ref{eq:A2reE}) is not unique and  only
real-valued \cffsreim\ are of interest. Moreover, one would
naturally utilize boundary conditions to select the desired
solution. Other disadvantages of using charge-even rather than
charge-odd observables are that the former asymmetries are expected
to be smaller than the latter ones and the $\mbox{\boldmath$c$}_{\mbox{\tiny\rm DVCS}}$ matrix is only known within some given
experimental uncertainty.
The gradient of the remaining \cffsreim\ is calculated from
\begin{eqnarray}
\label{eq:A2reE-gradient}
\frac{\partial \mbox{\boldmath ${\cal G}$}}{\partial \mbox{\boldmath $A$}} =  \frac{1}{N\!\left(\mbox{\boldmath $A$}\right)}\,
{\mbox{\boldmath  $c$}}_{\mbox{\tiny DVCS}}(\mbox{\boldmath $A$})\cdot \frac{\partial \mbox{\boldmath $A$}^{\mbox{\tiny DVCS}}}{\partial \mbox{\boldmath $A$}}  +
\frac{1}{N\!\left(\mbox{\boldmath $A$}\right)}\,   \frac{\partial {\bf c}_{\mbox{\tiny\rm DVCS}}(\mbox{\boldmath $A$})}{\partial \mbox{\boldmath $A$}}   \cdot \mbox{\boldmath $A$}^{\mbox{\tiny DVCS}}  -
\mbox{\boldmath ${\cal G}$}\otimes \frac{\partial N(\mbox{\boldmath $A$})}{N\!\left(\mbox{\boldmath $A$}\right)\,\partial \mbox{\boldmath $A$}}
\,,
\end{eqnarray}
where $\partial \mbox{\boldmath $A$}^{\mbox{\tiny DVCS}}/\partial \mbox{\boldmath $A$}$ is a $(n-m)\times n$ matrix that projects on the subspace of charge even asymmetries.

As we have shown, we can map, under certain
assumptions, the measured observables into the (sub)space of
 CFFs. We may collect these into the vector
$$\mbox{\boldmath $\cffF$}^\intercal = (\cffF_1^{\Im},\cdots,\cffF_n^{\Im},\cffF_1^{\Re},\cdots,\cffF_n^{\Re})\,,$$
where the \cffsreim\ ${\cal G}\,\cdots, {\cal G}_{n-m}$ are now a part of  ${\cal F}$ and $n$ is now replaced by the number $2n$.
Combining  the linear and  quadratic solutions (\ref{eq:A2reE}) allows us to find these  $2n$ \cffsreim\  from the asymmetries by means of
\begin{eqnarray}
\label{eq:A2F}
 \mbox{\boldmath ${\cal F}$} = \frac{1}{N\!\left(\mbox{\boldmath $A$}\right)}\, \mbox{\boldmath $c$}(\mbox{\boldmath $A$}|N\!\left(\mbox{\boldmath $A$}\right)) \cdot \mbox{\boldmath $A$}
\,,
\end{eqnarray}
which can be non-linear in the measured asymmetries $\mbox{\boldmath $A$}$, specified as vectors in (\ref{def-A-full}).
The  covariance matrix is calculated in the common manner as in (\ref{eq:cov}), replace there $\mbox{\boldmath $A$}^{\cal I}$ by  \mbox{\boldmath $A$}, where the gradient is now  given as a $2n\times 2n$ matrix
\begin{equation}
\label{eq:A2F-gradient}
\frac{\partial \mbox{\boldmath ${\cal F}$}}{\partial \mbox{\boldmath $A$}} =  \frac{1}{N\!\left(\mbox{\boldmath $A$}\right)}\left[ \mbox{\boldmath $c$}(\mbox{\boldmath $A$}) +
\frac{\partial \mbox{\boldmath $c$}(\mbox{\boldmath $A$})}{\partial \mbox{\boldmath $A$}}   \cdot \mbox{\boldmath $A$}  -
\mbox{\boldmath ${\cal F}$}\otimes \frac{\partial N(\mbox{\boldmath $A$})}{\partial \mbox{\boldmath $A$}}
\right].
\end{equation}
Let us recall that the normalization factor $N$ as a function of the $2n$ asymmetry measurements follows from
substituting the solution of (\ref{eq:A2F}) into the definition (\ref{eq:N}), which also allows us to calculate its gradient:
\begin{eqnarray}
\label{eq:cfftE2N}
N(\mbox{\boldmath $A$})\,, \quad  \frac{\partial N(\mbox{\boldmath $A$})}{N\,\partial \mbox{\boldmath $A$}}
\,.
\end{eqnarray}
Remaining observables that are not used for the extraction of CFFs can be evaluated from the solution (\ref{eq:A2F}) and may serve as a test for
the validity of assumptions.

\subsection{Local extraction of Compton form factors}
\label{sec:localExtraction}

As explained in the preceding section, the extraction of CFFs  can be considered as a map of
random variables from the space of observables to the space of \cffsreim\
rather than as a fitting problem in which one
relies on a given model and tries to find its parameters by means of statistical methods.
Of course, one may consider the theory as a model and  use the  least-squares method (or maximum likelihood  estimation) to extract locally CFFs from measurements.
In the following sections we use both points of view in local extraction procedures and confront the findings.

The fourteen H{\sc ermes} measurements, which we will utilize in different variations, together with the definition of asymmetries,
described in Sect.~\ref{sec:analyze-cffs2obs},  are listed in Tab.~\ref{tab:obs}. Naturally, we will arrange the selected
asymmetries as a vector, e.g., an eight dimensional one $\mbox{\boldmath $A$}^\intercal =(A_1,\cdots, A_8)$.
In our analyses we assume that experimental errors are normally distributed, and the error
propagation is performed as described in
Sect.~\ref{sec:maps}. The correlations of experimental
errors have not been analyzed to the full extent; however,
the results for the overall values as published by the
H{\sc ermes} collaboration may suggest that the
covariance matrix is mostly diagonal \cite{Airapetian:2012mq}. Thus, we may safely assume in the following that the experimental errors are uncorrelated,
e.g., a eight-dimensional covariance matrix
\begin{eqnarray}
\label{cov-A}
 {\rm cov}\!\left(\mbox{\boldmath $A$}\right) = \left(
                \begin{array}{ccccc}
                  \delta^2 A_1& 0 & \cdots & \cdots &0\\
                  0  &  \delta^2 A_2   & 0 &  \cdots &  0\\
                  \vdots  &  \cdots   & \ddots  &  \cdots &  \vdots\\
                  0   & \cdots  & \cdots & \delta^2 A_7  &0\\
                  0  & \cdots & \cdots & 0 &\delta^2 A_8\\
                \end{array}
              \right)
\end{eqnarray}
contains only diagonal entries, which for simplicity (actually,
due to missing information we cannot do better) are given as
the sums of squared statistical and systematic errors.

\begin{table}[t]
\centering
\begin{tabular}{|l|lc|l|}
\hline
Observable & Definition & Formulae & Data from ref.  \\
\hline\hline
$\Asin{1}{LU,I}$ &
$\int_{-\pi}^\pi\!\frac{d\phi\,\sin\phi}{\pi} A_{\rm LU,I}(\phi)\phantom{\Big|}$ &
\phantom{1}(\ref{A_LUI}) and (\ref{A_LUI^sinphi-BMK},\ref{C_unp^I}) &
\cite{Airapetian:2012mq}, see~Sect.~\ref{sec:HERMES}  \\
$\Acos{1}{C}$ &
$\int_{-\pi}^\pi\!\frac{d\phi\,\cos\phi}{\pi} A_{\rm C}(\phi) \phantom{\Big|}$ &
(\ref{A_C})  and (\ref{A_C^cosphi-BMK},\ref{C_unp^I}) &
\cite{Airapetian:2012mq}, see Sect.~\ref{sec:HERMES}    \\
\hline
$\Acos{0}{C}$ &
$\int_{-\pi}^\pi\!\frac{d\phi}{2\pi} A_{\rm C}(\phi)\phantom{\Big|}$ &
(\ref{A_C})  and (\ref{A_C^cos0phi-BMK},\ref{C_unp^I}) &
\cite{Airapetian:2012mq}, see Sect.~\ref{sec:HERMES}      \\
\hline\hline
$\Asin{1}{UL,+}$ &
$\int_{-\pi}^\pi\!\frac{d\phi\,\sin\phi}{\pi} A_{{\rm UL},+}(\phi) \phantom{\Big|}$ &
(\ref{A_UL+}) and (\ref{A_UL+^sinphi-BMK},\ref{C^I_LP})  &
Tab.~4 from \cite{Airapetian:2010ab}  \\
$\Acos{1}{LL,+}$ &
$\int_{-\pi}^\pi\!\frac{d\phi\, \cos\phi}{\pi} A_{{\rm LL},+}(\phi)\phantom{\Big|}$ &
(\ref{A_LL+}) and (\ref{A_LL+^cosphi-BMK},\ref{C^I_LP})   &
Tab.~4 from \cite{Airapetian:2010ab}  \\
\hline
$\Acos{0}{LL,+}$ &
$\int_{-\pi}^\pi\!\frac{d\phi}{2\pi} A_{{\rm LL},+}(\phi)\phantom{\Big|}$ &
(\ref{A_LL+}) and (\ref{A_LL+^cos0phi-BMK},\ref{A_LLI^cos0phi-BMK})  &
Tab.~4 from \cite{Airapetian:2010ab}     \\
\hline\hline
$\Asincos{1}{UT,I}$ &
$\int_{-\pi}^\pi\!\frac{d\varphi\,\sin\varphi}{\pi}\!\int_{-\pi}^\pi\!\frac{d\phi\,\cos\phi}{\pi}
A_{\rm UT, I}(\phi,\varphi)\phantom{\Big|}\phantom{\Big|}$ &
(\ref{A_UTI}) and (\ref{A_UTI^sinvarphicosphi-BMK},\ref{C^I_TP-})  &
Tab.~1b  from \cite{Airapetian:2008aa}  \\
$\Acossin{1}{UT,I}$ &
$\int_{-\pi}^\pi\!\frac{d\varphi\,\cos\varphi}{\pi}\!
\int_{-\pi}^\pi\!\frac{d\phi\,\sin\phi}{\pi}
A_{\rm UT, I}(\phi,\varphi)\phantom{\Big|}\phantom{\Big|}$ &
(\ref{A_UTI}) and (\ref{A_UTI^cosvarphisinphi-BMK},\ref{C^I_TP+})  &
Tab.~1b  from \cite{Airapetian:2008aa}   \\
$\Asincos{0}{UT,DVCS}$ &
$\int_{-\pi}^\pi\!\frac{d\varphi\,\sin\varphi}{\pi}\!\int_{-\pi}^\pi\!\frac{d\phi}{2\pi}A_{\rm UT, DVCS}(\phi,\varphi)\phantom{\Big|}$ &
(\ref{AUT_DVCS}) and (\ref{A_UTDVCS^sinvarphicos0phi-BMK})\phantom{10}   &
Tab.~1a  from \cite{Airapetian:2008aa}  \\
\hline
$\Asincos{0}{UT,I}$ &
$\int_{-\pi}^\pi\!\frac{d\varphi\,\sin\varphi}{\pi}\!\int_{-\pi}^\pi\!\frac{d\phi}{2\pi}
A_{\rm UT, I}(\phi,\varphi)\phantom{\Big|}$ &
(\ref{A_UTI}) and (\ref{A_UTI^sinvarphicos0phi-BMK},\ref{C^I_TP-})  &
Tab.~1b from \cite{Airapetian:2008aa}   \\
\hline\hline
$\Asinsin{1}{LT,I}$ &
$\int_{-\pi}^\pi\!\frac{d\varphi\,\sin\varphi}{\pi}\!\int_{-\pi}^\pi\!\frac{d\phi\,\sin\phi}{\pi}
A_{\rm LT, I}(\phi,\varphi)\phantom{\Big|}$ &
(\ref{A_LTI}) and (\ref{A_LTI^sinvarphisinphi-BMK},\ref{C^I_TP-})    &
Tab.~2 from \cite{Airapetian:2011uq}    \\
$\Acoscos{1}{LT,I}$ &
$\int_{-\pi}^\pi\!\frac{d\varphi\,\cos\varphi}{\pi}\!\int_{-\pi}^\pi\!\frac{d\phi\,\cos\phi}{\pi}
A_{\rm LT, I}(\phi,\varphi)\phantom{\Big|}$ &
(\ref{A_LTI}) and (\ref{A_LTI^cosvarphicosphi-BMK},\ref{C^I_TP+})   &
Tab.~2 from \cite{Airapetian:2011uq}    \\
$\Acoscos{0}{LT,BH+DVCS}$ &
$\int_{-\pi}^\pi\!\frac{d\varphi\,\cos\varphi}{\pi}\!\int_{-\pi}^\pi\!\frac{d\phi}{2\pi}A_{\rm LT,even}(\phi,\varphi)\phantom{\Big|}$ &
(\ref{ALTBHDVCS}) and (\ref{A_LTDVCS^cosvarphicos0phi-BMK})\phantom{10}  &
Tab.~3 from \cite{Airapetian:2011uq}   \\
\hline
$\Acoscos{0}{LT,I}$ &
$\int_{-\pi}^\pi\!\frac{d\varphi\,\cos\varphi}{\pi}\!\int_{-\pi}^\pi\!\frac{d\phi}{2\pi}
A_{\rm LT, I}(\phi,\varphi)\phantom{\Big|}$ &
(\ref{A_LTI}) and (\ref{A_UTI^cosvarphicos0phi-BMK},\ref{C^I_TP+})  &
Tab.~2 from \cite{Airapetian:2011uq}    \\
\hline
\end{tabular}
\caption{\small
Observables from H{\sc ermes} measurements that are utilized for the extraction of twist-two associated CFFs.}
\label{tab:obs}
\end{table}

As there is no broad consensus for the precise definition of CFFs, we adopt here
the conventions of \cite{Belitsky:2010jw} for unpolarized and longitudinally polarized target asymmetries and take for transversally polarized ones the $1/\Q$ expanded expressions from \cite{Belitsky:2001ns}.  Moreover, we will restrict ourselves here to the twist-two sector, i.e. to the CFFs ${\cal F}\in \{\cal H, \cal E, \widetilde{\cal H}, \widetilde{\cal E}\}$ related to the observables by setting the remaining eight CFFs ${\cal F}_{\rm eff}$ and ${\cal F}_{\rm T}$ to zero, which is justified by the fact that higher harmonics are compatible with zero or are difficult to interpret\footnote{In contrast
to other higher harmonics,
in the longitudinal single target asymmetry a relatively large
$\sin(2\phi)$ moment with large uncertainties has been observed, which
increases with growing $-t$  and which is maximally up to $\sim
2\sigma$ deviations away from zero \cite{Airapetian:2010ab}. However,
this large value is concentrated in a single bin in the
$x_{\mathrm{B}}$ projection, which is supportive of concluding that
the amplitude is simply a fluctuation.
}.
The imaginary and real parts of the twist-two associated CFFs $\cffF$ are collected in an eight dimensional vector
\begin{eqnarray}
\label{"CFFs-8"}
\mbox{\boldmath $\cffF$}^\intercal = \left(
\cffH^{\Im},\cfftH^{\Im},\cffE^{\Im}, \cffbE^{\Im},\cffH^{\Re},\cfftH^{\Re},\cffE^{\Re}, \cffbE^{\Re}
\right),
\end{eqnarray}
written below as $\mbox{\boldmath $\cffF$}=  \left(
{ \mbox{\boldmath $\cffF$}^{\Im}  \atop \mbox{\boldmath $\cffF$}^{\Re} }
\right)$  in terms of  two four dimensional column vectors
\begin{eqnarray}
\label{"CFFs"}
\mbox{\boldmath $\cffF$}^{\Im} = \Im\!\left(\!\!
\begin{array}{c}
\cffH \\
\cfftH \\
\cffE \\
\cffbE \\
\end{array}
\!\!\right)
\quad\mbox{and}\quad
\mbox{\boldmath $\cffF$}^{\Re} = \Re\!\left(\!\!
\begin{array}{c}
\cffH \\
\cfftH \\
\cffE \\
\cffbE \\
\end{array}
\!\!\right),
\quad\mbox{where we use   }
\cffbE = \frac{\xB}{2-\xB} \cfftE\,.
\end{eqnarray}
The (approximate) redefinition  of $\cfftE$ into  $\cffbE$ removes the factor $\xB/(2-\xB)$  that enters in the common form factor decomposition and makes the kinematical coefficients of $\cfftE$ in the cross section rather small. This redefinition restores also the common ``Regge'' behavior, i.e.,
$\cffbE \propto \xB^{-\alpha(t)}$ for small $\xB$ (see
\cite{Bechler:2009me}) and simplifies the discussion of the real photon limit for CFFs \cite{Belitsky:2012ch}. However, the common definition is better suited to relate GPDs to form factors.

\subsubsection{Maps of asymmetries to CFFs}
\label{sec:regression}

To determine the \cffsreim\ (\ref{"CFFs"}) by means of a map (\ref{eq:A2F}),
we select eight twist-two related asymmetries out of the measured first harmonics,
listed in Tab.~\ref{tab:obs}.
Thereby, we take the kinematical means from Tab.~\ref{tab:means}. In
any of the possible  maps we assume the twist-two dominance hypotheses
and we naturally employ for the access of the imaginary parts of the twist-two associated CFFs $\cffF$ (specified in (\ref{"CFFs"})), the dominant single spin asymmetries in the following sequence
\begin{eqnarray}
\label{Asin-HERMES}
\mbox{\boldmath ${A}$}^{\sin}\equiv
\left(\!\!
\begin{array}{c}
A_1\\
A_2\\
A_3\\
A_4
\end{array}
\!\!\right)=\left(\!\!
\begin{array}{c}
\Asin{1}{LU,I} \\
\Asin{1}{UL,+}\\
\Asincos{1}{UT,I}\\
\Acossin{1}{UT,I} \\
\end{array}
\!\!\right)
\quad \Rightarrow \quad
\mbox{\boldmath $\cffF$}^{\Im} =\Im\!\left(\!\!
\begin{array}{c}
\cffH \\
\cfftH \\
\cffE \\
\cffbE \\
\end{array}
\!\!\right),
\end{eqnarray}
see the expressions  (\ref{A_LUI^sinphi-BMK},\ref{A_UL+^sinphi-BMK},\ref{A_UTI^sinvarphicosphi-BMK},\ref{A_UTI^cosvarphisinphi-BMK}) for asymmetries in
terms of CFF combinations (\ref{C_unp^I},\ref{C^I_LP},\ref{C^I_TP-},\ref{C^I_TP+}).  The analogous relation for the corresponding
even harmonics and the real parts of CFFs reads
\begin{eqnarray}
\label{Acos-HERMES}
\mbox{\boldmath ${A}$}^{\cos}\equiv
\left(\!\!
\begin{array}{c}
A_5\\
A_6\\
A_7\\
A_8
\end{array}
\!\!\right)=
\left(\!\!
\begin{array}{c}
\Acos{1}{C} \\
\Acos{1}{LL,+}\\
\Asinsin{1}{LT,I}\\
\Acoscos{1}{LT,I} \\
\end{array}
\!\!\right)
\quad \Rightarrow \quad
\mbox{\boldmath $\cffF$}^{\Re} =\Re\!\left(\!\!
\begin{array}{c}
\cffH \\
\cfftH \\
\cffE \\
\cffbE \\
\end{array}
\!\!\right),
\end{eqnarray}
see (\ref{A_C^cosphi-BMK},\ref{A_LL+^cosphi-BMK},\ref{A_LTI^sinvarphisinphi-BMK},\ref{A_LTI^cosvarphicosphi-BMK}) and (\ref{C_unp^I},\ref{C^I_LP},\ref{C^I_TP-},\ref{C^I_TP+}).
In the following  we utilize, however, two  different analytic methods  for the extraction of the real parts.

First, we consider a linearized map (\ref{eq:A2F-generic}), where we take only charge odd asymmetries that arise from the interference of BH and DVCS processes.
Thus, we transform the single longitudinally polarized target spin asymmetry $\Asin{1}{UL,+}$ and the longitudinally double spin-flip asymmetry $\Acos{1}{LL,+}$
(both measured only with a positron beam), appearing in the four dimensional vectors (\ref{Asin-HERMES}) and (\ref{Acos-HERMES}), into the charge odd sector. In doing so, we eliminate the interference term in the denominator of  these asymmetries  by means of the known beam charge asymmetry,
i.e.~we find from the relation (\ref{A_UL-relations}):
\begin{eqnarray}
\Asin{1}{UL,I} &\!\!\! =\!\!\!& \Asin{1}{UL,+} \left[1+  \Acos{0}{C} -\frac{1}{2}\Acos{2}{C}\right] +
\frac{1}{2} \Asin{2}{UL,+} \left[\Acos{1}{C} -\Acos{3}{C}\right] + \cdots -  \Asin{1}{UL,DVCS} \,,
\nonumber\\
&\!\!\! \approx \!\!\!& \Asin{1}{UL,+} \left[1+  \Acos{0}{C}\right],
\phantom{\Bigg]}
\\
\Acos{1}{LL,I} &\!\!\!= \!\!\!&
\Acos{1}{LL,+} \left[1+ \Acos{0}{C} + \frac{1}{2} \Acos{2}{C}\right] +\Acos{0}{LL,+} \Acos{1}{C} + \frac{1}{2}\Acos{2}{LL,+}\left[\Acos{1}{C}  + \Acos{3}{C} \right],
\nonumber\\
&&+\cdots - \Acos{1}{LL,BH}- \Acos{1}{LL,DVCS} \,,
\phantom{\bigg]}
\nonumber\\
&\!\!\! \approx \!\!\!& \Acos{1}{LL,+} \left[1+ \Acos{0}{C} \right] - \Acos{1}{LL,BH}\,,
\end{eqnarray}
and we neglect the twist-three related asymmetries $\Acos{2}{C}$, $\Asin{2}{UL,+}$, $\Acos{2}{LL,+}$,  $\Asin{1}{UL,DVCS}$, $\Acos{1}{LL,DVCS}$,
and higher harmonics, indicated by ellipses.
Moreover, for the charge odd double longitudinal spin asymmetry, the first harmonic of the BH contribution turns out to be small and
can thus be safely subtracted (proton form factor uncertainties are neglected).
The imaginary and real parts of the CFFs follow from the solution of two linearized homogenous equation, see (\ref{eq:A2F-generic}) with $ \mbox{\boldmath $b$} =0$, which we write in the form of the map (\ref{eq:A2F}) as
\begin{eqnarray}
\label{A2CFF-lin}
\left({ \mbox{\boldmath $\cffF$}^{\Im} \atop \mbox{\boldmath $\cffF$}^{\Re} }\right) =
\frac{1}{N\!\left(\mbox{\boldmath $A$}\right)}  \left(
\begin{array}{cc}
 \mbox{\boldmath $c$}_{\Im} & {\bf 0}_{4\times 4} \\
 {\bf 0}_{4\times 4}  &  \mbox{\boldmath $c$}_{\Re}
  \end{array}
\right) \cdot  \left({
            \mbox{\boldmath $A$}^{\sin} \atop
            \mbox{\boldmath $A$}^{\cos} }
        \right)\,.
\end{eqnarray}
Here ${\bf 0}_{n\times m} $ denotes a $n$-by-$m$ zero matrix in the $8\times8$ coefficient matrix $\mbox{\boldmath $c$}$, calculated by the inversion of (\ref{c^{-1}-matrices}), and the first odd and even harmonics
\begin{eqnarray}
\label{obs-interference}
\mbox{\boldmath ${A}$}^{\sin}\equiv
\left(\!\!
\begin{array}{c}
A_1\\
A_2\\
A_3\\
A_4
\end{array}
\!\!\right)=
\left(\!\!
\begin{array}{c}
\Asin{1}{LU,I} \\
\Asin{1}{UL,I}\\
\Asincos{1}{UT,I}\\
\Acossin{1}{UT,I} \\
\end{array}
\!\!\right)
\quad\mbox{and}\quad
\mbox{\boldmath ${A}$}^{\cos}\equiv
\left(\!\!
\begin{array}{c}
A_5\\
A_6\\
A_7\\
A_8
\end{array}
\!\!\right)=
\left(\!\!
\begin{array}{c}
\Acos{1}{C} \\
\Acos{1}{LL,I}\\
\Asinsin{1}{LT,I}\\
\Acoscos{1}{LT,I} \\
\end{array}
\!\!\right),
\end{eqnarray}
respectively.
Consequently, in this linearized map, the cross talk of imaginary and real parts of \cffsreim{}
arises only via the overall normalization factor,  which we analytically determine from the cubic equation (\ref{eq:N}).
The gradient of this linearized \cffsreim\ solution can be analytically calculated by means of (\ref{partial_A F-hom}) and the  covariance
matrix follows from (\ref{eq:cov},\ref{eq:cov-hom}) and (\ref{cov-A}).

Unfortunately, the statistics in the double spin-flip
asymmetry measurements are rather limited, in particular, for
the $\Asinsin{1}{LT,I}$ and $\Acoscos{1}{LT,I}$ harmonics
that essentially constrain the real value of $\cffE$ and $\cffbE$. Hence, we try a more general map (\ref{eq:A2F}) where
we replace the two longitudinal-transverse double-flip asymmetries in the charge odd sector  by  the two DVCS-squared related asymmetries $\Asincos{0}{UT,DVCS}$ and $\Acos{0}{LL,+}$,
which are in general measured with smaller uncertainties, exemplified in Fig.~\ref{fig:A_...-HERMES}. We arrange our eight asymmetries now into three
parts,
\begin{eqnarray}
\label{obs-INTDVCS}
\mbox{\boldmath $A$}=
\left(\!\!
          \begin{array}{c}
            \mbox{\boldmath $A$}^{\sin} \\
            \mbox{\boldmath $A$}^{\cos} \\
            \mbox{\boldmath $A$}^{\mbox{\tiny DVCS}} \\
          \end{array}
        \!\!\right)
\;\;\mbox{with}\;\;
\mbox{\boldmath $A$}^{\cos} \equiv
\left(\!\!
\begin{array}{c}
A_5\\
A_6
\end{array}
\!\!\right)=
 \left(\!\!
\begin{array}{c}
\Acos{1}{C} \\
\Acos{1}{LL,I}
\end{array}
\!\!\right),\quad
\mbox{\boldmath $A$}^{\mbox{\tiny DVCS}}\equiv
\left(\!\!
\begin{array}{c}
A_7 \\
A_8
\end{array}\!\!
\right)=
\left(\!\!
\begin{array}{c}
\Asincos{0}{UT,DVCS}\\
\Acos{0}{LL,+}
\end{array}\!\!
\right),
\nonumber\\
\end{eqnarray}
and  $\mbox{\boldmath $A$}^{\sin}$ is the same vector as in (\ref{obs-interference}). For the first even harmonics of the charge odd asymmetries
$\mbox{\boldmath $A$}^{\cos}$ we take here the first two entries
of the corresponding four dimensional vector in (\ref{obs-interference}), which are mostly sensitive to the real parts of CFFs $\cffH$ and $\cfftH$.
Consequently,  we split the four-dimensional vector $\mbox{\boldmath $\cffF$}^{\Re}$ into two two-dimensional vectors,
\begin{eqnarray}
\mbox{\boldmath $\cffF$}^{\Re}=
\left(\!\!
\begin{array}{c}
\mbox{\boldmath $\cffH$}^{\Re} \\
\mbox{\boldmath $\cffE$}^{\Re}
\end{array}
\!\!\right)
\quad\mbox{with}\quad
\mbox{\boldmath $\cffH$}^{\Re}= \Re\! \left(\!\!
\begin{array}{c}
\cffH \\
\cfftH
\end{array}
\!\!\right)
\quad\mbox{and}\quad
\mbox{\boldmath $\cffE$}^{\Re} = \Re\! \left(\!\!
\begin{array}{c}
\cffE  \\
\cffbE
\end{array}
\!\!\right).
\end{eqnarray}
Here, $\mbox{\boldmath $\cffH$}^{\Re}$ ($=\mbox{\boldmath $\cffF$}$ in
the notation of Sect.~\ref{sec:maps}) is considered as
a solution of the inhomogeneous equation (\ref{eq:A2F-generic}) and $\mbox{\boldmath $\cffE$}^{\Re}$ ($=\mbox{\boldmath $\cal G$}$ in the notation of Sect.~\ref{sec:maps}) governs its inhomogeneous term, i.e.,
\begin{eqnarray}
\label{eq:A2reF-Hermes}
\mbox{\boldmath $\cffH$}^{\Re} =  \frac{1}{N(\mbox{\boldmath $A$})}  \mbox{\boldmath $c$}_{\Re}\cdot
\mbox{\boldmath $A$}^{\cos} -  \mbox{\boldmath $c$}_{\Re}\cdot  \mbox{\boldmath $b$}_{\Re} \cdot \mbox{\boldmath $\cffE$}^{\Re}\,.
\end{eqnarray}
The transverse single target spin asymmetry  in the charge even sector is proportional to the real parts of CFFs $\cffE$ and $\cfftE$,  see (\ref{A_UTDVCS^sinvarphicos0phi-BMK}), and provides us one linear
constraint
\begin{eqnarray}
\label{eq:AUTDVCS2ReF-HERMES}
A_7\equiv \Asincos{0}{UT,DVCS} = N(\mbox{\boldmath $A$})\, \mbox{\boldmath $ c$}^\intercal_{\cffE}(\mbox{\boldmath $A$}^{\sin},\mbox{\boldmath $A$}^{\cos})\cdot \mbox{\boldmath $\cffE$}^{\Re},
\end{eqnarray}
where the $ \mbox{\boldmath $ c$}^\intercal_{\cffE}$-coefficient
depends on the  six charge odd asymmetries $\mbox{\boldmath $A$}^{\sin}$ and $\mbox{\boldmath $A$}^{\cos}$ or \cffsreim\ $\mbox{\boldmath $\cffF^{\Im}$}$ and $\mbox{\boldmath $\cffH^{\Re}$}$  that are extracted from the set of linear equations. Our system of equations is completed by  the  measurements of the longitudinal  double spin flip asymmetry (\ref{A_LL+^cos0phi-BMK}),
\begin{eqnarray}
A_8 \equiv \Acos{0}{LL,+} = \Acos{0}{LL,BH}+\Acos{0}{LL,I}  + \Acos{0}{LL,DVCS} \,.
\nonumber
\end{eqnarray}
It contains (in addition to a large BH-squared term) contributions from the polarized interference term and the
twist-two DVCS-squared contribution in the numerator.
Here we cannot assume that the interference term overwhelms the DVCS-squared term;
rather we assume that the two terms enter kinematically on the same level. Hence we have a quadratic constraint  for the CFF, which we write as
\begin{eqnarray}
\label{eq:ALL2ReEtE-HERMES}
 A_8\equiv \Acos{0}{LL,+} = N(\mbox{\boldmath $A$}) \left[ a(\mbox{\boldmath $A$}^{\sin},\mbox{\boldmath $A$}^{\cos})+ {\bf b}^\intercal(\mbox{\boldmath $A$}^{\sin},\mbox{\boldmath $A$}^{\cos})  \cdot  \mbox{\boldmath $\cffE$}^{\Re} +
\mbox{\boldmath $\cffE$}^{\Re ^\intercal} \cdot  {\bf c}(\mbox{\boldmath $A$}^{\sin},\mbox{\boldmath $A$}^{\cos})\cdot \mbox{\boldmath $\cffE$}^{\Re}
\right],
\end{eqnarray}
where all coefficients depend again on the six measurements $\mbox{\boldmath $A$}^{\sin}$ and $\mbox{\boldmath $A$}^{\cos}$, used previously.
The linear constraint (\ref{eq:AUTDVCS2ReF-HERMES}) together with the quadratical one (\ref{eq:ALL2ReEtE-HERMES}) can be analytically solved, giving us the desired last two \cffsreim\ which we write as a matrix equation, too
\begin{eqnarray}
\label{eq:A2reE_1-HERMES}
 \mbox{\boldmath $\cffE$}^{\Re} = \frac{1}{N\!\left(\mbox{\boldmath $A$}\right)}\,
 \mbox{\boldmath $c$}_{\cffE}(\mbox{\boldmath $A$}|N(\mbox{\boldmath $A$})) \cdot
\mbox{\boldmath $A$}^{\mbox{\tiny DVCS}}\,,
\end{eqnarray}
where the $2\times 2$ matrix $\mbox{\boldmath $c$}_{\cffE}(\mbox{\boldmath $A$}|N(\mbox{\boldmath $A$}))$ is a non-linear function of
the eight asymmetries $\mbox{\boldmath $A$}$.
The final solution can be written as in (\ref{eq:A2F}) with a $8\times 8$ coefficient matrix $\mbox{\boldmath $c$}(\mbox{\boldmath $A$}|N(\mbox{\boldmath $A$}))$,
\begin{eqnarray}
\label{A2CFF-nonlin}
\left(\!\!
          \begin{array}{c}
            \mbox{\boldmath $\cffF$}^{\Im} \\
            \mbox{\boldmath $\cffH$}^{\Re} \\
            \mbox{\boldmath $\cffE$}^{\Re} \\
          \end{array}\!\!
        \right) = \frac{1}{N\!\left(\mbox{\boldmath $A$}\right)}
\left(
\begin{array}{ccc}
  \mbox{\boldmath $c$}_{\Im} & {\bf 0}_{4\times 2} &  {\bf 0}_{4\times 2} \\
 {\bf 0}_{2\times 4}  &  \mbox{\boldmath $c$}_{\Re} &   -  \mbox{\boldmath $c$}_{\Re}\cdot {\bf b}_{\Re} \cdot
 \mbox{\boldmath $c$}_{\cffE}(\mbox{\boldmath $A$}|N(\mbox{\boldmath $A$})) \\
 {\bf 0}_{2\times 4}  & {\bf 0}_{2\times 2}  &   \mbox{\boldmath $c$}_{\cffE}(\mbox{\boldmath $A$}|N(\mbox{\boldmath $A$}))
  \end{array}
\right) \cdot \left(\!\!
          \begin{array}{c}
            \mbox{\boldmath $A$}^{\sin} \\
            \mbox{\boldmath $A$}^{\cos} \\
            \mbox{\boldmath $A$}^{\mbox{\tiny DVCS}} \\
          \end{array}
        \!\!\right)
\,.
\end{eqnarray}
The normalization $N(\mbox{\boldmath $A$})$ is again determined by the consistency equation (\ref{eq:N}). The gradient of the \cffsreim\ solution is calculated by means of (\ref{eq:A2F-gradient}) and the covariance matrix
${\rm cov}(\mbox{\boldmath ${\cffF}$})$ is calculated from (\ref{eq:cov}), replaced there $\mbox{\boldmath $A$}^{\cal I}$ by  \mbox{\boldmath $A$},  and (\ref{cov-A}).

A few comments are in order.

It is clear that, within our hypothesis of twist-two
dominance, the matrix valued equations (\ref{A2CFF-lin},\ref{A2CFF-nonlin}) are solutions of well defined problems. However,
in practice we have to deal with the
fact that the eigenvalues of
$\mbox{\boldmath $c$}^{-1}$ (or its inverse $\mbox{\boldmath $c$}$) matrix can become very small (big).  Obviously, this unpleasant property arises from the fact
that in particular the helicity target flip CFFs $\cffE$ and $\cffbE$  are kinematically suppressed in the set of observables.

In addition we have six observables, namely the three lowest harmonics (\ref{A_C^cos0phi-BMK},\ref{A_UTI^sinvarphicos0phi-BMK},\ref{A_UTI^cosvarphicos0phi-BMK})
$$
\Acos{0}{C}\,, \quad  \Asincos{0}{UT,I}\,, \quad \Acoscos{0}{LT,I}\,,
$$
and three twist-two related asymmetries (\ref{A_UTDVCS^sinvarphicos0phi-BMK},\ref{A_LTDVCS^cosvarphicos0phi-BMK},\ref{A_LL+^cos0phi-BMK})
$$
 \Asincos{0}{UT,DVCS}\,,\quad \Acoscos{0}{LT,BH+DVCS}\,,\quad \Acos{0}{LL,+}\quad\mbox{ for our linearized map (\ref{A2CFF-lin}),}
$$
or alternatively the asymmetries (\ref{A_LTI^sinvarphisinphi-BMK},\ref{A_LTI^cosvarphicosphi-BMK},\ref{A_LTDVCS^cosvarphicos0phi-BMK})
$$
 \Asinsin{1}{LT,I}\,,\quad \Acoscos{1}{LT,I}\,, \quad \Acoscos{0}{LT,BH+DVCS}\quad\mbox{ for our  map (\ref{A2CFF-nonlin})}\,,
$$
which are predicted by the extracted twist-two associated CFFs
and serve as a consistency check of our extraction procedure.
However, we should bear in mind that the three lowest harmonics
(\ref{A_C^cos0phi-BMK},\ref{A_UTI^sinvarphicos0phi-BMK},\ref{A_UTI^cosvarphicos0phi-BMK})
may be more strongly contaminated by twist-three related CFFs.

For comparison we will also employ the brute force
method, where we solve eight quadratic equations numerically and numerically evaluate the variation of the solution. This allows us to
judge the validity of the approximation used to linearize the constraints.
Thereby, we can employ the original H{\sc ermes} data for the first even and odd harmonics (\ref{Asin-HERMES},\ref{Acos-HERMES})
rather than the transformed ones (\ref{obs-interference}).
As for the linearized map, we obviously have six observables for consistency checks available.

Let us now present our results.
Within the linearized map  we find that the consistency equation for
the normalization (\ref{eq:N}) has two non-trivial real-valued
solutions in eleven out of the twelve kinematic bins and the overall asymmetry values, shown in Fig.~\ref{fig:A_...-HERMES}. According to the experimental indications,
we pick the solution for the BH regime, where we find for the weighted cross section ratio
\begin{eqnarray}
0.75 \le \frac{\int_{-\pi}^\pi\! d\phi\,w(\phi)d\sigma_{\rm BH}(\phi)}{\int_{-\pi}^\pi\! d\phi\, w(\phi)\left[d\sigma_{\rm BH}(\phi)  d\sigma_{\rm DVCS}(\phi)\right]} \le 0.95\,.
\end{eqnarray}
The inverse transformation of the \cffreim{} solution back to the original asymmetries, where we use the set of non-linear equations,  reproduces in general the means and standard errors of the seven asymmetries
$$
\left\{\Asin{1}{LU,I}, \Asin{1}{UL,+},\Asincos{1}{UT,I},\Acossin{1}{UT,I},\Acos{1}{C}, \Acos{0}{LL,+}, \Asincos{0}{UT,DVCS}\right\}\,,
$$
(the derivatives of which we have used for the map to the
\cffsreim\ space) in most of the cases on the level of a few percent.
We observe in some bins a larger deviation only for $\Acos{1}{LL,+}$,
which is naturally explained by the fact that we neglected in our linearization procedure the DVCS-squared term, which becomes important if the asymmetry
is small. However, also in this case the original data can be considered as well-reproduced. This inverse map is shown for all considered observables in Figs.~\ref{fig:unp}--\ref{fig:LTP}, shown in Appendix \ref{app:HERMESvsExtraction}, as empty circles.
We add that a typical mapping example is presented for the overall asymmetry values in Fig.~\ref{fig:A_...-HERMES}.

A true one-to-one map of random numbers is reached if we use the brute force method, where the start values may be taken
from the solution of the linearized map. Moreover, the mean values of the remaining asymmetries, used for the consistency check, are well reproduced on the $\sim 1 \sigma$  level, see
stars in Figs.~\ref{fig:pred_cos0phi} and \ref{fig:pred_BH}.

Let us have a closer look to the problematic bin $\#3$.   The linear map (\ref{A2CFF-lin}) yields
\begin{eqnarray}
\label{cffF:sol3-lin}
\Im  \left(
\begin{array}{r}
\cffH\\ \cfftH\\ \cffE \\ \cffbE \\
\end{array}
\right)
                          =\frac{1}{N(\mbox{\boldmath$A$})} \left(
                           \begin{array}{r}
                              7.6  \pm 1.2\\
                              1.8 \pm 1.3\\
                             -4.5 \pm 6.0 \\
                             11.0 \pm 6.1 \\
                           \end{array}
                         \right)
\quad\mbox{and}\quad
\Re  \left(
\begin{array}{r}
\cffH\\ \cfftH\\ \cffE \\ \cffbE \\
\end{array}
\right) =\frac{1}{N(\mbox{\boldmath$A$})} \left(
                           \begin{array}{r}
                             0.4 \pm  2.1\phantom{0} \\
                              5.7 \pm 4.1\phantom{0}\\
                             -28.7\pm   21.2 \\
                             -24.8 \pm  22.2 \\
                           \end{array}
                         \right)\,,
\end{eqnarray}
where the uncertainties are calculated from (\ref{eq:cov-hom}) by neglecting the variation of the normalization.
The consistency equation (\ref{eq:N}) provides us two complex valued solutions $N= 0.5 \pm 0.06\, i $, where the
smallness of the  imaginary part tells us that the inconsistency is rather weak.
From (\ref{cffF:sol3-lin}) we may conclude that the large mean values for $\cffE^{\Re}$ and $\cffbE^{\Re}$, which suffer from a large uncertainty,
ruin in turn the normalization constraint (\ref{eq:N}). Indeed, setting one (both) of them to zero allows us to solve the normalization constraint, where
$N\sim 0.75\, (0.85)$.
There are other possibilities to turn around the normalization inconsistency. In our case the data mapping and/or the linearization procedure
imply this inconsistency. Applying the brute-force method
to the eight original observables provides the solution:
\begin{eqnarray}
\label{cffF:sol3-BForce}
\Im \left(
\begin{array}{r}
\cffH\\ \cfftH\\ \cffE \\ \cffbE \\
\end{array}
\right) = \left(
                           \begin{array}{r}
                              11.8 \pm 8.9 \\
                              2.6  \pm  2.4 \\
                             -8.1  \pm 13.2 \\
                             15.6  \pm  12.7\\
                           \end{array}
                         \right)
\quad\mbox{and}\quad
\Re  \left(
\begin{array}{r}
\cffH\\ \cfftH\\ \cffE \\ \cffbE \\
\end{array}
\right) = \left(
                           \begin{array}{r}
                             0.8 \pm 3.6\\
                              6.2  \pm  9.9\\
                            -41.9  \pm  52.4\\
                            -43.3  \pm  66.6\\
                           \end{array}
                         \right)\,,
\end{eqnarray}
where the errors are propagated as discussed above. From these findings we obtain the relatively small normalization factor\footnote{We neglected here correlation of \cffreim{} errors otherwise the normalization error increases to 0.48. In turn we also neglect the error correlation in utilizing this normalization error in the linear map (\ref{cffF:sol3-lin}).} $N=0.64\pm 0.28$.
Substituting this number into (\ref{cffF:sol3-lin}) shows us that the \cffreim\ mean values from the linear map are compatible with those of (\ref{cffF:sol3-BForce}), obtained with the brute-force method. However, in particular, the net error for the \cffreim\ $\Im\cffH$, calculated from the \cffreim\ errors (\ref{cffF:sol3-lin}) and the normalization uncertainty, turns out to be smaller than that in (\ref{cffF:sol3-BForce}).

\begin{figure}[t]
\begin{center}
\includegraphics[width=15cm]{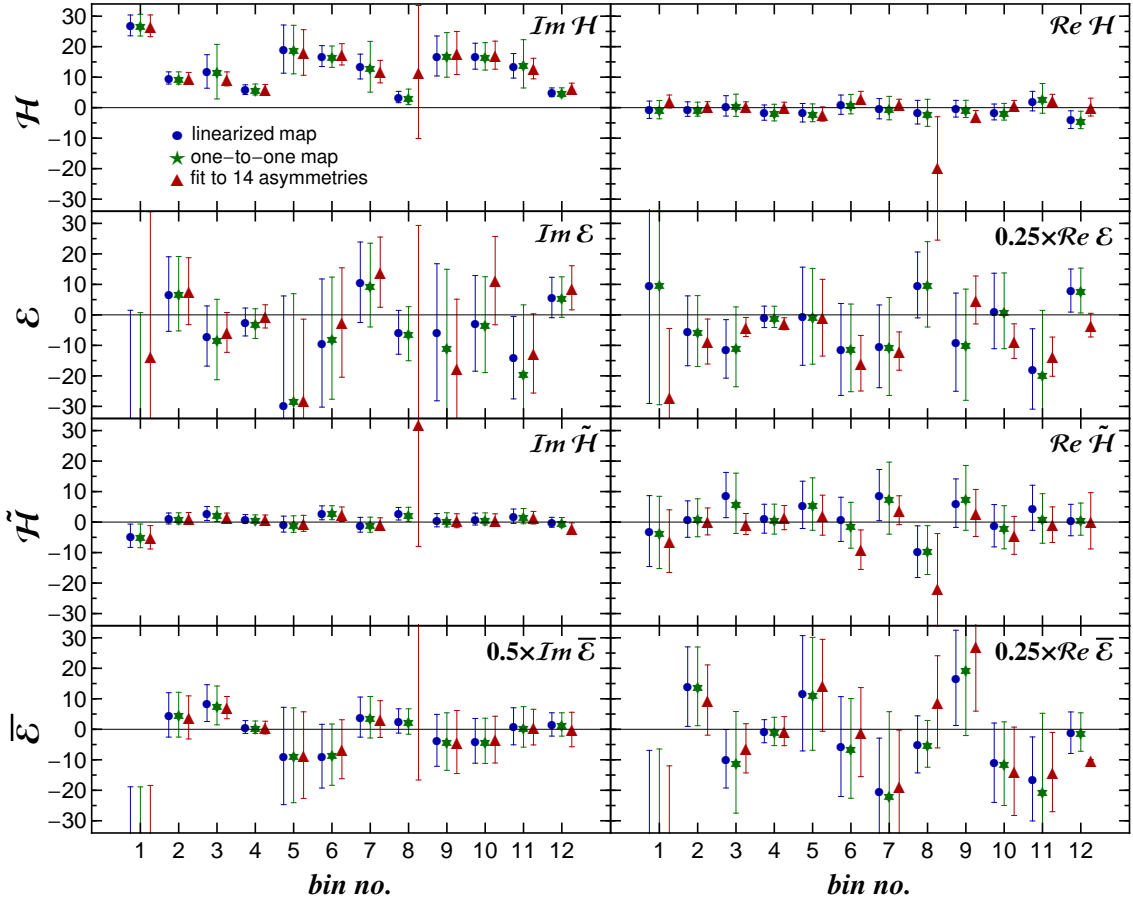}
\end{center}
\vspace{-0.5cm}
\caption{\small Resulting \cffsreim\ from a linearized (circles, shifted to the left) and a one-to-one map (stars)
of eight twist-two dominated charge odd asymmetries as well as from
a least squares fit (triangles,  shifted to the right)
to fourteen twist-two related observables for each of 12 H{\sc ermes} bins.}
\label{fig:methods}
\end{figure}

The resulting CFFs from the linearized (circles) and one-to-one (stars) maps  are presented for all twelve bins in Fig.~\ref{fig:methods}. Both maps
provide rather similar results, except that the errors for the \cffreim\
$\Im\cffH$ in bins $\#3$, $\#7$, and $\#11$ are approximately two times larger.
In these three bins the cross section ratio $N\lesssim 0.7$, calculated from the one-the-one map, is rather small.  We conclude that non-linear
effects in these circumstances are rather important for the error propagation (see above discussion for bin $\#3$).
Clearly, the imaginary part of $\cffH$ is (as expected) positive, rather large and incompatible with zero.
Note that the relatively large errors in bin $\#5$ and
$\#9$ are (partially) induced by the  fluctuations of $\Im
\cffE$; see for example the analytic expressions
for the beam spin (\ref{A_LUI^sinphi-BMK},\ref{C_unp^I}) and the $\cos\phi$  projection (\ref{A_UTI^sinvarphicosphi-BMK},\ref{C^I_TP-})
of the single transverse target  spin asymmetries.

All other \cffsreim\ can be considered within the uncertainties as compatible with zero. Surprisingly,  the imaginary part of $\cfftH$ possesses  even smaller absolute errors than $\Im \cffH$. Both of these \cffsreim\ are contaminated in the first place by proton helicity flip \cffsreim\
$\Im \cffE$ and   $\Im \cfftE$, respectively (see CFF combinations (\ref{C_unp^I}) and (\ref{C^I_LP})), which both suffer from larger uncertainties.
However, comparing the approximated expressions  (\ref{A_LUI^sinphi-BMK}) and (\ref{A_UL+^sinphi-BMK})  for the asymmetries $\Asin{1}{LU,I}$ and
$\Asin{1}{UL,+}$,
$$
\frac{\Asin{1}{LU,I}}{\Asin{1}{UL,I}} \backsimeq -\frac{y(2-y)}{2-2y+y^2}
\frac{\Im\big[F_1 \cffH - \frac{t}{4 M^2}F_2\cffE +\cdots \big]}{
\Im\big[F_1 \cfftH - \frac{t}{4 M^2}F_2\bar\cffE +\cdots \big]},
$$
one realizes that the beam spin asymmetry has an additional relative suppression factor,
$$
 \frac{y(2-y)}{2-2y+y^2} \sim  0.5\quad\mbox{with}\quad \xB \sim 0.1 \mbox{  and  } \Q^2 \sim 2.2\, \GeV^2
$$
for typical H{\sc ermes} kinematics. Hence, the larger errors of the polarized longitudinal target spin asymmetry are reduced in the propagation to the \cffsreim\ by a factor of two or so.  Another reason why the absolute error of $\Im \cffH$ is larger than of $\Im \cfftH$ is that the former \cffreim\ is more sizeable and so it  is also more important for the normalization ratio (\ref{eq:N}) than the latter. In return, $\Im \cffH$ suffers from a larger absolute error, see, e.g.~the explicit form of the covariance matrix (\ref{eq:cov-hom}) for a linearized map.  Note, however, that the twist-two hypothesis may induce here a bias, since we neglected here the twist-three induced $\sin\phi$ harmonic of the DVCS-squared term.

The imaginary parts of the proton helicity flip CFFs $\cffE$ and
$\cfftE$ are less constrained, partially due to larger errors of
the single transversely polarized target spin-flip
asymmetries --- however, also in part due to their being kinematically suppressed, see (\ref{A_UTI^sinvarphicosphi-BMK}, \ref{C^I_TP-}) and (\ref{A_UTI^cosvarphisinphi-BMK},\ref{C^I_TP+}).
Only the real part of  the  CFF $\cffH$, also compatible with zero, can be definitely considered as well constrained due to the precise data and since $\cffH$ dominantly enters in the charge asymmetry (\ref{A_C^cosphi-BMK},\ref{C_unp^I}) even without additional $y$ suppression. As one sees, the real parts of the remaining three \cffsreim\ are very noisy, in particular, those of the proton helicity flip CFFs.
A generic GPD model interpretation of our \cffreim\  findings is presented below in Sect.~\ref{sec:global}.

In the alternative map we replace the charge odd asymmetries $\Asinsin{1}{LT,I}$ and $\Acoscos{1}{LT,I}$ by $\Acos{0}{LL,+}$ and $\Asincos{0}{UT,DVCS}$
that contain the DVCS-squared term in the numerator. The normalization is now determined by the non-linear equation, see also (\ref{N-spin0-tw2}),
$$
N = N(A,N) \quad\mbox{with}\quad N\in \mathbb{R} \quad\mbox{and}\quad 0< N  < 1\,,
$$
the solution of which yields now four roots.
Literally taken, the mapping method is now only applicable in 7 out of 12 bins, namely, in
$$
\#1\,,\; \#2\,,\;  \#4\,,\;    \#9\,,\;  \#10\,,\;  \#11\,,\; \#12\,,
$$
and the overall bin.   The failure of the method in bins $\#3$
and $\#6$ is caused by the longitudinal double spin asymmetry
measurements $\Acos{0}{LL,+}$, which as mentioned above contradicts
the assumptions that the BH amplitude overwhelms the DVCS one, see
Fig.~\ref{fig:pred_cos0phi}. The failure in bin $\#8$ is related to
the large transverse spin  asymmetry $\Asincos{0}{UT,DVCS}$ (see Fig.~\ref{fig:pred_BH}) while the inconsistencies in the mapping method for the bins $\#5$ and $\#6$ are rather weak. Hence, we conclude that, even if the uncertainties of the both replaced charge odd asymmetries are rather large, these observables yield important constraints that ensure the consistency of the mapping method.

\subsubsection{Local least squares fits to asymmetries}
\label{sec:CFFfit}

Let us now employ the method  of least squares, where one looks for the minima
of the $\chi^2$ function
\begin{eqnarray}
\label{chi^2}
\chi^2(\mbox{\boldmath $\cffF$})
&\!\!\! = \!\!\!&
\left[\mbox{\boldmath $\hat{A}$} -\mbox{\boldmath $A$}(\mbox{\boldmath $\cffF$})\right]^\intercal \cdot {\rm cov}^{-1}(\mbox{\boldmath $\hat{A}$}) \cdot
\left[\mbox{\boldmath $\hat{A}$} -\mbox{\boldmath $A$}(\mbox{\boldmath $\cffF$})\right].
\end{eqnarray}
Here, the vector \mbox{\boldmath $\hat{A}$} contains the measured asymmetries and ${\rm cov}^{-1}(\mbox{\boldmath $\hat{A}$})$ is the inverse of the  covariance matrix.
For uncorrelated errors the covariance matrix (\ref{cov-A}) and its inverse are diagonal and the $\chi^2$-function (\ref{chi^2})
reduces to the most common form
$$
\chi^2(\mbox{\boldmath $\cffF$}) = \sum_{i=1}^n \frac{1}{\left(\delta A_i\right)^2}
\left[\hat{A}_i - A_i(\mbox{\boldmath $\cffF$})\right]^2,
$$
which is well-known.
The error propagation is often performed via the Hessian matrix
\begin{eqnarray}
\label{Hessian}
\mbox{\boldmath $H$} = \left(
                \begin{array}{ccc}
                  H_{11}&  \cdots &  H_{1n}\\
                  \vdots  &  \vdots  & \vdots\\
                  H_{n1}  & \cdots  & H_{nn}
                \end{array}
              \right)\quad\mbox{with}\quad H_{ij} =\frac{1}{2} \frac{\partial^2\chi^2}{\partial {\cffF}_i \partial {\cffF}_j}\Big|_{\chi=\chi_{\rm min}}\,,
\end{eqnarray}
where its inverse provides the covariance matrix for the \cffsreim
\begin{eqnarray}
\label{cov-fromH}
{\rm cov}(\mbox{\boldmath $\hat\cffF$}) =   \mbox{\boldmath $H$}^{-1}\,.
\end{eqnarray}

Let us first remind the reader that, instead of finding a one-to-one map of normally distributed random variables with the methods outlined in Sec.~\ref{sec:maps} and used in Sect.~\ref{sec:regression}, one may equivalently utilize the least squares method.
Obviously, if a  solution (\ref{eq:A2F}) exists, written as
$$\mbox{\boldmath $\hat \cffF$} =  \mbox{\boldmath $\cffF$}(\mbox{\boldmath $\hat A$})\quad \mbox{with}\quad
\mbox{\boldmath $\hat{A}$} = \mbox{\boldmath $A$}(\mbox{\boldmath $\hat \cffF$}),$$
the $\chi^2$ function (\ref{chi^2}) takes exactly the value
$\chi^2=0$.  However, using blindly a ``black box'' fitting routine to
extract the CFF values from DVCS asymmetries, one may find {\em only one} global minimum with $\chi^2\approx 0$ due to numerical
errors. This solution may be associated  to an unphysical root,
e.g., in H{\sc ermes} kinematics it may be associated with a
solution where the DVCS cross-section dominates the BH cross-section.
Hence, in  such ``fits'' one must search for all $\chi^2\approx 0$ minima and decide then by means of the cross section ratio (\ref{eq:N}) which of them is to be considered as the physical one. Alternatively,
one may implement the constraint $N(\mbox{\boldmath $\hat \cffF$})>
1/2$ for the BH dominated regime $\Big[ N(\mbox{\boldmath $\hat
  \cffF$})< 1/2$ for the DVCS dominated regime$\Big]$ or one can constrain the
value of a single \cffreim\ that is not very well determined by the data.
It is easy to realize from (\ref{chi^2}--\ref{cov-fromH}) that, if a one-to-one map exists,
the covariance matrix (\ref{cov-fromH}) can be written in the form
\begin{eqnarray}
{\rm cov}(\mbox{\boldmath $\cffF$}) =
\left[\frac{\partial\mbox{\boldmath $A$}(\mbox{\boldmath $\cffF$})}{\partial \mbox{\boldmath $\cffF$}}\right]^{-1}
\cdot {\rm cov}(\mbox{\boldmath $A$}) \cdot
\left[\frac{\partial\mbox{\boldmath $A$}^\intercal (\mbox{\boldmath $\cffF$})}{\partial \mbox{\boldmath $\cffF$}}\right]^{-1} \;,
\end{eqnarray}
or by means of the inverse function theorem in the form of (\ref{eq:cov}),  constructed from the Jacobian (\ref{eq:A2F-gradient}). We emphasize that the use of constraints (particularly if the fitted parameter ends on the boundary)  may influence the results and so a true one-to-one map cannot be obtained.

We used the equivalence of brute-force and least squares methods, employed for the set of charge odd asymmetries (\ref{obs-interference}), for a numerical cross check between two independent software tools. Utilizing the popular minimization routine {\sc minuit} \cite{James:1975dr} in one code and the brute-force method in the other, we obtain (except for bins $\#3$ and $\#4$)  the same one-to-one map that is shown in Fig.~\ref{fig:methods} (stars).
In these two bins {\sc minuit}  finds  global minima in the DVCS
dominated regime rather than the BH one, exemplifying that taking here
the global minima gives an answer that we consider
to be wrong. The small deviation of the total
$$
\chi^2 = 2.95\times 10^{-4} \quad\mbox{or}\quad  \chi^2/n_{\rm d.o.f.}=\chi^2/(12 \times 8 - 8)  = 3.35 \times 10^{-6}
$$
value for all 12 bins from zero is here to be considered as a measure of the numerical accuracy rather than reflection of statistical fluctuations.

Obviously, we can employ the least squares method to an ill-posed mapping problem, e.g. as discussed in Sec.~\ref{sec:regression} for the alternative map with the set of eight observables (\ref{obs-INTDVCS}), where an inconsistency appeared in a few bins. Of course, in such a case the $\chi^2$ value will differ from zero and can be taken as a measure of the inconsistency, which may originate from the statistical fluctuation of means in data.

Moreover, we can use the least squares method also in the case of an overcomplete set of equations.
In such a case the $\chi^2$ value of a global minima will also differ from zero. Including the two twist-two associated observables
$\Asincos{0}{UT,DVCS}$ and $\Acoscos{0}{LT,BH+DVCS}$ in the set of the interference term dominated asymmetries (\ref{Asin-HERMES},\ref{Acos-HERMES}), the total
$\chi^2 = 14.3$ value differs now significantly from zero. Note that
$$\chi^2/n_{\rm d.o.f.} = 14.3/(12 \times 10 - 8) \approx  0.13$$
is naturally  a small number and that the $\chi^2$ value in a given bin can be considered as a measure of the size of statistical fluctuations and/or the validity of the utilized twist-two dominance hypothesis. The resulting \cffsreim\ are entirely compatible with those from the one-to-one map (not shown).

Extending the twist-two dominance hypotheses also to the lowest asymmetry harmonics, we have 14 observables available and
we may consider the twist-two associated \cffsreim\ as eight separate
independent parameters that we would like to extract. Formulating
the extraction problem in such a manner makes closer contact to the
work in \cite{Guidal:2010de}, where all 24 asymmetries that were
available at that time were used. In our fit, the total $\chi^2$ value increases  to
$$\chi^2 = 67.7  \quad\mbox{or}\quad \chi^2/n_{\rm d.o.f.}  = 67.7/(12\times 14 - 8) \approx 0.42\,,$$
the
mean values remain in general stable (see  triangles in Fig.~\ref{fig:methods}) except for bin \#8 and \#12, in which the solution in the DVCS
dominated regime is again obtained. For  \#8  we present the {\sc minuit} outcome and one realizes that the solution in the  DVCS
dominated regime possesses a  very large uncertainty. For bin \#12 we used the constraint $|\Re\bar{\cffE}| < 40$,
to find the local minimum that provides the BH dominated solution. Thereby, the fit ends on a boundary  $\Re\bar{\cffE}=-40$, and, hence, the error propagation for $\Re\bar{\cffE}$ completely fails---see the corresponding triangle in bin \#12 on Fig.~\ref{fig:methods}.
Compared to the one-to-one map, the uncertainties naturally decrease and become even
smaller than in our linearized map  (filled circles).  As a
consequence of this decrease, one may view now the real part of
$\cffE$ as not any more entirely compatible with zero. In conclusion,
extending the twist-two dominance hypothesis also to observables that
are potentially more contaminated by the remaining sixteen \cffsreim\ and/or the inclusion of constraints may underestimate
the errors and can potentially result in an overinterpretation of the resulting  \cffsreim.

Finally, let us compare our findings with those obtained from a regression
analysis.  Here, one would consider the  $\chi^2$ value as a statistical
measure and would consider a solution with the value of $\chi^2/{\rm
d.o.f.}\approx 1$ as optimal. As we have seen, the precision of data and the
fact that for any given observable most of the \cffsreim\ are kinematically
suppressed, presently prevents us from accurately determining all eight \cffsreim.  Although
we have already answered the question {\it `Which \cffsreim\ can be
extracted with some reliability?'},
in the rest of this section we attempt to address this same question using the
method of stepwise regression.
Thereby, one first performs
eight separate single-\cffreim\ fits and sees which \cffreim\ alone describes the data best (measured by
$\chi^2$ value). Then one proceeds to two-\cffreim\ fits, where the best \cffreim\ from the first step
is combined with each of the seven remaining \cffsreim\ and the best-fitting pair of \cffsreim\ is
retained. This procedure is continued until there is either no improvement in the
description of the data or new \cffsreim\ are not extracted with any statistical significance.

\begin{figure}[t]
\begin{center}
\includegraphics[scale=0.6]{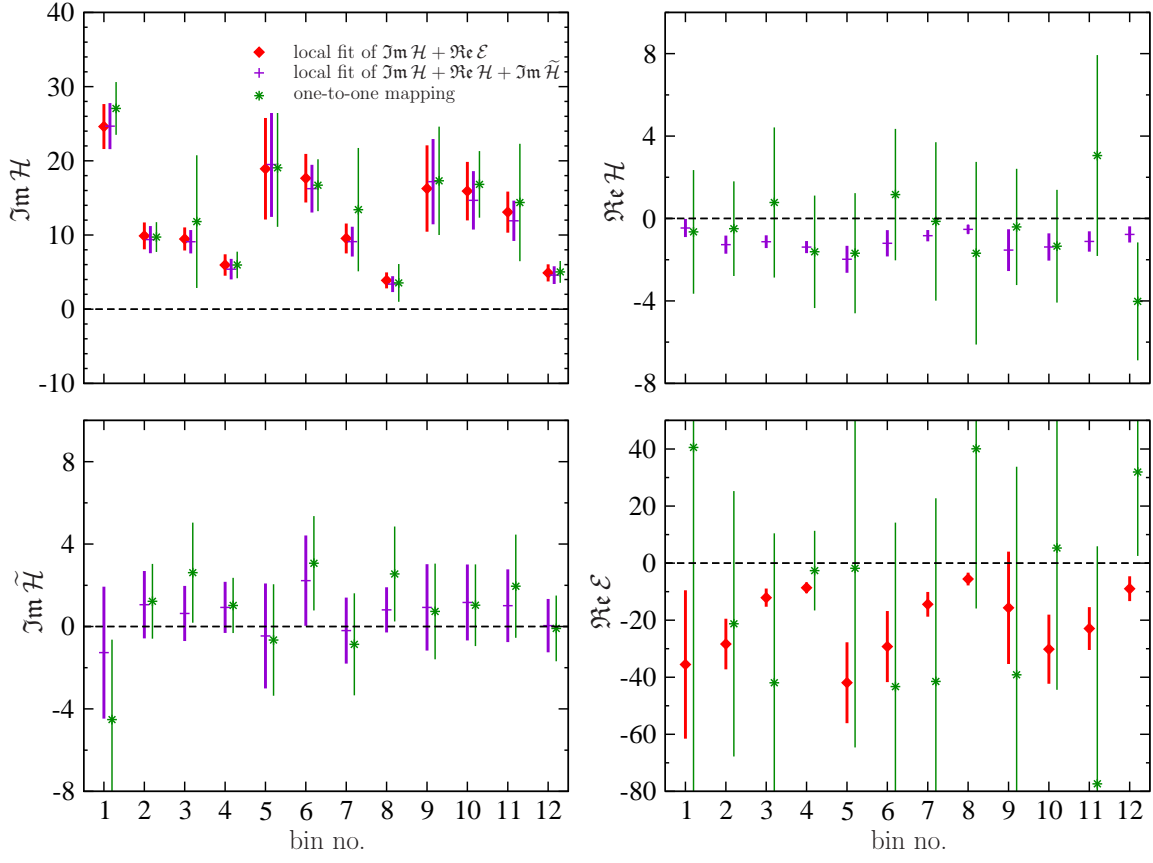}
\end{center}
\caption{\small Results of least-squares fits in two scenarios with only a small
number of CFFs locally fitted to
data separately for each of 12 H{\sc ermes} bins. First,  with
only $\Im\mathcal{H}$ and $\Re\mathcal{E}$ fitted  (red diamonds) and, second, with
$\Im\mathcal{H}$, $\Re\mathcal{H}$ and $\Im\widetilde{\mathcal{H}}$ (purple pluses).
For comparison, result of one-to-one mapping
procedure from section \ref{sec:regression} is also shown (green stars).
\label{fig:regr}}
\end{figure}

To make this stepwise procedure more reliable we temporarily remove
from consideration the observables
$\Asincos{0}{UT,DVCS}$ and $\Acoscos{0}{LT,BH+DVCS}$ because, for these harmonics, squared combinations of \cffsreim\ play a dominant
role so a) they cannot be reasonably described
in first-step single-\cffreim\ fits and b) their inclusion would introduce strong
correlations between \cffsreim\ in the second and further steps
thus potentially introducing bias for some \cffsreim.

Fitting was performed by standard minimization of the $\chi^2$ function separately in each kinematic
bin, using the \textsc{minuit} package.
As expected, it is $\Im\mathcal{H}$ that gives clearly the best description of all
data. The single-\cffreim\ fit of $\Im\mathcal{H}$ gives $\chi^2/n_{\rm d.o.f.}$ =
198.4/132, with second best being $\Re\mathcal{H}$ with $\chi^2/n_{\rm d.o.f.}$ =
472.7/132.

In the second step, we have two equally good fits:
\begin{itemize}
\item  fit of $\Im\mathcal{H}$ and $\Re\mathcal{H}$ with $\chi^2/n_{\rm d.o.f.}$ =
    102.3/120, and
\item  fit of $\Im\mathcal{H}$ and $\Re\mathcal{E}$ with $\chi^2/n_{\rm d.o.f.}$ =
    103.0/120.
\end{itemize}
(Next best being fits to
$\Im\mathcal{H}$ and $\Re\widetilde{\mathcal{H}}$ with $\chi^2/n_{\rm d.o.f.}$ =
122.4/120 and to
$\Im\mathcal{H}$ and $\Re\widetilde{\mathcal{E}}$ with $\chi^2/n_{\rm d.o.f.}$ =
185.4/120.)
Trying now the third step, for several choices of the third \cffreim{}
quality of the fits
improve somewhat (measured by their p-value) but the values of this
third \cffreim\ cannot be extracted with any statistical significance. Furthermore,
after adding the $\Asincos{0}{UT,DVCS}$ data, the fits
of $\Im\mathcal{H}$ and $\Re\mathcal{E}$ improve much more than
the other scenarios
due to the dominant contribution of terms involving the
$\Im\mathcal{H} \cdot \Re\mathcal{E}$ product to this observable, cf.~(\ref{A_UTDVCS^sinvarphicos0phi-BMK}).
Thus, as our final results we present two scenarios: one selected
by this stepwise regression procedure and another, more in agreement
with common expectations, where fits are done with the predetermined set of
\cffsreim\ $\Im\mathcal{H}$, $\Re\mathcal{H}$ and
$\Im\widetilde{\mathcal{H}}$. In both scenarios we now perform fits to
the complete set of 14 H{\sc ermes} observables.

\begin{itemize}
\item Scenario 1: Fit of $\Im\mathcal{H}$ and $\Re\mathcal{E}$.  $\chi^2/n_{\rm d.o.f.}$
= 134.2/144, when adding $\chi^2$ values for all 12 bins.
Fits are bad for bin $\#3$ ($\chi^2/n_{\rm d.o.f.}$ = 19.9/12) and
bin $\#8$ ($\chi^2/n_{\rm d.o.f.}$ = 21.1/12). Other bins are fine.

\item Scenario 2: Fit of $\Im\mathcal{H}$, $\Re\mathcal{H}$ and
$\Im\widetilde{\mathcal{H}}$.  $\chi^2/n_{\rm d.o.f.}$ = 148.8/144.
Bins $\#3$ and $\#8$ again show a bad fit result, with
$\chi^2/n_{\rm d.o.f.}$ = 21.8/11 and $\chi^2/n_{\rm d.o.f.}$ = 21.1/11.
\end{itemize}

We note that, in both scenarios, the fit in bin $\#8$ is fine if the
$\Asincos{0}{UT,DVCS}$ and $\Acoscos{0}{LT,BH+DVCS}$ data points are removed.
The resulting \cffsreim\ from both scenarios are plotted on Fig.~\ref{fig:regr} and are compared with the one-to-one map. One realizes that the \cffreim\ $\Im \cffH$
is quite robust and that the real part of CFF $\cffE$ differs on some $2\sigma$ level from zero.
The results from the second scenario
essentially agree with those in \cite{Guidal:2010de}, showed there for
three selected bins $\#2$, $\#3$, and $\#4$.
We add that those results were obtained by means of model-independent
least squares fits to {\em twenty-three} asymmetries. However, most of the asymmetries
were not related to twist-two dominated quantities and the \cffreim\
parameters were deliberately chosen with reference to GPD model constraints.
How successfully fits from our two scenarios (and a global world fit
presented in Sect.~\ref{sec:global}) describe particular observables
is visible on Figs.~\ref{fig:unp}--\ref{fig:LTP}
in Appendix \ref{app:HERMESvsExtraction}.

\subsection{Uses of HERMES data for model builders and in global fits}
\label{sec:global}

Let us discuss the constraints from H{\sc ermes} DVCS data, presented in terms of CFFs on Fig.~\ref{fig:methods},
for GPD model builders. We  consider here the GPD framework
in the perturbative leading order (LO) approximation and we can safely
restrict ourselves to the qualitative aspects. We adopt and refine  here some older discussions, given in Sect.~5.1 of \cite{KirMue03} and  illuminated with BMK model predictions in Sect.~5.2 there.
The imaginary part of CFFs is, in this approximation, given by the GPDs on the cross over line:
$$
F(x,x,t,\mu^2=\Q^2) \stackrel{\rm LO}{=} \frac{1}{\pi} \Im {\cal F}(\xB,t,\Q^2)\Big|_{\xB=\frac{2x}{1+x}}\,.
$$
As known, the valence quark part of GPD $H$ is essentially
governed by the asymmetry $\Asin{1}{LU}$, which is almost  saturated
by the valence quark content of the forward PDF, decorated with some $t$-dependence. On the other hand, rather generic model estimates tell us that sea quark contributions  are important in H{\sc ermes} kinematics, i.e. the partonic  decomposition%
\footnote{Our decompositions of CFFs, which are charge even, in
  valence and sea quark parts contain squared quark charges and the
  terms  {\it valence} and {\it sea} are adopted from the common
  terminology as used in global  parton distribution function fits.} %
reads
\begin{equation}
\label{cffH-estimate}
\Im \cffH = \Im \cffH_{\rm val} + \Im \cffH_{\rm sea}\quad \mbox{with}\quad \Im \cffH_{\rm sea} \sim  \Im \cffH_{\rm val}\,.
\end{equation}
Consequently, the H{\sc ermes} data, taken in terms of $\Im\cffH$, require a small skewness effect for GPD $H$, which is also
required for the LO description of H{\sc era} collider data. To get an easy handle on the real part of CFF $\cffH$, one may use instead of
the LO convolution formula a signature-even GPD dispersion relation, see \cite{Diehl:2007jb,Kumericki:2008di,Kumericki:2007sa,Teryaev:2005uj} and
references therein%
\begin{eqnarray}
\label{DR-even}
\Re \left\{ \cffH \atop \cffE \right\}(\xB,t,\Q^2) \stackrel{\rm LO}{=} {\rm PV}\int_{0}^1\!dx  \frac{2x}{\xi^2-x^2}  \left\{ H \atop E \right\}(x,x,t,\mu^2=\Q^2)
 \pm {\cal D}(t,\Q^2) \Big|_{\xi=\frac{\xB}{2-\xB}}\,.
\end{eqnarray}
For H{\sc ermes} kinematics, we can take for granted that the sign of
the resulting real part is determined by the ``Regge''-behaviour,
inherited from the PDF behavior. Hence, valence quarks provide a
large positive real part while sea quarks contribute a negative
part and, in addition, there is a subtraction
constant that is related to the ``D-term'' \cite{Polyakov:1999gs}. The H{\sc ermes} data, taken in the form of $\Re\cffH$,
tell us that the modulus of this quantity is rather small and, hence,
we would interpret it as representing a cancelation between the three contributions.  Note that the experimental  constraint on
the subtraction constant depends also on details of the GPD model and it is therefore rather weak.

Going along the same line, we can now discuss the CFF $\cfftH$. Its real part arises from a signature-odd GPD dispersion relation,
\begin{eqnarray}
\label{DR-odd}
\Re \left\{ \cfftH \atop \cfftE \right\}(\xB,t,\Q^2) \stackrel{\rm LO}{=} {\rm PV}\int_{0}^1\!dx  \frac{2\xi}{\xi^2-x^2}
\left\{ \widetilde{H} \atop \widetilde{E} \right\}(x,x,t,\mu^2=\Q^2) \Big|_{\xi=\frac{\xB}{2-\xB}}\,,
\end{eqnarray}
where no subtraction is needed. From phenomenological PDF parameterizations, one expects that the  GPD $\widetilde{H}$, taken  on the cross-over line
and as it enters in the DVCS amplitude, is (much) smaller than  GPD
$H$. Hence, one expects that both the real and imaginary parts
of CFF $\cfftH$ are relatively small. This is entirely
compatible with H{\sc ermes}  data, which surprisingly provide us also
with a rather strong constraint for $\Im \cfftH$.

A generic discussion can be also given for the CFF $\cffE$. However, here only form factor information is available. Hence, one would assume that
the zero-skewness GPD has a simple functional form in which nodes are absent that arises from Regge and large-$x$ arguments. The normalization of such  valence quark GPDs is adopted from the anomalous magnetic moments
$$
\kappa_{u_{\rm val}} = 1.673\quad \mbox{and}\quad  \kappa_{d_{\rm val}} =  -2.033\,,
$$
taken from the nucleon.
A crude approximation of the size of the valence part of $\Im \cffE$
is obtained if we assume that the same functional
form holds for all valence quark contributions,
$$
\frac{\Im\cffE_{\rm val}}{\Im\cffH_{\rm val}} \sim \frac{e_u^2 \kappa_{u_{\rm val}} + e_d^2 \kappa_{d_{\rm val}}}{e_u^2 + e_d^2} \sim 1
\quad\Rightarrow\quad  \Im \cffE_{\rm val} \sim  \frac{1}{2} \Im \cffH\,,
$$
where we used (\ref{cffH-estimate}).  However, we should keep in mind that the functional form w.r.t.~both the $t$- and $x$-dependences may alter our estimates. It is expected that the
$t$- and $x$-fall off for $E$ GPDs is steeper than for $H$ GPDs \cite{Diehl:2004cx}, i.e. GPD-model-refined estimates would give even smaller predictions for $\Im \cffE_{\rm val}$.
Considering the CFF data in Fig.~\ref{fig:methods},
one realizes that the noise of $\Im \cffE$ is of the order of the means of $\Im \cffH$.
Hence, contrary to GPD-model-based claims, e.g. that H{\sc ermes} data
provide a constrain on the quark orbital angular momentum
decomposition \cite{Airapetian:2008aa} or the suggestion that negative sea
quark contributions to $\Im\cffE$ are favored
\cite{Kroll:2012sm}, our generic arguments tell us that a partonic
interpretation of H{\sc ermes} data in this specific case is entirely
biased by model assumptions. Let us add that the dispersion relation
(\ref{DR-even}) that we used for an estimate of $\Re \cffE$,
together with the standard model assumptions, tells us
that the large negative $\Re \cffE$ scenario that we obtained from
the regression method is difficult to understand from the GPD model perspective
(positive subtraction constant for negative $D$-term, positive contribution for valence quarks, and positive contribution for negative sea quarks).

To complete our short examination of the CFF data, we mention
that the CFF $\cfftE$ should contain a pion pole contribution that should be large at small $-t$ \cite{Mankiewicz:1998kg,Frankfurt:1999xe}. Sometimes,
GPD model builders believe that this is the most important
contribution and neglect for that reason the imaginary part. If one
wishes, one can see sizeable and negative $\cfftE$ \cffsreim\  at small
$-t$ --- see the first bin in Fig.~\ref{fig:methods} where the significance for the imaginary part is
even more pronounced than for the real part. This clearly
  contradicts the common GPD model assertion. However,
since these data are very noisy,  a definite conclusion cannot
be drawn and we consider the CFF $\cfftE$ as essentially unconstrained
from H{\sc ermes} DVCS data.

For comparison purposes, we also performed one global model fit to the
world DVCS off-the-proton data with a version of the hybrid model
used in \cite{Kumericki:2009uq} that we initially used to access
the GPD $H$ from unpolarized proton DVCS data. The hybrid model
  comprises a full GPD model in the flavor singlet sector (dominated
at small $\xB$ by sea quark and gluon contributions), while in
the flavor non-singlet (or valence quark) sector dispersion relations
are used and, hence, only the GPDs on the cross-over line are
needed. Keeping in mind that, apart from H{\sc ermes}, no other
experiment with a proton target could provide information on the
full separation of the various CFF contributions, we neglect in our
fit the $\Im \cffE$  and $\Im \cfftE$ contributions, which are
compatible with zero for H{\sc ermes} kinematics. However we do
include $\Re \cffE$  and $\Re \cfftE$, which are related to
subtraction constants in the dispersion relation (\ref{DR-even}) and
in the  oversubtracted analog of (\ref{DR-odd}), respectively. The
reason for doing so is that HALL A unpolarized cross section
measurements \cite{Munoz_Camacho:2006hx} indicate a rather large
unpolarized DVCS cross section, which suggest that the DVCS amplitude
contains a large real part.  In our hybrid model, the contributions of sea quarks and gluons are modelled using
conformal moments of GPDs $H^{\rm sea}$ and $H^{\rm G}$, respectively,
and LO QCD evolution is taken into account, while
the contributions of valence quarks are described by
directly modelling $\Im\mathcal{H}$ and $\Im\mathcal{\widetilde{H}}$
and using dispersion relations to obtain $\Re\mathcal{H}$ and
$\Re\mathcal{\widetilde{H}}$ (and evolution is neglected). Details of the model are given in ref. \cite{Kumericki:2009uq}, to which we
refer the interested reader.

In particular, we used the following data sets:
\begin{itemize}

\item The H{\sc ermes} combined data on $\Asin{1}{LU,I}$,
$\Acos{0}{C}$ and $\Acos{1}{C}$ \cite{Airapetian:2012mq};
on $\Asin{1}{UL,+}$ and $\Acos{0}{LL,+}$ \cite{Airapetian:2010ab};
and on $\Asincos{1}{UT,I}$
\cite{Airapetian:2008aa}. To work with statistically independent data we
considered only the projection of the data along the
$-t$ axis, i.e. just the first
third of the published 18 \cite{Airapetian:2012mq} or
12 \cite{Airapetian:2008aa,Airapetian:2010ab} kinematic bins.
This gives $3\times 6 + 2\times 4 + 4 = 30$ data points.

\item The first ($\sin\phi$) harmonics of the CLAS data on
a) the beam spin asymmetry with an unpolarized target \cite{Girod:2007aa},
where we used only data with $\mathcal{Q}^2 > 2\,{\rm GeV}$ (4
points), and b) the longitudinal target spin asymmetry with
a  polarized target \cite{Chen:2006na} (6 points).

\item Fourier transforms of the Hall A measurements of beam spin
  difference (12 points) and beam spin sum (8 points)
  \cite{Munoz_Camacho:2006hx}, where cross-sections were weighted with  the inverse product of the Bethe-Heitler propagators.

\item Measurements by the H1 collaboration of
  \begin{itemize}
   \item The DVCS cross section differential in $t$, \cite{Aktas:2005ty}, Table 1,
         1996-1997 data (4 points) and 1999-2000 data (4 points).  
   \item The DVCS cross section differential in $t$, \cite{Aaron:2009ac}, Table 3a,
         (12 points). 
  \end{itemize}

\item Measurements by the ZEUS collaboration of
  \begin{itemize}
   \item The DVCS cross section differential in $t$, \cite{Chekanov:2008vy}, Table 1, only
    $\mathcal{Q}^2 > 4\,{\rm GeV}^2$ points (5 points). 
   \item The total DVCS cross section, \cite{Chekanov:2008vy}, Table 4 (4 points). 
   \item The total DVCS cross section, \cite{Chekanov:2003ya}, Table 1 (6 points). 
  \end{itemize}

\end{itemize}
In total, we have 95 data points. Fitted to these points was a version of the model used in \cite{Kumericki:2009uq}.
Here we only describe differences, and list the free fitting parameters.
The first difference is that, beside leading partial wave in SO(3) expansion
of conformal moments of GPDs $H^{\rm sea}$ and $H^{\rm G}$ (the normalization
of which is fixed by DIS $F_2$ data, and residual $t$-dependence of $H^{\rm sea}$ being
determined by free dipole mass parameter $M^{\rm sea}$) and subleading partial wave
with relative strength parameters $s_2^{\rm sea}$ and $s_2^{\rm G}$,
we take here into account also the third partial wave with two new strength parameters
$s_4^{\rm sea}$  and   $s_4^{\rm G}$. The second difference to
\cite{Kumericki:2009uq} is that $\Re\mathcal{\widetilde{E}}$ is here
modelled by a shape suggested by the pion pole contribution, but with
normalization $r_{\pi}$ and additional $t$-slope dipole mass $M_{\pi}$ as two additional free parameters.
Together with parameters $M^{\rm val}$, $r^{\rm val}$, and  $b^{\rm val}$
parametrizing $\Im\mathcal{H}$,  $C$ and  $M_C$  parametrizing subtraction constant%
\footnote{Note that the subtraction constant here is given by -$\cal D$, appearing in (\ref{DR-even}).},
and $\tilde{M}^{\rm val}$,  $\tilde{r}^{\rm val}$, and  $\tilde{b}^{\rm val}$
parametrizing $\Im\mathcal{\widetilde{H}}$ in the same way as in
\cite{Kumericki:2009uq},
this brings the number of parameters to a total of 15. The
fit of this model to all of the above
data results in $\chi^2/n_{\rm d.o.f.} = 124.1/80$, which is strictly speaking not
a good fit, but it is acceptable for a global fit to data coming
from such a variety of experiments and observables. Parameters of
the fitted model are given in Tab.~\ref{tab:ps-KMM12}.
\begin{table}[t]
\begin{center}
\begin{tabular}{|cccccccccc|}
\hline
$M^{\rm val}$ &  $r^{\rm val}$ &  $b^{\rm val}$ &  $C$ &  $M_C$ &
$\tilde{M}^{\rm val}$ &  $\tilde{r}^{\rm val}$ &  $\tilde{b}^{\rm val}$ &  $r_{\pi}$ &  $M_{\pi}$ \\ \hline
0.95 & 1.12 & 0.40 & 1.00 & 2.08 & 3.52 & 1.30 & 0.40 & 3.84 & 4.00 \\ \hline \hline
$(M^{\rm sea})^2$ &  $s_2^{\rm sea}$ &  $s_4^{\rm sea}$ & $s_2^{\rm G}$  & $s_4^{\rm G}$ & & & & & \\ \hline
0.46 & 0.31 & -0.14 & -2.77 & 0.94  & & & & & \\ \hline
\end{tabular}
\end{center}
\caption{\small Valence (top) and sea quark (bottom) related hybrid model parameters, extracted from the global DVCS fit {\it KMM12}.}
\label{tab:ps-KMM12}
\end{table}
and the values resulting from this global fit for the 10 H{\sc ermes} observables used
for local fits in previous sections are given by solid lines on Figs.~\ref{fig:unp}--\ref{fig:LTP} in Appendix \ref{app:HERMESvsExtraction}.

We now finally consider
the possible
tension between our simple GPD model fit and the data. As one can see in
Fig.~\ref{fig:unp}, our 
fit reasonably describes the
beam spin and beam charge asymmetries, but shows a
slight tendency for the mean values to slightly overshoot the data values. This, as
discussed above, is related to the problem of overshooting the beam
spin asymmetry measurements, taken from events selected by the missing
mass technique, with GPD models which employ a leading order
description that have skewness ratios $r\gtrsim 1$. The tension that
appears in the lowest $\xB$ and $\Q^2$ bin for the first harmonic of
the beam charge asymmetry may be
related to a technicality; namely, our input scale for evolution is chosen to be $4\,\GeV^2$ and backwards evolution to $\Q^2 \lesssim 1.5\, \GeV^2$
in the flavor singlet sector is a delicate procedure that is rather
sensitive to the initial conditions.  The dominant longitudinal proton
spin-flip asymmetry is also overvalued by the model,
see Fig.~\ref{fig:LP}, which just reflects the fact that the description of unpolarized HALL A cross section measurements requires a large real part
in the DVCS amplitude which is partially also obtained via a GPD dispersion relation from  the imaginary part of CFF  $\cfftH$.  Such a parametrization,
which is effective for unpolarized DVCS measurements, is clearly disfavored if longitudinal proton spin-flip asymmetry data are included.
The transverse proton spin-flip asymmetries, shown in
Fig.~\ref{fig:TP} are well described by our $\Im \cffE=\Im \cfftE=0$
models, within the large experimental uncertainties. Mostly, this is also the case for the transverse proton spin-flip asymmetries, presented in Fig.~\ref{fig:LTP}.
As an exception on might view $\Acoscos{1}{LTI}$, given in
(\ref{A_LTI^cosvarphicosphi-BMK}), where one may see some hint that
the sign of the real part of $\cfftE$, adopted from the pion pole,
contradicts data. Let us emphasize that the basic modeling of $\cfftE$
in terms of pion pole contribution is oversimplified and already
contradicts the GPD interpretation of $\pi^+$ electroproduction data
in the collinear framework
\cite{Bechler:2009me,Airapetian:2007aa}.

Compared to previous good DVCS world data fits to data from an
unpolarized proton, having $\chi^2/{\rm d.o.f.} \approx 1$, we may
interpret our findings here as a slight tension between a very simple
model and fixed target data. One may argue that this tension is
induced by the attempt to explain the unpolarized photon
electroproduction cross section data from JLab Hall A by means of the
common four twist-two GPDs, which have a oversimplified functional
form. A definite conclusion cannot be given, however, since one may
use more intricate parameterizations of these four twist-two GPDs,  or
use an improved framework by inclusion of perturbative corrections,
certain twist-three \cite{Anikin:2000em,Belitsky:2000vx,Kivel:2000cn}
and twist-four contributions
\cite{Braun:2011zrBraun:2011dg,Braun:2012bgBraun:2012hq}, or extend the number of
GPDs and thus CFFs. Certainly, there are also experimental
uncertainties related to the issue of exclusivity, in particular the
potential inclusion of events including a $\Delta$-resonance within the
experimental data set. We emphasize, however, that the inclusion of polarized
proton data does not contradict the conclusion that GPD $H$ plays the dominant
role in the description of present DVCS data.

\section{Summary and outlook}
\begin{figure}[t]
\begin{center}
\hspace{-1cm}
\includegraphics[width=14cm]{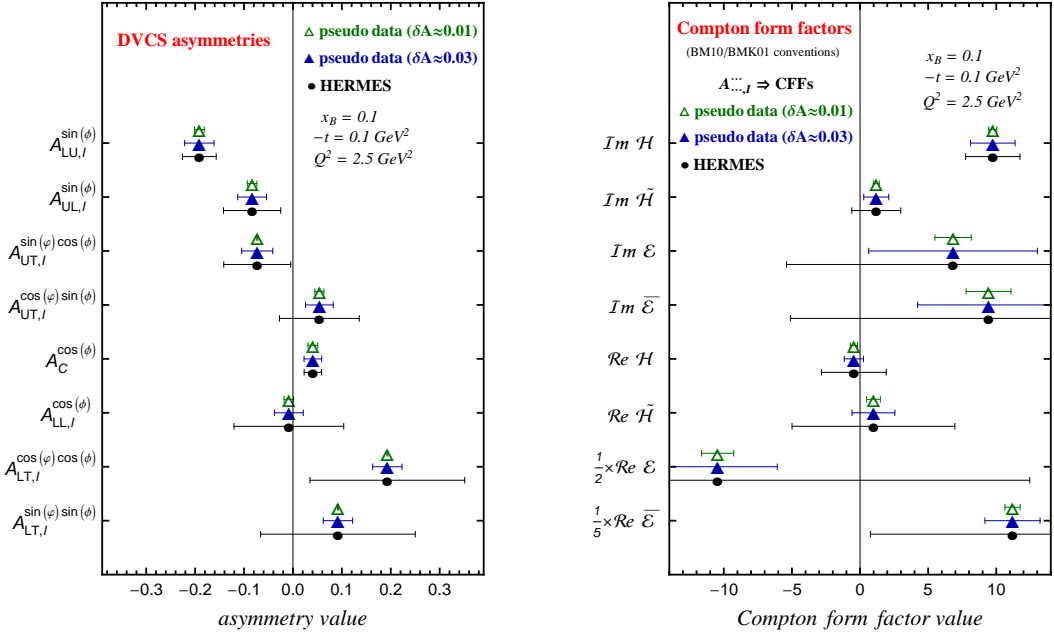}
\end{center}
\caption{\small Left: Two projections with total errors $0.01 \lesssim \delta A \lesssim 0.03$ (filled triangles) and  $\delta A \approx 0.01$ (empty triangles) of charge odd asymmetries (left) for  $\xB=0.1$, $t=-0.1\,\GeV^2$, and $\Q^2=2.5\,\GeV^2$
 are compared with H{\sc ermes} measurements in bin \#2 (solid circles). Right: the resulting CFFs from a one-to-one map, based on the twist-two dominance hypothesis.}
\label{fig:A2CFF-projection}
\end{figure}

In this article we analyzed, by means of mapping and regression methods, the final set of DVCS off-the-proton data from the
H{\sc ermes} collaboration extracted using a missing-mass event
selection method. Thereby, we still utilized the twist-two dominance hypothesis and, thus, we restricted ourselves to an overcomplete set of {\em fourteen} asymmetries. We showed that the H{\sc ermes} collaboration provided an experimental proof of principle that, with an (almost) complete measurement of {\em eight} first harmonic asymmetries in the charge-odd sector, all four twist-two associated CFFs can be accessed in a BH dominated regime. In the remaining {\em six} zero harmonic asymmetries the CFFs are kinematically suppressed and we used them to check the twist-two dominance hypothesis or, alternatively, employed them to access twist-two associated CFFs. Higher harmonics, in the first place related by the remaining set of {\em eight} CFFs, were not considered.

Apart from restricting ourselves to the twist-two sector, our
analysis is rather general. We tested different methods to
access CFFs: mapping, local regression analysis, and model
fits. In any CFF (or GPD model) extraction procedure from a set of
asymmetries, one {\em must} bear in mind that separate solutions
may exist for the BH dominated and the DVCS dominated
regimes. Based on experimental evidence and model expectations,
we took in our analysis the BH dominated solution. A one-to-one
map of asymmetries to the space of CFFs can be either found by picking
up the appropriate roots of a non-linear equation system (in our
case eight equations that can be linearized) or, equivalently,
by the least squares method. However, numerical noise in a
blind ``fitting'' technique can provide an unphysical solution. Constraining the value of CFFs to force a ``black box'' fitting routine to provide the physical solution, a rather popular method,
may have the disadvantage that a true one-to-one map cannot be obtained. The inclusion of more (twist-two related and unrelated) asymmetries in
a fitting procedure will increase the noise and the physical solution is not necessarily obtained from the global minimum.

In a one-to-one map of interference dominated, mostly charge-odd asymmetries, it turns out that only the imaginary part of CFF $\cffH$ is not compatible with zero, and all other seven \cffsreim\ can be considered as compatible with zero.
Surprisingly, it turns also out that the best constrained quantity is
the imaginary part of CFF $\cfftH$, followed by the real part
and imaginary parts of $\cffH$, and to a much lesser
extent also by the imaginary part of CFF $\cffE$. The
remaining four \cffsreim\ are very poorly constrained.
One may be tempted to consider the extraction of CFFs as a regression
problem, and thus use selection criteria for filtering out the
noise. As expected, in this approach it turns out that $\Im \cffH$ is
a robust quantity, the method does not necessarily yield a
unique solution. We present two solutions: one obtained by
the strict use of selection criteria and the other a more
hand-picked solution informed by standard GPD model
considerations. The first one suggests that the negative real
part of CFF $\cffE$ can be considered as large and the second
solution suggests that one may consider the real part of CFF $\cfftH$ as negative and the imaginary part of $\cfftH$ as  positive. Certainly, from the GPD model point of view one would give the hand-picked standard solution preference. However,
we emphasize that our one-to-one map only shows that this well
constrained quantity $\Im \cfftH$ is small and does not allow a
further partonic interpretation, e.g., extraction of any
$t$-dependence. A rather analogous situation appears for
the imaginary part of CFF $\cffE$ --- its magnitude is only
loosely constrained by the data. The relation of GPD $E$ to the spin sum rule inspires both experimentalists and theoreticians
to make definite GPD model statements that certainly cannot, at
present, be justified by simply describing experimental DVCS data
with a given model. This we have illuminated in our global GPD model
fit example where the GPD $E$ (in the standard double distribution
representation) has been set to zero, giving us the perhaps best
(however, not perfect) DVCS world data description that is
presently available.

Let us illustrate that with the H{\sc ermes} experiment, switched off 2007, our knowledge about CFFs could have been much better.
Supposing that the longitudinally polarized proton  asymmetries $A_{\rm UL,I}$ and $A_{\rm LL,I}$ would have been measured in the charge-odd sector with
the same statistics and slightly better systematics than
the existing data taken with positron beam, the proton helicity
conserved CFFs $\cffH$ and $\cfftH$ could have been accessed with
roughly the same accuracy. Alternatively, a decrease of the
large statistical uncertainties for the $A_{\rm LT,I}$ quantities to a
typical value of other H{\sc ermes} asymmetries would have drastically
improved the constraints for the real parts of  proton helicity flip
CFFs $\cffE$ and $\cfftE$ (note, however, that these CFFs naturally
suffer from a larger uncertainty). The effect of a decrease in
the uncertainties that would be feasible at a fixed target experiment such as
H{\sc ermes}  is illustrated by the filled triangles in
Fig.~\ref{fig:A2CFF-projection}, where we naively assumed for all polarized
charge odd asymmetries a $\delta A\approx 0.03$ error 
(statistical and systematic uncertainties are added in quadrature).

The H{\sc ermes} collaboration provided a total set of {\em thirty-four} asymmetries that includes an (almost) complete
measurement of the second harmonic
asymmetries in the charge odd sector, which are primarily
associated with {\em four} twist-three and  {\em four}  twist-two
related CFFs. In principle, {\em twenty-four} asymmetries  can be
utilized in a one-to-one map, which leaves us ten
asymmetries for a consistency check. Due to kinematic
suppressions and noise  it is expected that access to the {\em
  eight} twist-tree and transversity associated CFFs is not
achievable; however one may hope for a test of whether our
(and other) results are robust in an unbiased map, i.e.~where one drops the twist-two dominance hypothesis. Before one undertakes such an attempt,  several improvements and technicalities should be taken into account, which are partially also needed on more general grounds:
\begin{itemize}
\item A  physically motivated parametrization of the DVCS tensor in terms of {\it twelve} CFFs.
\item A code  which relates the set of {\it twelve} well defined CFFs to observables in an {\it exact} manner, perhaps for different conventions.
\item The (small) differences between proton polarization vector, defined in the lab frame, and those in the rotated frame, see Fig.~\ref{fig:frame-Trento},
should be taken into account, at least from the principled point of view.
\item  It would be desirable that the  covariance matrix for experimental measurements would be available for each given kinematical bin.
\end{itemize}
Let us add to the first two points that various codes are presently used that relate (twist-two associated) CFFs (in most cases understood as quantities that  are expressed by conventionally defined GPDs) that are based on different conventions and approximations. At present numerical differences do not matter, nevertheless, it is not desirable to proceed in this manner to the next generation of DVCS experiments. The remaining obstacles can be overcome by means of the
parameterization for the virtual Compton scattering tensor and the analytic cross section results that are presented in \cite{Belitsky:2012ch}.

In planned DVCS experiments at JLAB@12GeV, having an electron beam and fixed proton target, it is expected that experimental uncertainties will
become much smaller, as already demonstrated by the {\sc hall a} collaboration in measurements of DVCS cross sections.
 Altogether in such an  experimental set up one can measure
{\em eight} ({\em seven}) azimuthal angular dependent cross section combinations  (asymmetries), compared to {\em sixteen} ({\em fifteen}) for both charges of electrons. Having precise cross section measurements, one can form, at least in principle,  $4\times 4$ even and $4\times 3$ odd harmonics, giving us
{\em twenty-four} independent and four dependent (constant term)
observables. Hence, one may have a handle on the CFF separation via a harmonic analysis \cite{Belitsky:2001ns}.  Qualitatively or semi-quantitatively, one may also employ Rosenbluth separation to address the three separate parts of the photon electroproduction cross section. To which extent this separation is feasible requires detailed studies, where, unfortunately, the results will depend on assumptions.
However, the possibility of obtaining any unbiased extraction of
CFFs cannot be clearly stated at present. Certainly, a clean separation of the charge odd (interference term) and charge even part (sum of BH and DVCS cross sections) requires a high-luminosity positron beam, too.
Supposing that in a (next-to-) next generation of lepton-proton
scattering experiments both kinds of electrons are available and
the total experimental uncertainties for all asymmetry (or, better, cross section difference) measurements are
three times smaller than those of the beam-spin asymmetry measurements at H{\sc
ermes}, see empty triangles on Fig.~\ref{fig:A2CFF-projection}, we may expect
that the CFF $\cffE$ can be accessed in the BH-dominated DVCS regime with an
accuracy that is presently available for CFF $\cffH$.

\subsection*{Acknowledgements}
 K.~K.~and D.~M.~are indebted to the Nuclear Physics group at the
 University of Glasgow for the warm hospitality during their stay,
 where this project has been initialized. For discussions we would like to
 thank M.~Guidal, P.~Kroll, H.~Moutarde, F.~Sabatie, and G.~Schnell.  This work was partly supported by the Scottish
 Universities Physics Alliance, the UK's Science and Technology
 Facilities Council, by the Joint Research Activity
''Study of Strongly Interacting Matter'' (acronym HadronPhysics3, Grant Agreement No.~283286) under the Seventh Framework Program of the European Community,
and by Croatian Ministry of Science, Education and Sport, contract no.
119-0982930-1016.

\newpage
\appendix
\section{Visualization of HERMES data descriptions }
\label{app:HERMESvsExtraction}
\begin{figure}[th]
\begin{center}
\includegraphics[scale=0.6]{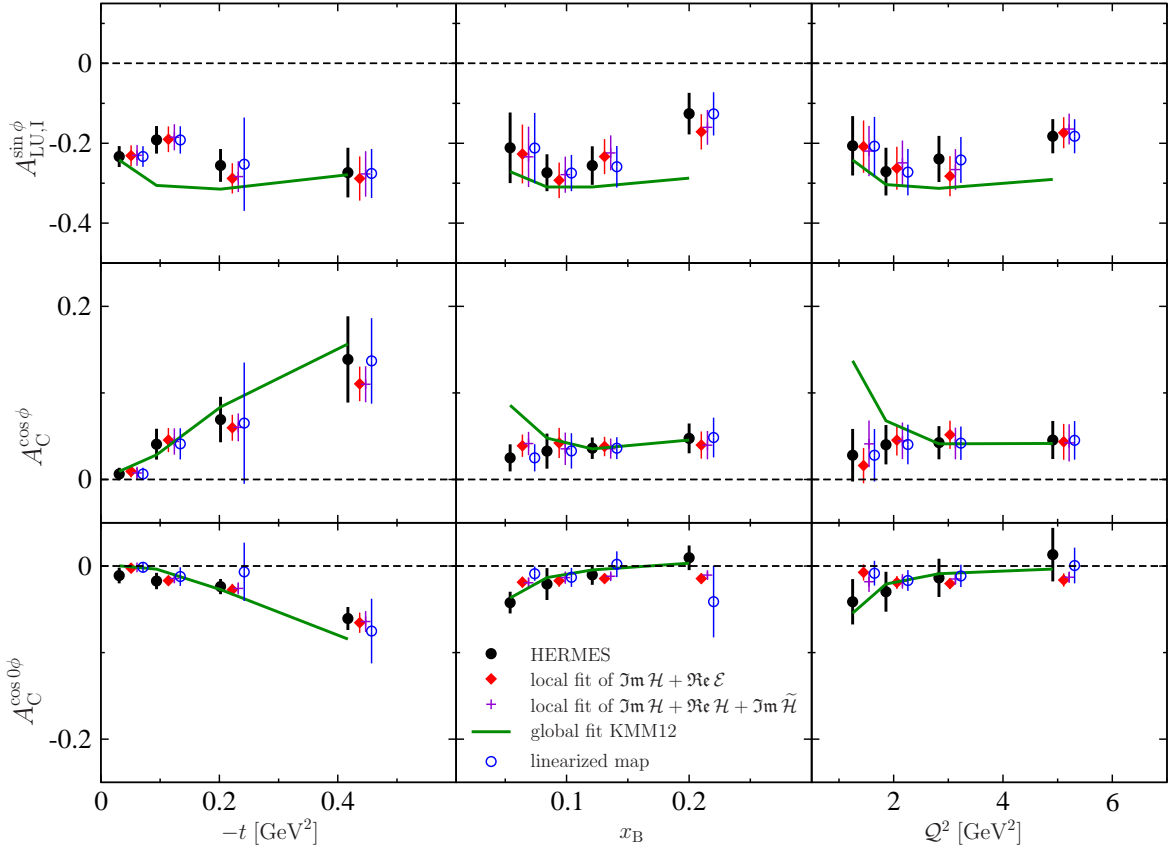}
\end{center}
\caption{\small Fits  to harmonics of asymmetries of scattering on an \emph{unpolarized} target.
Black dots are H{\sc ermes} data with systematic errors added in quadrature. Local fits in two
different scenarios are shown as red diamonds (fit to $\Im\mathcal{H}$ and $\Re\mathcal{E}$) and
blue pluses (fit to $\Im\mathcal{H}$, $\Re\mathcal{H}$, and $\Im\widetilde{\mathcal{H}}$),
slightly displaced to the right for legibility.
For comparison, we also show the result of a global fit to world DVCS data as a green solid line.}
\label{fig:unp}
\end{figure}
\newpage
\begin{figure}[h]
\begin{center}
\includegraphics[scale=0.6]{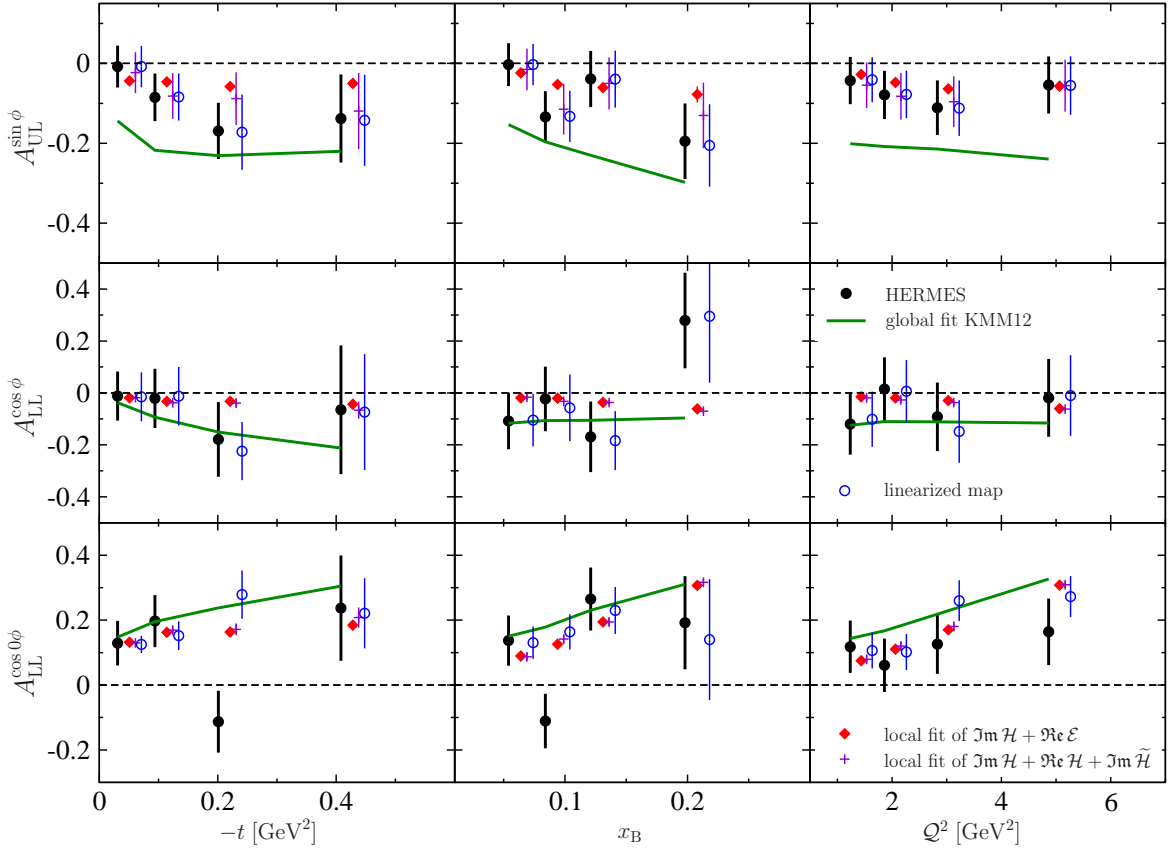}
\end{center}
\vspace{-0.7cm}
\caption{\small Fits to \emph{longitudinally
polarized} target asymmetry harmonics. Legend is as for Fig.~\ref{fig:unp}}
\label{fig:LP}
\end{figure}
\newpage
\begin{figure}[h]
\begin{center}
\includegraphics[scale=0.6]{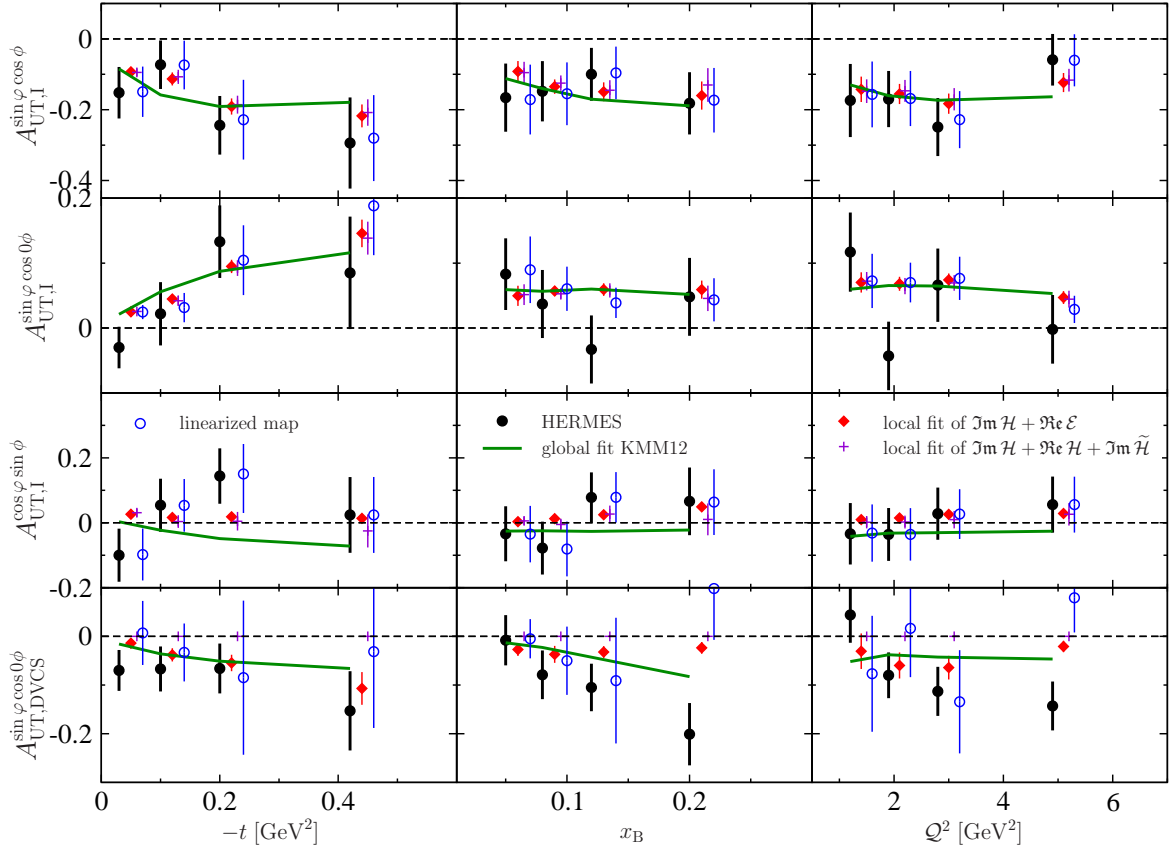}
\end{center}
\vspace{-0.7cm}
\caption{\small Fits to \emph{transversely
polarized} target asymmetry harmonics. Legend is as for Fig.~\ref{fig:unp}}
\label{fig:TP}
\end{figure}
\newpage
\begin{figure}[th]
\begin{center}
\includegraphics[scale=0.6]{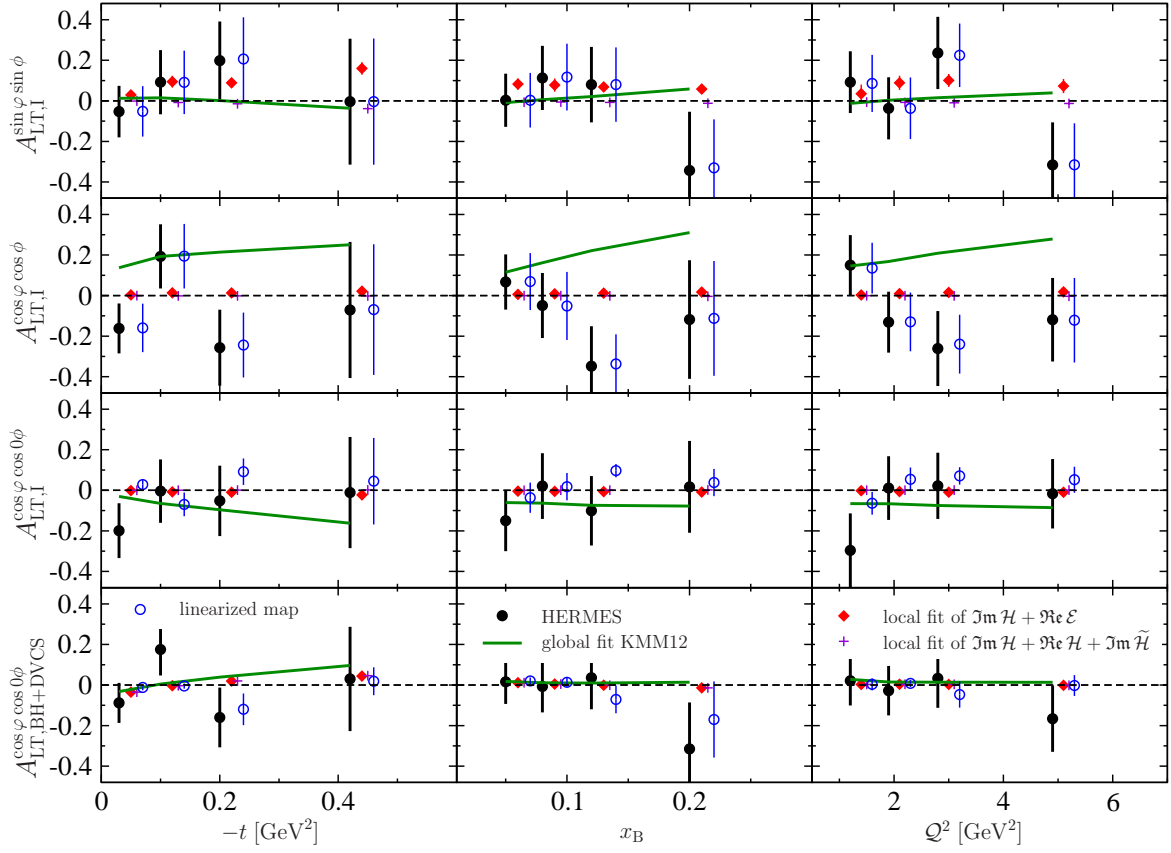}
\end{center}
\caption{\small Fits to asymmetry harmonics of  \emph{polarized}
  electron/positron beam scattering on a \emph{transversally
polarized} target. Legend is as for Fig.~\ref{fig:unp}}
\label{fig:LTP}
\end{figure}


\begin{thebibliography}{10}

\bibitem{Compton:1923zz}
\textit{Compton A.~H.},
\newblock Phys.Rev. {\bf 21}, 1923, 483.

\bibitem{Guichon:1995pu}
\textit{Guichon P.~A., Liu G., Thomas A.~W.},
\newblock Nucl.Phys. {\bf A591}, 1995, 606, nucl-th/9605031.

\bibitem{Mueller:1998fv}
\textit{M{\"u}ller D., Robaschik  D., Geyer B., Dittes F.-M., Ho\v{r}ej{\v s}i J.},
\newblock Fortschr. Phys. {\bf 42}, 1994, 101, hep-ph/9812448.

\bibitem{Radyushkin:1996nd}
\textit{Radyushkin A.~V.},
\newblock Phys. Lett. {\bf B380}, 1996, 417, hep-ph/9604317.

\bibitem{Ji:1996nm}
\textit{Ji X.},
\newblock Phys. Rev. {\bf D55}, 1997, 7114, hep-ph/9609381.

\bibitem{Ji:1996ek}
\textit{Ji X.},
\newblock Phys. Rev. Lett. {\bf 78}, 1997, 610, hep-ph/9603249.

\bibitem{Burkardt:2000za}
\textit{M.~Burkardt},
\newblock Phys. Rev. {\bf D62}, 2000, 071503, hep-ph/0005108,
\newblock Erratum-ibid.D66, 2002, 119903.

\bibitem{Ralston:2001xs}
\textit{Ralston J.~P., Pire B.},
\newblock Phys. Rev. {\bf D66}, 2002, 111501, hep-ph/0110075.

\bibitem{Saull:1999kt}
ZEUS, \textit{Saull P.},
\newblock ``Prompt photon production and observation of deeply virtual compton
scattering'', 2000,
\newblock hep-ex/0003030.

\bibitem{Amarian_2000vx}
HERMES, \textit{Amarian M.} ,
\newblock AIP Conf. Proc. 570, pp. 428-432
\newblock DESY-HERMES-00-47

\bibitem{Airapetian:2001yk}
HERMES, \textit{Airapetian A., et~al.},
\newblock Phys. Rev. Lett. {\bf 87}, 2001, 182001, hep-ex/0106068.

\bibitem{Stepanyan:2001sm}
CLAS, \textit{Stepanyan S., et~al.},
\newblock Phys. Rev. Lett. {\bf 87}, 2001, 182002, hep-ex/0107043.

\bibitem{Airapetian:2008aa}
HERMES, \textit{Airapetian A., et~al.},
\newblock JHEP {\bf 06}, 2008, 066, arXiv:0802.2499.

\bibitem{Belitsky:2001ns}
\textit{Belitsky A.~V., M{\"u}ller D., Kirchner A.},
\newblock Nucl. Phys. {\bf B629}, 2002, 323, hep-ph/0112108.

\bibitem{Airapetian:2011uq}
HERMES, \textit{Airapetian A., et~al.},
\newblock Phys. Lett. {\bf B704}, 2011, 15, arXiv:1106.2990.

\bibitem{Belitsky:2010jw}
\textit{Belitsky A.~V., M\"uller D.},
\newblock Phys. Rev. {\bf D82}, 2010, 074010, arXiv:1005.5209.

\bibitem{Belitsky:2012ch}
\textit{Belitsky A.~V., M\"uller D., Ji Y.},
Compton scattering: from deeply virtual to quasi-real, 2012,
\newblock arXiv:1212.6674.

\bibitem{Airapetian:2012pg}
HERMES, \textit{Airapetian A., et~al.},
\newblock JHEP {\bf 1210}, 2012, 042, arXiv:1206.5683.

\bibitem{Brasse:1976bf}
\textit{Brasse F., et~al.},
\newblock Nucl.Phys. {\bf B110}, 1976, 413.

\bibitem{Airapetian:2009aa}
HERMES, \textit{Airapetian A., et~al.},
\newblock JHEP {\bf 11}, 2009, 083, arXiv:0909.3587.

\bibitem{Airapetian:2010ab}
HERMES,  \textit{Airapetian A., et~al.},
\newblock JHEP {\bf 06}, 2010, 019, arXiv:1004.0177.

\bibitem{Airapetian:2012mq}
HERMES, \textit{Airapetian A., et~al.},
\newblock JHEP {\bf 1207}, 2012, 032,  arXiv:1203.6287.

\bibitem{Guidal:2008ie}
\textit{Guidal, M.},
\newblock Eur. Phys. J. {\bf A37}, 2008, 319, arXiv:0807.2355.

\bibitem{Guidal:2010ig}
\textit{Guidal, M.},
\newblock Phys. Lett. {\bf B689}, 2010, 156, arXiv:1003.0307.

\bibitem{Guidal:2010de}
\textit{Guidal, M.},
\newblock Phys. Lett. {\bf B693}, 2010, 17, arXiv:1005.4922.

\bibitem{Kumericki:2011rz}
\textit{Kumeri{\v c}ki K., M{\"u}ller D., Sch{\"a}fer A.},
\newblock JHEP {\bf 1107}, 2011, 073, arXiv:1106.2808.

\bibitem{Rad97}
\textit{Radyushkin A.~V.},
\newblock Phys. Rev. {\bf D56}, 1997 5524, hep-ph/9704207.

\bibitem{GoePolVan01}
\textit{Goeke K., Polyakov M.~V., Vanderhaeghen M.},
\newblock Prog. Part. Nucl. Phys. {\bf 47}, 2001, 401, hep-ph/0106012.

\bibitem{VanGuiGui98}
\textit{Vanderhaeghen M., Guichon P.~A.~M., Guidal M.},
\newblock Phys. Rev. Lett. {\bf 80}, 1998, 5064.

\bibitem{Freund:2001rk}
\textit{Freund A., McDermott  M.},
\newblock Phys. Rev. {\bf D65}, 2002, 074008, hep-ph/0106319.

\bibitem{Goloskokov:2005sd}
\textit{Goloskokov S.~V.,  Kroll P.},
\newblock Eur. Phys. J. {\bf C42}, 2005, 281, hep-ph/0501242.

\bibitem{Goloskokov:2007nt}
\textit{Goloskokov S.~V.,  Kroll P.},
\newblock Eur. Phys. J. {\bf C53}, 2008, 367, arXiv:0708.3569.

\bibitem{Freund:2002qf}
\textit{Freund A., McDermott M., Strikman, M.},
\newblock Phys. Rev. {\bf D67}, 2003, 036001, hep-ph/0208160.

\bibitem{Goldstein:2010gu}
\textit{Goldstein G.~R., Hernandez J.,Liuti S.},
\newblock Phys.Rev. {\bf D84}, 2011, 034007, arXiv:1012.3776.

\bibitem{Diehl:2007jb}
\textit{Diehl M., Ivanov D.~Y.},
\newblock Eur. Phys. J. {\bf C52}, 2007, 919, arXiv:0707.0351.

\bibitem{Kumericki:2008di}
\textit{Kumeri{\v c}ki K., M{\"u}ller D., Passek-Kumeri{\v c}ki K.},
\newblock Eur. Phys. J. {\bf C58}, 2008 193, arXiv:0805.0152.

\bibitem{Kumericki:2007sa}
\textit{Kumeri{\v c}ki K., M{\"u}ller D., Passek-Kumeri{\v c}ki K.},
\newblock Nucl. Phys. B {\bf 794}, 2008, 244, hep-ph/0703179.

\bibitem{Kumericki:2009uq}
\textit{Kumeri{\v c}ki K., M{\"u}ller D.},
\newblock Nucl. Phys. {\bf B841}, 2010, 1, arXiv:0904.0458.

\bibitem{Mueller:2005ed}
\textit{M{\"u}ller D., Sch{\"a}fer A.},
\newblock Nucl. Phys. {\bf B739}, 2006, 1, hep-ph/0509204.

\bibitem{Shuvaev:1999fm}
\textit{Shuvaev A.~G.},
\newblock Phys. Rev. {\bf D60}, 1999, 116005, hep-ph/9902318.

\bibitem{Polyakov:2002wz}
\textit{Polyakov  M.~V., Shuvaev A.~G.},
\newblock ``{On} `dual' parametrizations of generalized parton distributions'', 2002,
\newblock hep-ph/0207153.

\bibitem{Kirch:2005tt}
\textit{Kirch M., Manashov A., Sch{\"a}fer A.},
\newblock Phys. Rev. {\bf D72}, 2005, 114006, hep-ph/0509330.

\bibitem{Bechler:2009me}
\textit{Bechler C., M{\"u}ller  D.},
\newblock ``Generic modelling of non-perturbative quantities and a description
  of hard exclusive $\pi^+$ electroproduction'', 2009,
\newblock arXiv:0906.2571.

\bibitem{James:1975dr}
\textit{James F., Roos M.},
\newblock Comput. Phys. Commun. {\bf 10}, 1975, 343.

\bibitem{KirMue03}
\textit{Kirchner A., M{\"u}ller, D.},
\newblock Eur. Phys. J. {\bf C32}, 2003, 347, hep-ph/0302007.

\bibitem{Teryaev:2005uj}
\textit{Teryaev O.~V.},
\newblock Analytic properties of hard exclusive amplitudes, 2005,
\newblock hep-ph/0510031.

\bibitem{Polyakov:1999gs}
\textit{Polyakov  M.~V., Weiss C.},
\newblock Phys. Rev. {\bf D60}, 1999, 114017, hep-ph/9902451.

\bibitem{Diehl:2004cx}
\textit{Diehl M., Feldmann T., Jakob R., Kroll P.},
\newblock Eur. Phys. J. {\bf C39}, 2005, 1, hep-ph/0408173.

\bibitem{Kroll:2012sm}
\textit{Kroll P., Moutarde H., Sabatie F.},
From hard exclusive meson electroproduction to deeply virtual Compton scattering,  2012,
\newblock arXiv:1210.6975.

\bibitem{Mankiewicz:1998kg}
\textit{Mankiewicz L., Piller G., Radyushkin A.},
\newblock Eur. Phys. J. {\bf C 10}, 1999, 307, hep-ph/9812467.

\bibitem{Frankfurt:1999xe}
\textit{Frankfurt L.~L., Polyakov M.~V., Strikman M., Vanderhaeghen M.},
\newblock Phys. Rev. Lett. {\bf 84}, 2000, 2589, hep-ph/9911381.

\bibitem{Munoz_Camacho:2006hx}
Jefferson Lab Hall A, \textit{Camacho C.~M., et~al.},
\newblock Phys. Rev. Lett. {\bf 97}, 2006, 262002, nucl-ex/0607029.

\bibitem{Girod:2007aa}
CLAS, \textit{Girod F.~X., et~al.},
\newblock Phys. Rev. Lett. {\bf 100}, 2008, 162002, arXiv:0711.4805.

\bibitem{Chen:2006na}
CLAS, \textit{Chen S., et~al.},
\newblock Phys. Rev. Lett. {\bf 97}, 2006, 072002, hep-ex/0605012.

\bibitem{Aktas:2005ty}
H1, \textit{Aktas A., et~al.},
\newblock Eur. Phys. J. {\bf C44}, 2005, 1, hep-ex/0505061.

\bibitem{Aaron:2009ac}
H1, \textit{Aaron F., et~al.},
\newblock Phys.Lett. {\bf B681}, 2009, 391, arXiv:0907.5289.

\bibitem{Chekanov:2008vy}
ZEUS, \textit{Chekanov S., et~al.},
\newblock JHEP {\bf 05}, 2009, 108, arXiv:0812.2517.

\bibitem{Chekanov:2003ya}
ZEUS, \textit{Chekanov S., et~al.},
\newblock Phys. Lett. {\bf B573}, 2003, 46, hep-ex/0305028.

\bibitem{Airapetian:2007aa}
HERMES, \textit{Airapetian A., et~al.},
\newblock Phys. Lett. {\bf B659}, 2008, 486, 0707.0222 [hep-ex].

\bibitem{Anikin:2000em}
\textit{Anikin I.~V., Pire B., Teryaev O.~V.},
\newblock Phys. Rev. {\bf D 62}, 2000, 071501, hep-ph/0003203.

\bibitem{Belitsky:2000vx}
\textit{Belitsky A.~V., M{\"u}ller D.},
\newblock Nucl. Phys. {\bf B589}, 2000, 611, hep-ph/0007031.

\bibitem{Kivel:2000cn}
\textit{Kivel N., Polyakov M.~V.},
\newblock Nucl. Phys. {\bf B600}, 2001, 334, hep-ph/0010150.

\bibitem{Braun:2011zrBraun:2011dg}
\textit{Braun V.~M.,  Manashov A.~N.},
\newblock Phys. Rev. Lett. {\bf 107}, 2011, 202001, arXiv:1108.2394;
\newblock JHEP {\bf 01}, 2012, 085, arXiv:1111.6765;

\bibitem{Braun:2012bgBraun:2012hq}
\textit{Braun V., Manashov A., Pirnay B.},
\newblock Phys.Rev. {\bf D86}, 2012, 014003, arXiv:1205.3332;
``Finite-$t$ and target mass corrections to deeply virtual Compton scattering'', 2012,
\newblock arXiv:1209.2559.

\end{thebibliography}

\end{document}